\documentclass[english,
]{aa}
\usepackage{epsf,amsfonts,amssymb,graphicx,fancyheadings,caption}
\usepackage{natbib}
\usepackage{longtable}
\usepackage{babel}
\usepackage{txfonts}
\usepackage{subfig}
\usepackage{rotating}
\usepackage{enumitem}

\bibpunct{(}{)}{;}{a}{}{,}   

\begin{document}

\title{HdC and EHe stars through the prism of Gaia DR3:}
\subtitle{3D distribution and Gaia's chromatic PSF effects}

\author{
P.~Tisserand\inst{1},
C.~L.~Crawford\inst{2},
J.~Soon\inst{3},
G.~C.~Clayton\inst{4},
A.~J.~Ruiter\inst{5},
I.~R.~Seitenzahl\inst{5}
}

\institute{
Sorbonne Universit\'es, UPMC Univ Paris 6 et CNRS, UMR 7095, Institut d'Astrophysique de Paris, IAP, F-75014 Paris, France \and
Sydney Institute for Astronomy (SIfA), School of Physics, University of Sydney, NSW 2006, Australia \and
Research School of Astronomy and Astrophysics, Australian National University, Cotter Rd, Weston Creek ACT 2611, Australia \and
Department of Physics \& Astronomy, Louisiana State University, Baton Rouge, LA 70803, USA \and
School of Science, University of New South Wales, Canberra, ACT 2600, Australia
}


\offprints{Patrick Tisserand; \email{tisserand@iap.fr}}

\date{}


\abstract {Upon its release the Gaia DR3 catalogue has led to tremendous progress in multiple fields of astronomy by providing the complete astrometric solution for nearly 1.5 billion sources.}
{We analysed the photometric and astrometric results for Hydrogen-deficient Carbon (HdC), Extreme Helium (EHe), and DYPer type stars to identify any potential biases. This analysis aimed to select stars suitable for kinematic and spatial distribution studies.}
{We investigated the information obtained from the Gaia image parameter determination (IPD) process, which was cross-matched with Gaia light curves. One main objective was to understand the impact of photometric declines in R Coronae Borealis (RCB) stars on Gaia astrometry.}
{Based on the evidence gathered, we have reached the conclusion that the astrometric fits for numerous RCB stars, including R CrB itself, are not valid due to the Gaia point spread function (PSF) chromaticity effect in both shape and centroid. The astrometric results of all stars with a significant time-dependent colour variation should be similarly affected. RCB stars might thus be promising sources to correct this effect in future Gaia releases. Furthermore, after validating the Gaia astrometric results for 92 stars, we observed that the majority of HdC and EHe stars are distributed across the three old stellar structures, the thick disk, the bulge and the halo. However, we have also uncovered evidence indicating that some HdC and EHe stars exhibit orbits characteristic of the thin disk. This is also particularly true for all DYPer type stars under study. Finally, we have produced a list of star memberships for each Galactic substructure, and provided a list of heliocentric radial velocities and associated errors for targets not observed by Gaia DR3.}
{We are beginning to observe a relationship between kinematics, stellar population, and metallicity in RCB and EHe stars. That relation can be explained, within the double degenerate scenario, by the large range in the delay time distribution expected from population synthesis simulations, particularly through the HybCO merger channel.}


\authorrunning{Tisserand, P.}
\titlerunning{HdC and EHe stars through the prism of Gaia DR3: 3D distributions and Gaia's chromatic PSF effect}

\maketitle

\section{Introduction \label{sec_intro}}

The Gaia space observatory has opened a new door to revolutionise our understanding of the tumultuous formation history and substructures of our galaxy. Its third release \citep{2021A&A...649A...1G,2021A&A...649A...2L,2021A&A...649A...3R} includes astrometric, photometric (Gaia bands: G, BP and RP), and spectroscopic measurements of nearly 1.5 billion sources observed over a span of 34 months (25 July 2014 to 28 May 2017). This release represents only a quarter of the entire dataset expected to be collected by 2025, the anticipated end date of the mission. Rare stellar populations are expected to benefit significantly from this extensive wealth of new information.

Hydrogen-deficient Carbon (HdC) stars are rare supergiant stars with an effective temperature (T$_{eff}$) ranging between 3500 and 8500 K \citep{2023MNRAS.521.1674C}. Accumulating evidence suggests a cataclysmic origin for these stars, which was originally theorised by \citet{1984ApJ...277..355W} and \citet{1998MNRAS.296.1019H}. Indeed, they are strongly suspected to result from the merger of one CO- plus one He-rich white dwarf (WD; double degenerate scenario) \citep{2012JAVSO..40..539C, 2011MNRAS.414.3599J} with a combined total mass between $\sim$0.6 and $\sim$1.05 solar masses \citep[see Fig. 13]{2022A&A...667A..83T} as predicted by population synthesis simulations of close binary systems \citep{2015ApJ...809..184K} after various mass transfer phases.

Historically, HdC stars have been divided into two groups: the R Coronae Borealis (RCB) and the dustless HdC (dLHdC) stars. By and large, they share the same spectroscopic characteristics, but the former is famously known for its heavy dust production, leading to large photometric declines (up to 9 mag in V) due to obscuration by newly formed dust clouds. In contrast, dLHdC stars do not form any dust. The number of known dLHdC stars has recently increased significantly thanks to the Gaia survey \citep{2022A&A...667A..83T}. They have been found to be less luminous than their dusty counterparts by an average of $\sim$2 mag, indicating that they could originate from WD mergers of a lower total mass compared to that of RCB stars. Furthermore, dLHdC stars present possibly even lower O$^{16}$/O$^{18}$ isotopic ratios \citep{2022A&A...667A..84K} than those measured in RCB stars \citep{2007ApJ...662.1220C, 2010ApJ...714..144G}. However, the boundary between RCB and dLHdC stars is not clear-cut, as a few dLHdC stars are found to be of a comparable brightness to RCB stars and some (F75, F152, C526, and A166) even have been found to produce small amounts of dust as a weak infra-red (IR) excess was detected. Therefore, some dLHdC stars could indeed be RCB stars that have temporarily ceased dust production.

On the warmer end of the scale (T$_{eff}>$8500 K), there is another group of supergiant stars known as Extreme Helium (EHe) stars, whose atmospheres are also nearly devoid of hydrogen. These stars are believed to be evolutionarily linked to HdC stars, representing the contracting phase following the helium shell-burning giant stage (i.e. HdC stars) before becoming heavy WDs \citep{2002MNRAS.333..121S, 2014MNRAS.445..660Z, 2019ApJ...885...27S}. Some atypical WDs, the sub-class of carbon-polluted DQ WDs, were recently proposed to be the best candidates for resulting from such double-degenerate mergers that failed to explode as Type Ia supernovae \citep{2015ASPC..493..547D,2023MNRAS.520.6299K}. On the cooler side (T$_{eff}<$3500 K), the DYPer type stars (with DY Persei as the prototype) are increasingly thought to be associated with HdC stars \citep{2007A&A...472..247Z,2018ApJ...854..140B,2023ApJ...948...15G}. However, spectroscopic studies of DY Persei in the visible and IR show that it is less hydrogen-deficient than HdC stars \citep{2009ARep...53..187Y,2013ApJ...773..107G}. These stars also produce dust, but their photometric declines are slower and shallower compared to those observed in RCB stars.

For a long time, spatial distribution studies of HdC stars had been conducted with the assumption of an absolute magnitude ranging between M$_V\sim$-5 and -4 mag, thanks to the few RCB stars found in the Large Magellanic Cloud (LMC). Radial velocity (RV) studies on a small sample of HdC and EHe stars suggested their potential membership in the bulge, the halo, or an old disk/Population I group of stars for the HdC stars, while the EHe stars appeared to be associated with the bulge/Population II stars, indicating a different spatial distribution \citep{1986hdsr.proc..9Drilling}. However, \citet{1998PASA...15..179C} demonstrated, using the proper motion dataset released by the Hipparcos astrometric mission, that HdC and EHe stars exhibit similar velocities dispersions, indicating a predominantly bulge distribution for both groups of stars. Unfortunately, none of the Hipparcos parallax measurements yielded significant results for the 21 observed HdC stars. Subsequently, it was discovered that the absolute magnitude of RCB stars varies with their effective temperature \citep{2001ApJ...554..298A,2009A&A...501..985T, 2022A&A...667A..83T}, and that numerous RCB stars exist within the Galactic bulge \citep{2005AJ....130.2293Z,2008A&A...481..673T, 2020A&A...635A..14T}. Finally, based on the Gaia early data early release 3 (EDR3), \citet{2021ApJ...921...52P} found that RCB stars have tightly bounded orbits, placing them in the inner regions of the Milky Way with a radius of less than or about 6.0 kpc, while most EHe stars have orbits extending beyond the 6 kpc radius. However, they note that there are exceptions in both groups of stars.

This article is published in association with another study that focuses on the intrinsic radial velocity variations and diverse dust production rate observed in HdC, EHe, and DYPer type stars using Gaia data release 3 (DR3) \citep{Tiss2023a}. 

In Section~\ref{sec_HdC_DR3}, we first explore the photometry provided by Gaia DR3 for our targets, with a specific focus on RCB stars. We investigate how their characteristic photometric decline phases have influenced the automatic astrometric fits. Following this, in Section~\ref{sec_AstromSelection}, we describe our selection process for stars that will be used in the subsequent 3D kinematic and spatial distribution studies (Section~\ref{sec_spatialdistrib}). In that section, we provide also a list of heliocentric radial velocity measurements and associated errors for those not published in Gaia DR3. We then discuss in Section~\ref{sec_discussion} the connection between the membership of HdC, EHe, and DYPer type stars in Galactic substructures and their metallicity within the context of the double degenerate scenario using delay time distributions resulting from population synthesis simulations. Finally, we summarise our findings in Section~\ref{sec_Concl}.

\section{HdC stars and the Gaia DR3 datasets \label{sec_HdC_DR3}}

\begin{figure}
\centering
\includegraphics[scale=0.32]{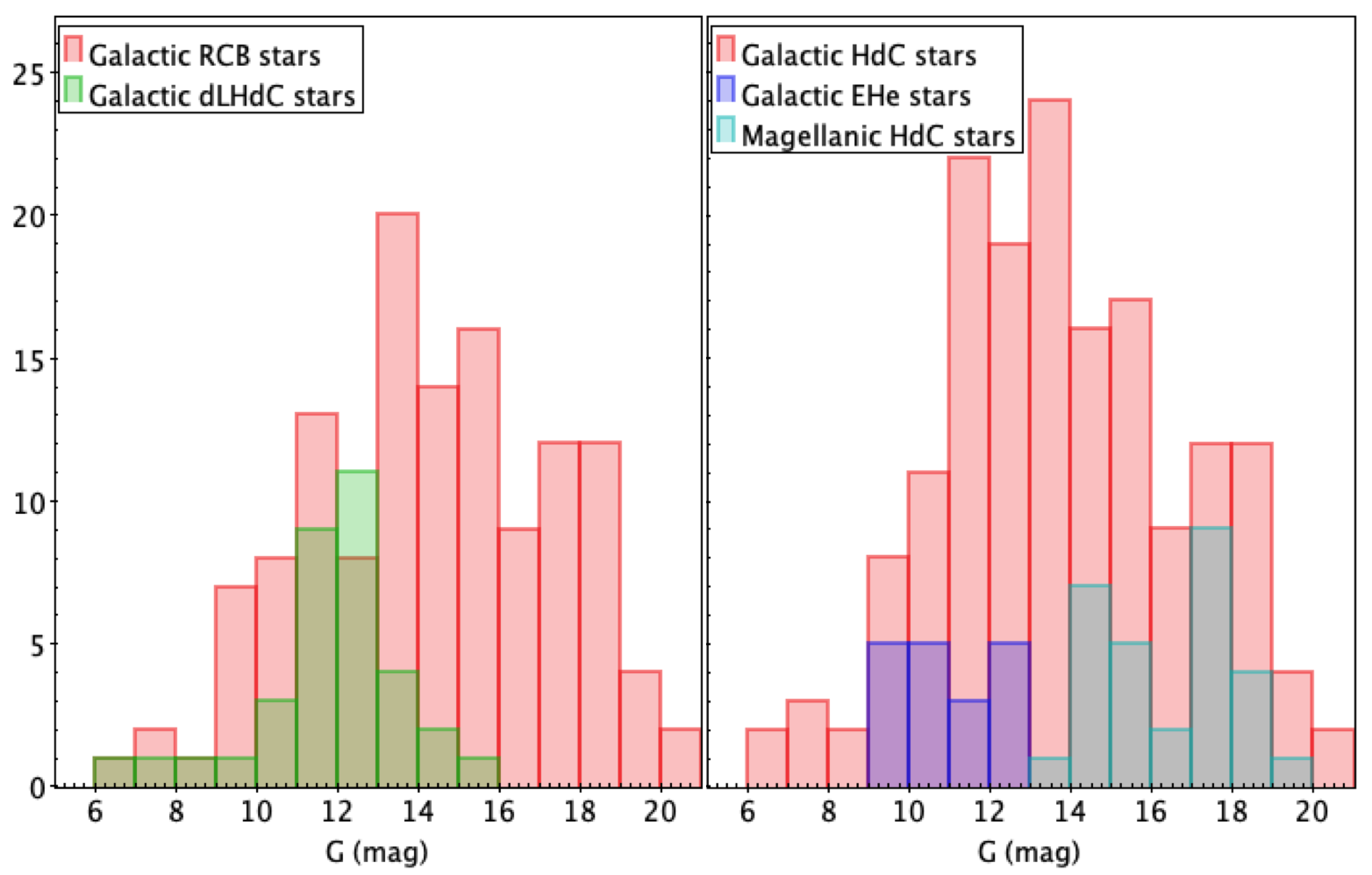}
\caption{Distributions of the G-band Gaia DR3 magnitudes for our targets. Left: Galactic RCB and dLHdC stars. Right: Galactic HdC and EHe stars and Magellanic HdC stars.} \label{fig_DistribGband}
\end{figure}

\begin{figure}
\centering
\includegraphics[scale=0.32]{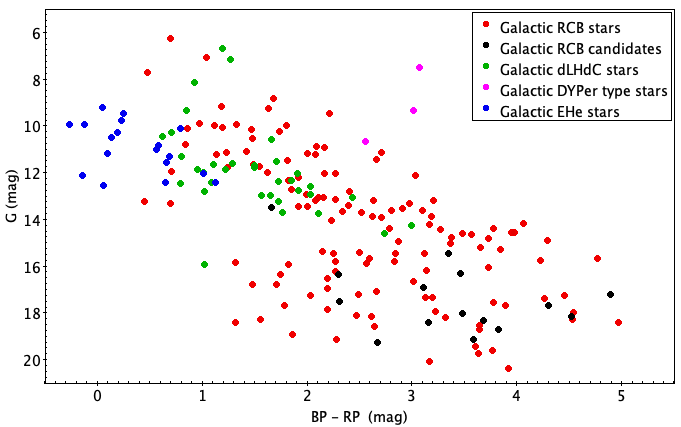}
\caption{Colour magnitude diagram G vs (BP-RP) for all Galactic targets colour-coded with their type.} \label{fig_GAIA_HR}
\end{figure}

In Gaia, the Image Parameter Determination (IPD) process performs fits on stars passing through each astrometric CCD to determine their centroid position and integrated flux. This is aided by a library of 1D line spread functions (LSF) and point spread functions (PSF) which, in Gaia DR3, are colour-dependent \citep{2021A&A...649A...2L,2021A&A...649A..11R}. For bright sources (G$<$13 mag) a full two-dimensional window is read, while pixels are collapsed in the across-scan direction during the reading process for fainter targets, leading to a one-dimensional set of data \citep{2023A&A...674A..25H}. Furthermore, for sources brighter than 12 mag in G, a gating scheme reduces the integration time to prevent and/or minimise the number of pixels affected by saturation.

We first studied the impact of blending and confusion as a majority of our targets are located towards crowded fields. Then, our study focused on the effect of variability and reddening on astrometry results, using IPD parameters, photometry and published light curves. We validated criteria to select targets suitable for further studies of distances and proper motions.

 \subsection{Targets studied}

We studied the results published in the Gaia DR3 release for the same list of known Galactic and Magellanic HdC, EHe and DYPer type stars given in \citet{Tiss2023a}. We added to that list 15 strong Galactic RCB candidates, most of them being listed in \citet{2020A&A...635A..14T}.

Our search was conducted in the Gaia DR3 database using a 1 arcsec radius to find our Galactic and Magellanic targets. We did not find a counterpart for only three Magellanic stars: the known RCB star MACHO 6.6575.13, and the strong RCB candidates EROS2-SMC-RCB-4, and [KDM2001] 2730. They remained in a faint phase during the Gaia DR3 time range, which was below its magnitude limit. All Galactic targets have an entry in Gaia DR3, and five of them had close secondary entries that were removed: EROS2-GC-RCB-11, EROS2-GC-RCB-9, MSX LMC 1795, EROS2-LMC-DYPer-4 and MACHO 81.8394.1358. The closest neighbour of the first target was even located at an angular distance of only 0.34 arcsec. As the window size used on the Gaia astrometric fields is 1.1" $\times$ 2.1", we should expect some effects for those objects.

 \subsection{Gaia DR3 photometry and RCB stars' declines}

\begin{figure}
\centering
\includegraphics[scale=0.33]{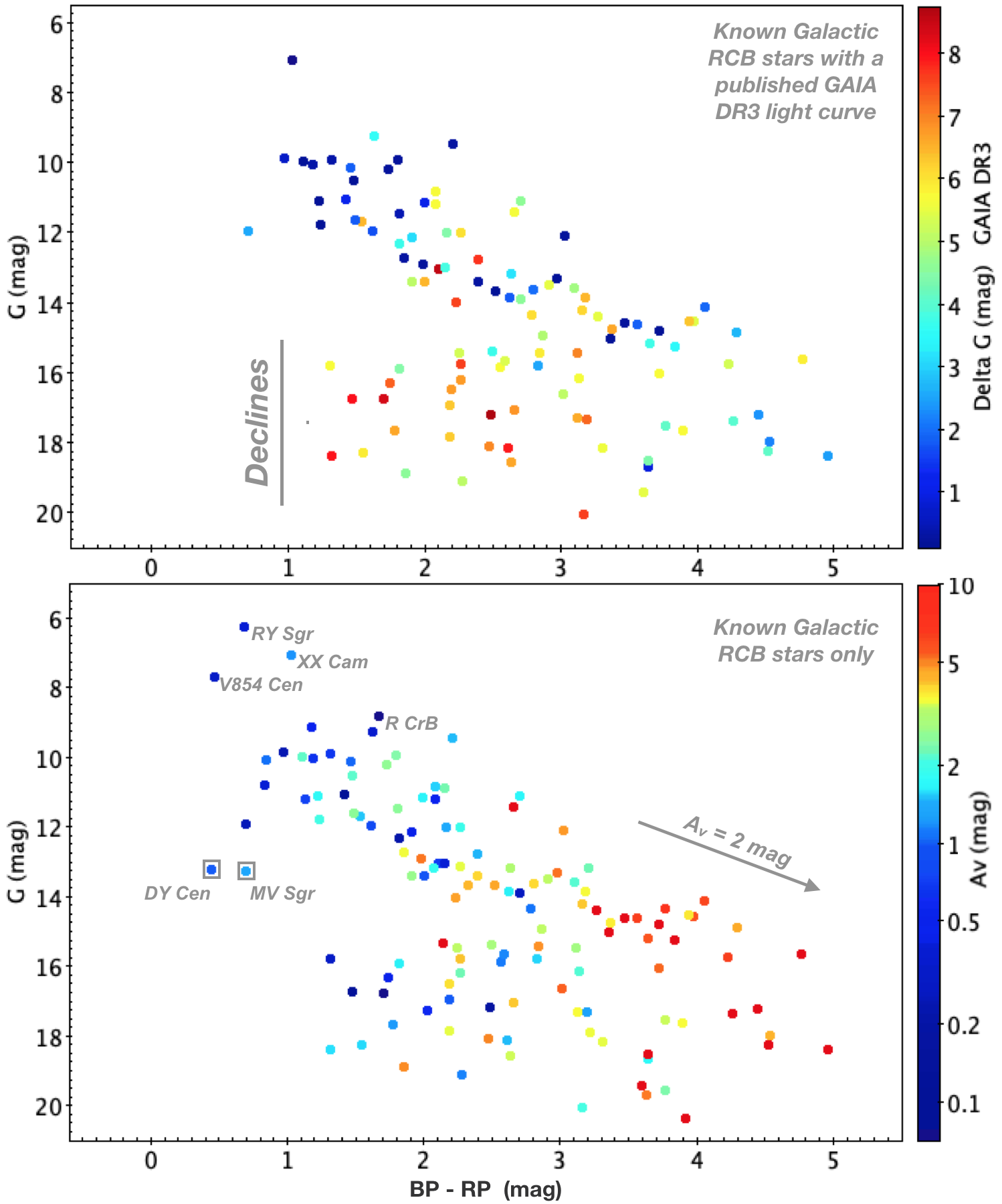}
\caption{Colour magnitude diagram G vs (BP-RP) with known RCB stars. Top: Diagram colour-coded by the maximum G-band variation observed in their respective Gaia light curve. The luminosity region containing RCB stars that went under a long decline events is indicated. Bottom: Diagram colour-coded by the visual interstellar extinction Av that was limited at a value of 10 mag. The arrow indicates the reddening effect of a 2 mag extinction.}
\label{fig_GAIA_HR_DGandAv}
\end{figure}

The median Gaia G-band apparent magnitude of known Galactic EHe and HdC stars is around 14 mag, as shown in figures~\ref{fig_DistribGband} and \ref{fig_GAIA_HR}. EHe stars have a bluer Gaia BP-RP median colour index, followed by dLHdC stars, and then RCB stars. As an apparent magnitude of 14 roughly corresponds to the magnitude of the brightest Magellanic HdC stars, two factors explain the magnitude and colour distributions of Galactic HdC stars. Firstly, the interstellar extinction for the 20 reddest Galactic stars is very high (A$_V>$6 mag), as indicated in the bottom colour-magnitude diagram in Figure~\ref{fig_GAIA_HR_DGandAv}. In calculating interstellar extinction parameters, one important data source was the 3D dust map created by \citet{2019ApJ...887...93G}. However, in cases where data from Green et al. was not accessible, we relied on E(B-V) reddening values provided by \citet{2011ApJ...737..103S} covering the entire sky. We used a value of 3.1 for the R$_V$ extinction factor. Secondly, about two-thirds of known HdC stars are the heavy dust producers RCB stars and some of them underwent a decline phase during the Gaia survey. As the Gaia G-band magnitudes are calculated as the mean of the G flux series, the declines cause the mean Gaia G magnitude to be biased towards the fainter side. Using the Gaia light curves, we calculated the overall maximum brightness variation ($\bigtriangleup$Gmax) observed in the 34 months of Gaia DR3 data. This is shown in the top colour-magnitude diagram of Figure~\ref{fig_GAIA_HR_DGandAv}. Most of the fainter RCB stars (G $>$ 15 mag) experienced significant luminosity changes, with a slightly reddened BP-RP colour index between 1.2 and 3.5 mag. More than 65\% of known RCB stars showed $\bigtriangleup$Gmax variations higher than 3 mag, with a peak around $\sim$6 mag. In contrast, only 18\% remained bright, with a maximum variation less than 0.5 mag.

\begin{figure}
\centering
\includegraphics[scale=0.33]{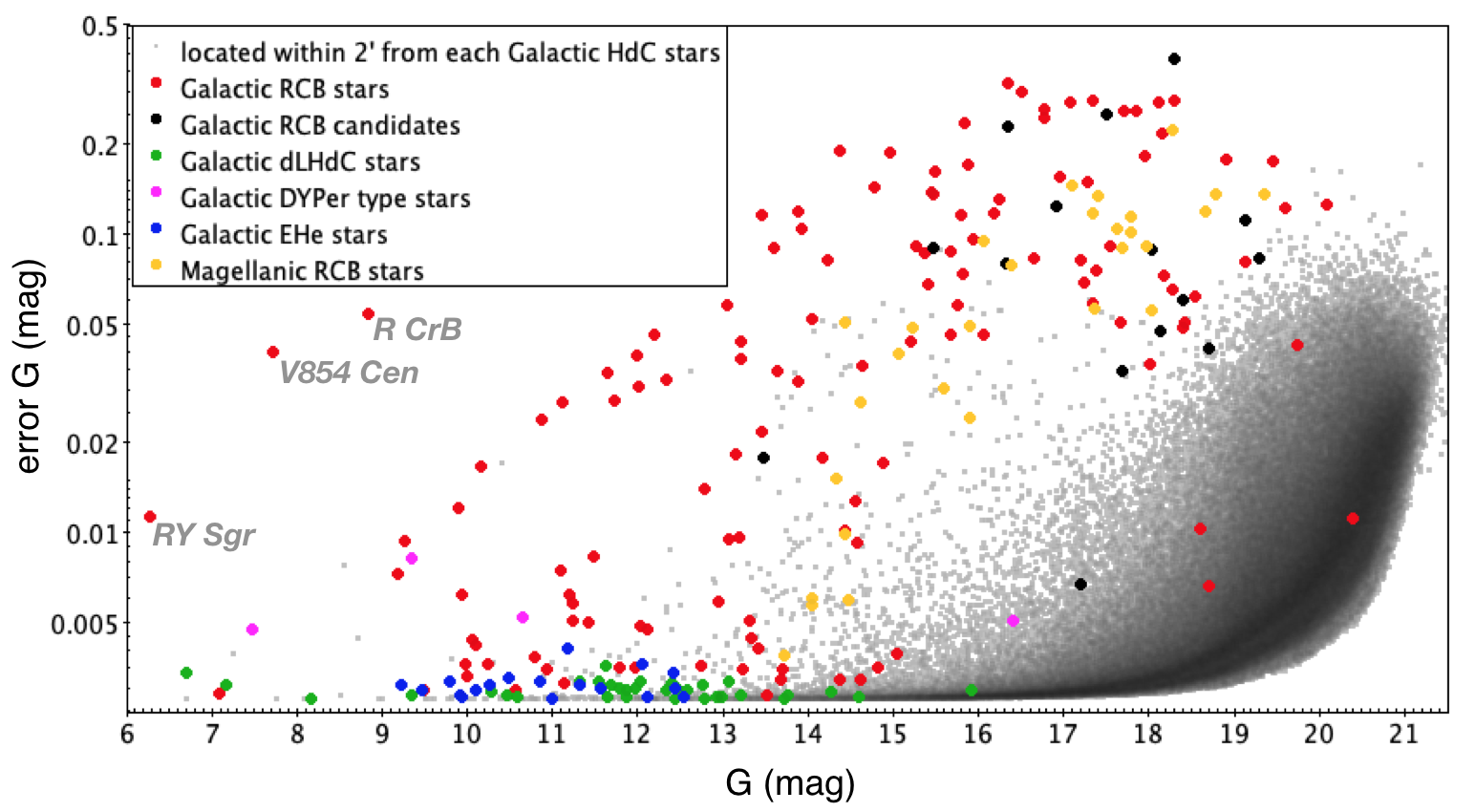}
\caption{Diagram representing the standard error (log scale) versus the mean G-band magnitude for all our Galactic and Magellanic targets (large coloured dots) and for stars located within 2 arcmins of any known Galactic HdC stars (small grey dots). The names of three of the brightest RCB stars are indicated.}
\label{fig_ErrGvsG}
\end{figure}

Photometric declines can also be detected in the G-band photometric errors. These errors are the standard error calculated over the entire Gaia light curve. We compared them to those of classical stars in the surrounding area. Figure~\ref{fig_ErrGvsG} shows the results and highlights the large variability of RCB stars. While EHe and dLHdC stars have G-band errors lower than 3 mmag, similar to the comparison sample at the same brightness, over 55\% of Galactic RCB stars have errors higher than 30 mmag, corresponding to the maximum expected error for pulsations of 0.2-0.4 mag peak-to-peak amplitude. RCB stars stand out from the classical stars distribution, with one having a standard error as high as 0.4 mag. The faintest RCB star in the figure, EROS2-CG-RCB-9, has a standard error similar to classical stars at its brightness, because it remained in a faint phase throughout Gaia DR3.

\subsection{High density fields \label{subsec_HighDensityFields}}

\begin{figure}
\centering
\includegraphics[scale=0.26]{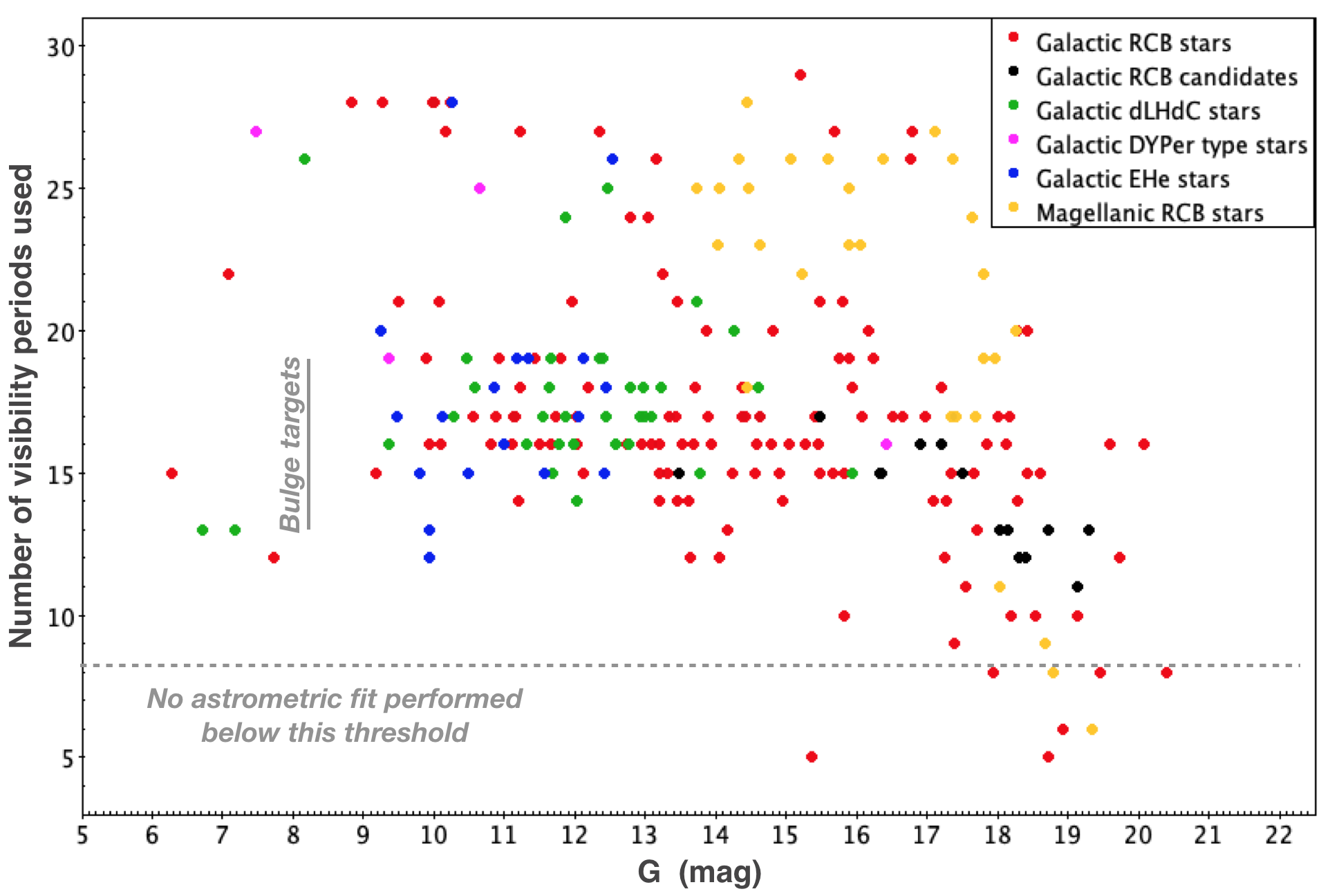}
\caption{Diagram representing the number of visibility periods versus the mean G-band magnitude for all our targets colour-coded with their types.}
\label{fig_NbVisvsG}
\end{figure}

Most known HdC stars are located towards the Galactic bulge. The spatial distributions of RCB and dLHdC stars are shown in \citet[Fig. 16]{2020A&A...635A..14T} and in \citet[Fig. 3]{2022A&A...667A..83T}, respectively. The Gaia astrometric instrument has been designed to cope with object densities up to some 750,000 stars per square degree. Consequently, in crowded fields, only the brightest stars were observed and some RCB star astrometric measurements might have been missed during decline phases. We performed a study to assess the effects of blending and confusion on a few Gaia parameters.

We analysed two parameters from the IPD module in the core processing, ipd\_frac\_multi\_peak and ipd\_frac\_odd\_win, which correspond respectively to the percentage of successful IPD windows for which more than one peak was found, and to the percentage of transits on the astrometric CCDs with truncated windows or multiple gates. The first one is generally used to reveal double stars, while the second indicates the presence of a bright close neighbour. We found that six targets have ipd\_frac\_multi\_peak values higher than 20\%, that is, the RCB stars ASAS-RCB-5, WISE J174257.19-362052.1, EROS2-CG-RCB-11 and -13, and the DYPer type stars EROS2-LMC-DYPer-4 and -6. These objects are most certainly affected by a blend due to a star located very close to their line of sight that cannot be resolved at the Gaia resolution. The second parameter has shown high values for the RCB stars ASAS-RCB-5, EROS2-CG-RCB-9, -11 and -13, but also for the DYPer type star 2MASS J17524872-2845190. A visual study of their respective charts confirms the presence of a star as bright as the target, located within 2 arcsec from it.

One main parameter that influences the final astrometric fit is the number of visibility periods used for each target. It corresponds to the number of groups of observations separated by at least 4 days. The higher the number, the higher the astrometric accuracy achieved would be. This number is presented as a function of the G magnitude in Figure~\ref{fig_NbVisvsG} for each group of targets studied. It reaches a median value of 16 for a large majority of our targets as many are located around the Galactic bulge, and increases up to 28 for the few located away from that Galactic region (for example: R CrB, S Aps, Y Mus, AO Her, V482 Cyg) and towards the Magellanic Clouds. The number of visibility periods used decreases with the mean G magnitude as some measurements were not considered for the RCB stars located in a crowded field that went through a faint phase due to the technical limitation related to stellar density mentioned above. An astrometric fit was performed only for the stars for which more than 8 visibility periods were accumulated. We found that six Galactic and three Magellanic RCB stars did not pass that threshold: EROS2-CG-RCB-9, MACHO 301.45783.9, WISE J163450.35-380218.5, WISE J172553.80-312421.1, WISE J174645.90-250314.1, WISE J184246.26-125414.7, EROS2-LMC-RCB-6, MSX SMC 014 and WISE J005113.58-731036.3. Only their position coordinates were thus released.

\subsection{Astrometric fit and the effect of PSF chromaticity}

The parallax measurements made by Gaia DR3 for stars in crowded sky areas such as the Galactic bulge have an uncertainty of less than 0.04 mas for 90\% of monitored bright stars (G$<$14 mag). A 3 sigma parallax detection could then be expected up to the centre of our galaxy ($\sim$8.2 kpc). At this distance, an HdC star with absolute magnitude ranging from -5 to -3 mag would have an apparent magnitude of 9.6 to 11.6 mag, to which one needs to add the contribution due to interstellar extinction.

Our study focuses on the potential astrometric biases occurring with RCB stars caused by their sudden and large photometric declines. Despite evidence that some active RCB stars could remain in a bright phase for up to 10 years (i.e., R CrB between 1923-1933), and that a few others (Y Mus, XX Cam, and HD 175893) have never or rarely been observed to go through decline phases for the past century despite undergoing active dust production\footnote{A circumstellar dust shell is detectable in the mid-IR \citep{2012A&A...539A..51T}}, we found that $\sim$80\% of RCB stars went through some decline events during the 34 months period covered by Gaia DR3 (Fig.~\ref{fig_GAIA_HR_DGandAv}, top). These events are therefore not uncommon.

Most Gaia sources underwent a 5-parameter astrometric fit (position, parallax, proper motion) using a fixed BP-RP colour as input. However, for fainter sources with limited colour information, a 6-parameter fit was performed by adding a pseudo-colour as a free parameter. Overall, up to one-third of our targets received a 6-parameter astrometric fit: one Galactic EHe star and 54 (16) Galactic (Magellanic) HdC and DYPer type stars.

The colour information used in the fit is crucial because the PSF changes in position and shape based on chromaticity. A study by \citet[Fig. 7]{2006MNRAS.367..290J} showed a shift of up to 1 milli-arcsec between two stars with a 2 mag difference in visual colour index, due to a change in coma-like optical aberration in the Gaia's spacecraft model used at the time. Consequently the chosen PSF model to measure star position depends on the star's spectral energy distribution. We do not know the scale of the above discussed shift for the actual Gaia telescope. Gaia DR3 assumes the colour remains the same in all observations, which is indeed a good approximation for most sources, but that choice should create astrometric biases for variable objects that, similarly to RCB stars, do change in colour \citep{2021A&A...649A...2L}. 

\subsubsection{Colour changes observed in Gaia DR3 light curves}

The Gaia DR3 release provided light curves in G, BP, and RP bandpasses for sources that passed quality criteria and were selected as variable stars \citep{2022arXiv220606416E,2022arXiv220701946G}. Out of the 183 Galactic and 58 Magellanic HdC and DYPer type stars, light curves are available for 130 and 46, respectively. Among the dLHdC and EHe stars, only the most variable have light curves available, i.e., 5 in each group. They display small-scale photometric variability with a peak-to-peak amplitude that is at maximum $\sim$0.1 mag.

\begin{figure}
\centering
\includegraphics[scale=0.52]{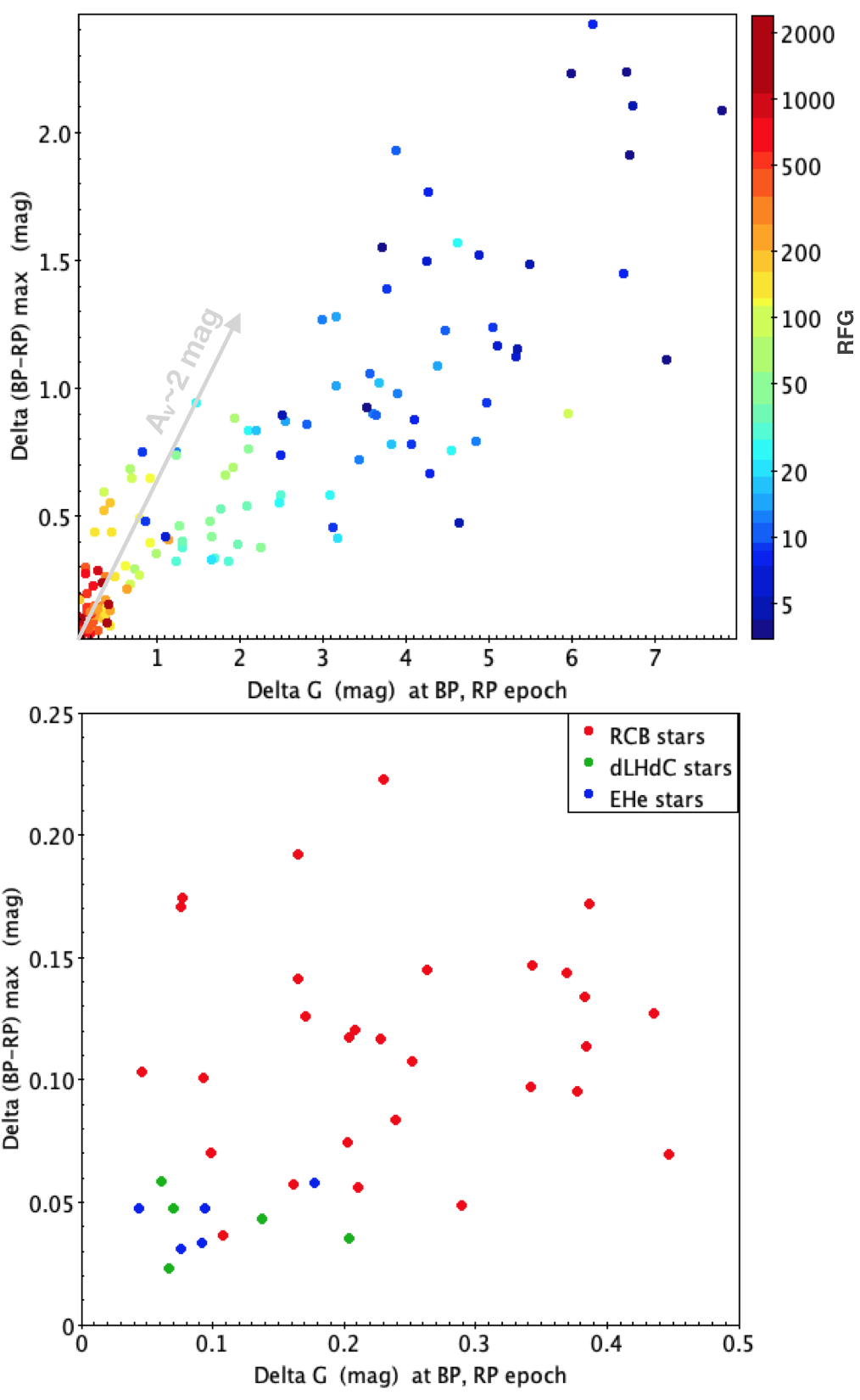}
\caption{Maximum colour variation BP-RP versus the corresponding G-band variation as observed in the Gaia DR3 light curve. Top: All our targets are represented and the diagram is colour-coded by the ratio G mean flux to its standard error (RFG). For comparison, we represented the effect of a 2 mag visual interstellar extinction with a grey arrow. The reddening effect occurring during RCB stars' photometric declines is on average two times higher. Bottom: Zoom on the targets with the least variation ($\bigtriangleup$G $<$ 0.5 mag) colour-coded with their types.}
\label{fig_DeltaBPRP_DeltaG}
\end{figure}

For all targets with a Gaia light curve, we calculated the maximum BP-RP colour variation ($\Delta$BPRPmax) using simultaneous BP and RP measurements and the corresponding $\Delta$G variation. The result, shown in Figure~\ref{fig_DeltaBPRP_DeltaG}, confirms the strong reddening effect caused by dust clouds made of amorphous carbon grains during the decline phases of RCB stars. This reddening is in average two times higher than the one resulting from interstellar extinction. However, it is also not uniform\footnote{Some occasional off-measurements at fainter magnitudes exist especially in the BP bandpass. Those have made our calculation of $\Delta$BPRPmax difficult. We remove the targets affected for that particular study.} among RCB stars and may spread up to 1.5 mag in $\Delta$BPRPmax for a maximum observed drop $\Delta$G of $\sim$7 mag. The spread could be due to differences in the section and fraction of chromosphere obscured. Additionally, close blending with bluer stars in crowded fields may mitigate the reddening effect. The largest reddening variation observed reached $\sim$2.4 mag.

For the targets presenting the lowest photometric variations ($\Delta$G $<$ 0.5 mag), we are seeing the effect of pulsations at maximum brightness. The total amplitude of the pulsations ranges mostly between 0.1 to 0.4 mag for RCB stars and is around 0.1 mag for the most variable dLHdC and EHe stars. The change in brightness during pulsations is also chromatic, with maximum BP-RP colour variations ranging from 0.05 to 0.2 mag for RCB stars, and around 0.04 mag for the few dLHdC and EHe stars represented.

\subsubsection{The effect on Gaia IPD parameters of RCB stars reddening during photometric declines}

We examined the impact of significant colour variations on the first IPD parameter, the IPDgofha (abbreviation for ipd\_gof\_harmonic\_amplitude), which reflects the amplitude in goodness-of-fit values ($\tilde{\chi}^2$) from PSF fitting of retained transit data versus scan direction angle. To calculate its value, a sinusoidal model fit to the natural logarithm of reduced $\chi^2$ (determined for each CCD observation) as function of scan angle (i.e., the ipd\_gof\_harmonic\_phase parameter) was performed \citep{2023A&A...674A..25H}. The equation\footnote{The reference level C$_0$ of the equation is not published in Gaia DR3.} is given by \citet[Eq.1]{2023A&A...674A..25H} who studied possible biases observed with this parameter values due to special circumstances. For example, a high amplitude often indicates a double source, with worse fit results towards the second star's direction, and the phase angle gives the pair's position angle. IPDgofha was thus used in an analysis to select Quasars by \citet{2021A&A...649A...5F} to filter out non-single objects. They used a threshold of 0.1.

\begin{figure}
\centering
\includegraphics[scale=0.30]{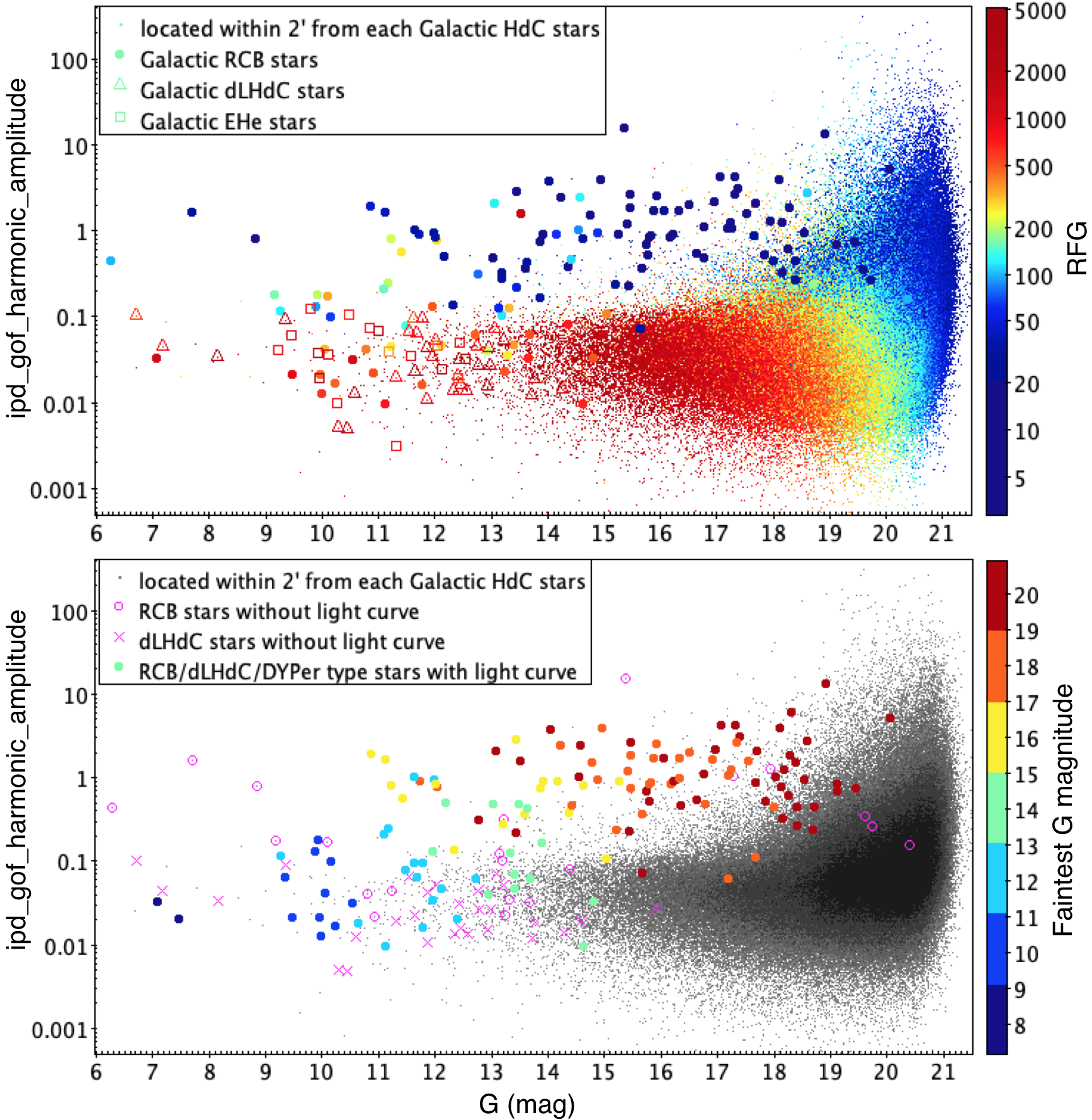}
\caption{IPD parameter IPDgofha versus the mean G-band magnitude for our Galactic targets (large colour dots and circles) and stars located within 2 arcmins from them. Top: Diagram colour-coded by the ratio G mean flux to its standard error (RFG). Bottom: Our targets of interest are colour-coded by their faintest G-band magnitude as observed in their respective light curves. The targets with and without a published Gaia DR3 light curve are also indicated.}
\label{fig_IPDgofhavsG}
\end{figure}

In Figure~\ref{fig_IPDgofhavsG}, top, IPDgofha versus mean G magnitude, with colour coding by RFG (ratio of G mean flux to error), was plotted for all targets. It shows that only previously reported RCB stars of low signal-to-noise ratio (S/N) have significantly higher IPDgofha values compared to nearby classical stars. All RCB stars that remained bright, along with dLHdC and EHe stars, have low IPDgofha values and minimum RFG of approximately 150.

Our primary explanation for this observation is that changes in RCB star colours led to changes in the optical PSF shape (likely primarily due to changes in coma) across transits. Using a single PSF model for all transits, with either a fixed or fitted colour index, caused variations in goodness-of-fit values as the PSF shape changed. For RCB stars with high IPDgofha values, we also analysed the associated ipd\_gof\_harmonic\_phase parameter, which gives the resulting phase angle discussed earlier. The distribution of these angles is similar to that found for classical stars.

Another explanation we considered is the high natural noise at fainter magnitudes during declines. As Figure~\ref{fig_IPDgofhavsG} shows, the spread of IPDgofha values increases below the 18th magnitude, so an RCB star that becomes fainter during a decline phase could result in high goodness-of-fit values for PSF fittings and consequently in high IPDgofha values. To test this, we created a similar graph colour-coded by the faintest recorded G magnitude for each light curve (Fig.~\ref{fig_IPDgofhavsG}, bottom), and found that while many RCB stars with high IPDgofha values did go fainter than 18th magnitude, others remained brighter.

We have indirect evidence supporting the report made by \citet{2006MNRAS.367..290J} that chromatic effects impact the Gaia PSF shape. They also showed that a shift effect is expected based on the PSF SED. Hence, we need to pay special attention to the astrometric results for RCB stars that underwent photometric declines, as they likely also changed in colour. We should exercise caution when interpreting results for RCB stars with RFG values lower than 150.

\subsubsection{Astrometric fit quality and the effect of chromatic PSF on centroid position \label{subsec_RUWE}}

\begin{figure*}
\centering
\includegraphics[scale=0.46]{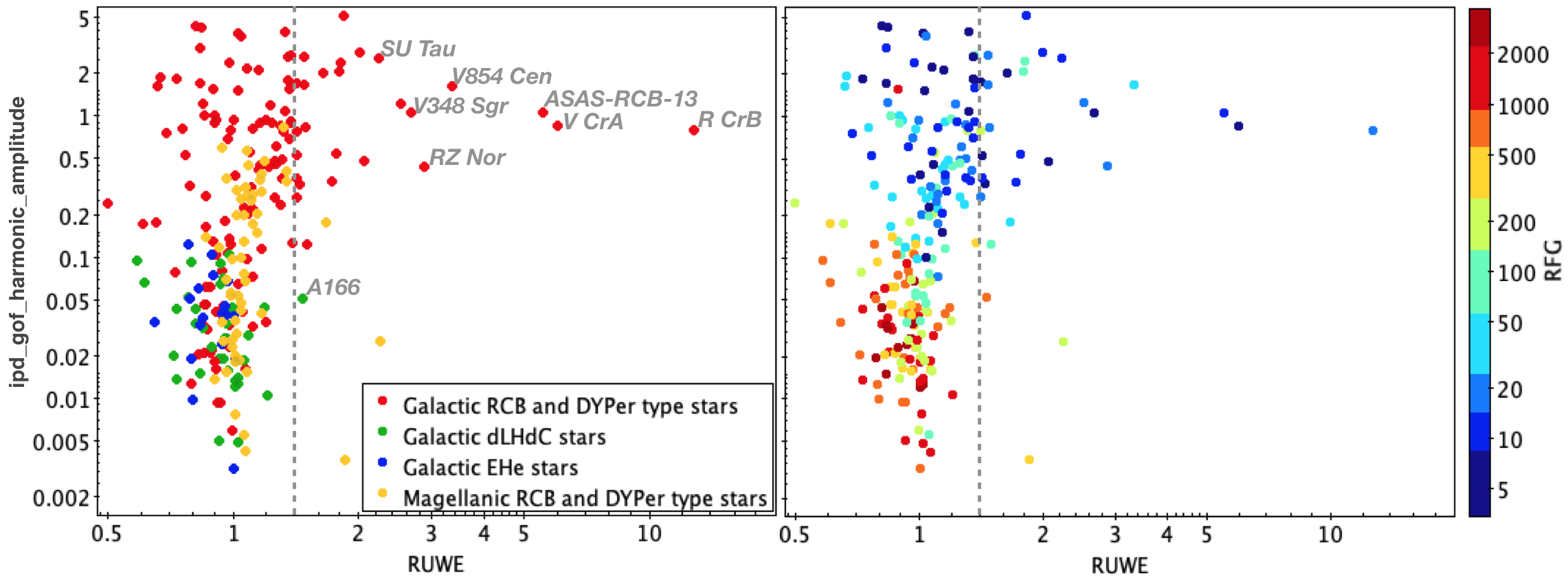}
\caption{IPD parameter ipd\_gof\_harmonic\_amplitude (IPDgofha) versus the astrometric fit quality parameter RUWE in log-log scale. The vertical grey lines indicate the RUWE threshold value above which the results of the astrometric fit were invalidated. Left: Targets colour-coded depending on their respective classification (EHe, dLHdC, RCB or DYPer type stars) and their location. The names of a few prominent targets are indicated. Right: Targets colour-coded by the ratio G mean flux to its standard error (RFG) on a log scale.}
\label{fig_IPDgof_vs_RUWE}
\end{figure*}

RUWE (renormalised unit weight error) is a Gaia parameter used to evaluate the quality of astrometric fit results. It's value is directly related to the fit's chi-square value which has been adjusted to account for dependence found on brightness and colour. A value of 1.0 is expected for well-behaved sources, with a threshold of 1.4 often used to reject results \citep{2021A&A...649A...5F}. In our study, we plotted the IPDgofha vs. RUWE in Figure~\ref{fig_IPDgof_vs_RUWE} and found a median RUWE value that is indeed around 1.0, with a smaller spread for Magellanic targets. The RUWE distribution of EHe stars, our warmer subgroup, has a lower median value of around 0.9.

At the distance of the Magellanic Clouds, only the proper motion should drive astrometric fits, as parallax values are expected to be insignificant. Our results on Magellanic targets confirm this, as no significant parallax values were found. Only three of them, RCB star MSX LMC 1601 and DYPer type stars EROS2-LMC-DYPer-4 and -6, had RUWE values that invalidated their astrometric fit results.

Of the Galactic targets, 23 have high RUWE values above 1.4. One is the dLHdC star A166 and the rest are RCB stars, including R CrB, V854 Cen, V348 Sgr, SU Tau, ASAS-RCB-16, and WISE J005128.08+645651.7. During the time frame of Gaia DR3, all six RCB stars have undergone at least one significant decline phase. They have very low RFG values, except for ASAS-RCB-16 that had only two of 21 measurements taken during its decline phase. Despite significant published parallax values for all these stars, no results from their astrometric fits should be further considered.

The chromatic PSF effect on centroid position is likely causing the issue of the above astrometric outcome. We suppose that a systematic shift away from the true position, correlated to the star's colour variation, was almost certainly added due to the Gaia DR3's computational strategy of using a single PSF model derived from a single colour index for all astrometric measurements. Figure~\ref{fig_IPDgof_vs_RUWE} reveals that RCB stars that have undergone decline events, recognizable by their high IPDgofha ($>$0.1) and low RFG ($<$150) values, resulted in astrometric fits with median RUWE values of around 1.2, worse than normal. This highlights the difficulty of performing astrometric fits for stars that have changed in colour. Astrometric results for these stars in Gaia DR3 should not be trusted.

Additionally, we analysed the pseudo-colour derived from the 6-parameter astrometric fits and found that for RCB stars that underwent declines, the results are often non consistent with their average BP-RP colour and/or even the series of colours in their light curve. This indicates poor performance of the fits for these stars and the need for a dedicated strategy to eliminate colour biases. As an illustration, we provide examples of two RCB stars with RUWE less than 1.4:

\begin{description}[font=-]
\item \textit{WISE J170343.87-385126.6 }(RUWE$\sim$1.27): the fit resulted with a pseudo-colour index of 1.01$\pm$0.03 corresponding to an extreme colour value of $\sim$17.3 mag \citep[Eq.4]{2021A&A...649A...2L}. This is completely outside the range of usual colour and is not consistent with the series of colours from 36 measurements given by its light curve. A median BP-RP colour of $\sim$5.0 mag is thus observed. 
\item \textit{ASAS-RCB-14} (RUWE$\sim$1.08): a pseudo-colour of $\sim$4.0 mag was determined by the astrometric fit, while its average BP-RP colour index reported by the Gaia DR3 database is $\sim$2.2 mag. Its light curve shows a large spread between 2.0 and 3.4 mag with a median value of $\sim$2.9 mag, due to two large decline events.

\end{description}

\subsection{Binarity}

One could wonder if the Gaia DR3 astrometric outcomes on RCB stars might result from binary or multiple stellar systems. Past researches on this matter have always shown negative results: no known HdC stars \citep{1993PASP..105..574R,2020MNRAS.493.3565J} nor EHe stars \citep{1987MNRAS.226..317J,2008ASPC..391...53J} show evidence of any binary companion. Gaia DR3 is now making some contribution to this with its work on unresolved binaries \citep{2022gdr3.reptE...7P}. Indeed, it provides a 'non-single star' solution \citep{2022yCat.1357....0G} using astrometric, spectroscopic and even photometric information. About 800,000 stars were released with a validated orbital solution and none of them is a known HdC or EHe star. That result is consistent with the double-degenerate scenario as HdC and EHe stars are expected to be post-merger products.

\section{Selected samples for parallax and proper motion studies \label{sec_AstromSelection}}

\subsection{Main selection}

To ensure accurate astrometric fits, we first removed all targets with RUWE values above the threshold of 1.4. Additionally, we removed any targets that may have experienced significant changes in colour. For this, we required an RFG value of higher than 150 for all targets without a Gaia light curve. For others, we set a maximum brightness variation of $\Delta$Gmax of 2 mag. We then examined the selected light curves for colour variability, keeping only those which have over 90\% of astrometric measurements made when the star's true colour remained within 0.25 mag from the colour used in its respective astrometric fit. It is unknown how the use of an incorrect PSF model may affect the astrometric results, and more specifically how the systematic shift added varies in relation to the difference between the model colour and the true colour indexes. A dedicated study is needed to answer this question. However, we consider pragmatically that astrometric fit results with over 90\% of measurements free of significant colour biases are likely to be only marginally impacted by a few outlier measurements.

We have selected a total of 74 Galactic HdC and DYPer type stars, as well as all 18 Galactic EHe stars for further study. Out of these, 28 RCB, 31 dLHdC, 3 DYPer type, and all 18 EHe stars show a parallax signal with a signal-to-noise ratio greater than three. In the Magellanic Clouds, similarly, we selected 8 RCB and 16 DYPer type stars. 

\subsection{Bright HdC stars status \label{sec_brightHdC}}

We analysed the Gaia DR3 results for 15 bright HdC stars studied by \citet[Tab.2]{1998PASA...15..179C} during their study of the Hipparcos dataset, to which we added the prototype DY Persei and RCB star V854 Cen. We expected their parallaxes to be detected by Gaia, and indeed, with the only exception of V CrA\footnote{If we consider an absolute magnitude M$_G$ between -5 and -4 mag for this warm RCB star, its parallax value should range between 0.13 and 0.20 mas. This falls within the range of possible parallax detections by Gaia DR3 for over 95\% of stars of similar brightness.}, significant parallaxes with a S/N higher than 3 were published. However, seven of these bright stars did no meet our selection criteria. We discuss the results of their respective automatic astrometric fit below.

Among them, four RCB stars (R CrB, V854 Cen, V CrA, and SU Tau) had astrometric RUWE values above the threshold, suggesting problems with their results. They experienced large ($>$4 mag) photometric variations during the Gaia DR3 period. No light curves were produced for the first two stars (possibly due to calibration issues in high luminosity range), but light curves from AAVSO\footnote{AAVSO: American Association of Variable Star Observers} reveal declines of more than 7 mag. SU Tau was in a rising phase during the entire Gaia DR3 survey, causing continuous colour changes. V CrA had a rising and declining phase, with a maximum brightness lasting only 300 days, producing a change of $\sim$1.2 mag in BP-RP colour indexes. 

For these four RCB stars, whose temperature classes are between HdC4 and HdC1 \citep{2023MNRAS.521.1674C}, one could reasonably expect an absolute magnitude M$_V$ ranging between -3.5 and -5 mag \citep{2022A&A...667A..83T}. Using the apparent magnitudes of 5.8, 7.1, 9.75 and 9.8 mag in the V band, measured at maximum brightness, and associated interstellar extinctions A$_V$ of 0.07, 0.3, 0.3 and 1.56. mag, respectively for R CrB, V854 Cen, V CrA, and SU Tau, we assume the following intervals of possible distances in kiloparsecs: R CrB (0.7 - 1.4), V854 Cen (1.1 - 2.3), V CrA (3.9 - 7.8), SU Tau (2.2 - 4.4). Finally, we note that R CrB was published in Gaia DR2 with an astrometric fit whose quality flags did not show signs of possible issues. It corresponded to a plausible distance of 1.2$\pm$0.1 kpc \citep{2018AJ....156...58B}. However, during the Gaia DR2 monitoring period of 22 months, R CrB underwent two large changes of luminosity: an increase of $\sim$6 mag followed 4 months later by a rapid decline phase of 7 mag that lasted up to the end of Gaia DR2's coverage. The PSF chromaticity issues discussed above must have also impacted the previous data release. Therefore, we cannot trust the distance estimate released by Gaia DR2 for R CrB.

The remaining 3 RCB stars (S Aps, UW Cen, and RY Sgr) have RUWE values below the quality threshold but still above 1.0. An investigation of their Gaia and AAVSO light curves showed brightness variations for S Aps and UW Cen, but none was detected for RY Sgr. We summarise below their respective photometric and astrometric status.

\textit{S Aps}: its Gaia DR3 light curve showed a 3.5 mag rising phase over 300 days (8 measurements) before reaching a maximum phase for the remaining 700 days (23 measurements). A 6-parameter astrometric fit was used and the resulting BP-RP colour of 1.66 mag is only 0.06 mag redder than the median colour for the maximum phase. During that phase, BP and RP magnitudes were measured for all 23 observations, but no colour information is available for the previous 8 observations. If the reddening rate during the final rising stage was extrapolated, the total colour variation during the first quarter of the observations would be $\sim$1.4 mag, which likely affected the astrometric fit.

\textit{UW Cen}: it had a slow rising phase during the 34 months studied in Gaia DR3, increasing its brightness by approximately 1.8 mag in the G-band while its total colour changed by $\Delta$BPRPmax$\sim$0.65 mag. Out of 100 photometric measurements, 80\% were within 0.25 magnitude of the colour used in the 5-parameter astrometric fit. This falls below our selection threshold of 90\%. The astrometric fit resulted in a RUWE value of $\sim$1.07. 

\textit{RY Sgr}: it was not selected in our study due to its low RFG ratio of 99, which falls below our established threshold for selection of stars without Gaia DR3 light curves. However, its AAVSO light curve reveals that it remained bright throughout the entire period covered by Gaia DR3, with some pulsations of about 0.4 mag in total amplitude. AAVSO's observations were taken at least weekly and the only existing gaps lasted about 100 days during the southern hemisphere winter seasons. The same observational constraints are applicable to Gaia. RY Sgr's high IPDgofha value of 0.44 and RUWE value of 1.21 indicate that the astrometric fit has not gone smoothly. At the brightness of RY Sgr, G$\sim$6.3 mag, and above, photometric calibrations are known to be challenging for Gaia \citep{2021A&A...649A...3R}. The high dynamic range obtained by the survey is due to reduction in exposure time for the pixels containing bright stars but this reactive strategy resulted in more difficult calibration.

We compared the proper motion results for these three RCB stars found by Gaia DR3 with those found by the Hipparcos survey \citep{1998PASA...15..179C}. Results were consistent in both directions within 1 sigma, except for UW Cen where Hipparcos found a proper motion along the declination axis of -12.6$\pm$3.9 mas/yr while Gaia DR3 found -1.83$\pm$0.01 mas/yr. Gaia DR3 parallax values for the 3 stars are over 0.2 mas with an S/N over 10; RY Sgr is reaching 0.59 mas with an S/N of 21. These strong astrometric signals suggest that the effect of colour changes on a fraction of the measurements was only marginal. These three RCB stars were reincorporated into the final selection. Furthermore, using apparent magnitudes of 9.7, 9.33 and 6.36 mag in V and an interstellar extinction of 0.36, 1.03, 0.41 mag respectively for S Aps, UW Cen and RY Sgr, we observed that the mean distances found by Gaia DR3 are consistent with an absolute magnitude M$_V$ of -5 mag for the warm RCB stars RY Sgr and UW Cen, and $\sim$-3.7 mag for S Aps, a colder RCB star whose temperature class is HdC5. These absolute magnitudes are consistent with the ones already observed in RCB stars of similar temperature \citep{2022A&A...667A..83T}.

Lastly, we note that Gaia DR3 does not have light curves available for the five bright HD stars - HD 137613, HD 148839, HD 173409, HD 182040, and HD 175893. The last was identified as a heavy dust producing RCB star in \citet{2012A&A...539A..51T}. The lack of significant photometric variations likely explains the absence of a light curve.

\section{Galactic HdC and EHe stars' spatial distribution, distances, and velocities \label{sec_spatialdistrib}}

\subsection{Heliocentric radial velocities and associated errors \label{subsec_helioRV}}

Heliocentric RV measurements and associated errors were published in Gaia DR3 for approximately 57\% and 36\% of our Galactic and Magellanic targets, respectively. However, these fractions vary depending on the type of stars being observed. Only four ($\sim$22\%) Galactic EHe stars have RV measurements available, as most of them are warmer than the upper range limit used for RV determination. Similarly, only one out of the four Galactic DYPer type stars has a published RV measurement, namely ASAS-DYPer-2. Although the DYPer type stars are the coldest targets, they should still be warm enough to be analysed. However, we suspect that for two of them, DY Persei and ASAS-DYPer-1, their RVS spectra may have had more than 40 pixels saturated, which is the threshold for rejection \citep{2023A&A...674A...5K}. These two targets are published with RP magnitudes of $\sim$8.1 and 6.2 mag, respectively. Conversely, the last DYPer type star, 2MASS J175248, was too faint, as its G magnitude was reported at $\sim$16.4 mag.

Among the Galactic dLHdC and RCB stars, we have RV measurements for 82\% and 57\% of them, respectively. The high success rate in the former group was expected, as they are photometrically stable and have an effective temperature range ideal for the RVS spectrograph. However, six dLHdC stars are missing RV measurements: saturation is most likely the culprit for the four brightest HD\footnote{HD 137613, HD 148839, HD 173409, HD 182040} stars, while HE 1015-2050 is too faint. The last one missing is A166, which is the coldest of all known dLHdC stars. However, based on its photometric Gaia colour and optical spectrum, A166 should still be warmer than the DYPer type stars and the 3100 K lower threshold for RV measurement. As A166 is the only dLHdC star with a Renormalised Unit Weight Error (RUWE) value higher than the astrometric quality threshold, it is possible that it did not pass the filter on low-quality astrometry applied before the scientific processing of RVS spectra \citep{2023A&A...674A...5K}. Similarly, this criterion has most certainly rejected many RCB stars. Indeed, among the RCB stars without RV measurements, but within the good brightness and temperature range, a large majority have high RUWE values, caused mainly by PSF astrometric shift occurring during the chromatic decline phases. For example, we note that the following bright RCB stars, namely R CrB, V854 Cen and SU Tau, have no RVS information. Furthermore, because of these photometric declines, many RCB stars' median GRVSmag\footnote{A narrow-band Vega-system magnitude defined from the transmittance of the Radial Velocity Spectrometer (RVS) \citep{2023A&A...674A...6S}} magnitude happened to be fainter than the 14th magnitude threshold, which corresponds roughly to the 16th magnitude in G.

We used the median heliocentric radial velocities reported in Gaia DR3 for the majority of our targets. However, since some stars lacked information from the Gaia RVS spectrograph, we compiled measurements from the literature. Additionally, we measured the RV for stars observed over the past decade using our mid-resolution survey conducted with the Wide Field Spectrograph (WiFeS) instrument mounted on the ANU 2.3m telescope at Siding Spring Observatory \citep{2007Ap&SS.310..255D}. This instrument has a mid-resolution of R$\sim$3000. Our wavelength calibration was performed using the RASCAL libraries \citep{2020ASPC..527..627V} with the sky emission lines as calibrators.

\begin{figure}
\centering
\includegraphics[scale=0.35]{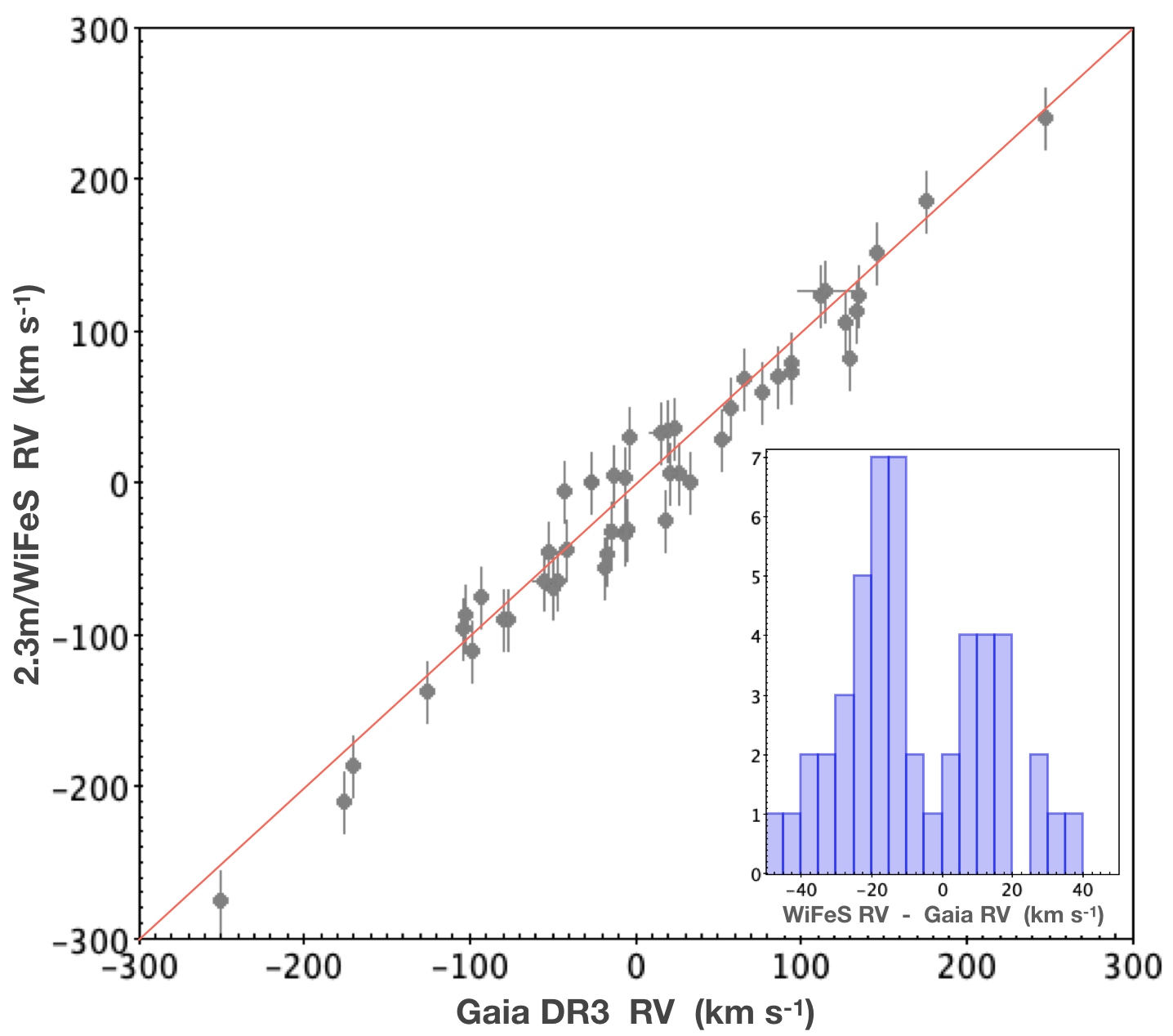}
\caption{Comparison of the heliocentric radial velocities measured with the 2.3m/WiFeS spectrograph to the ones published by Gaia DR3. Most of the Gaia DR3 errors are smaller than the dots. The error on the WiFeS RV was estimated to be $\sim$20 km s$^{-1}$. The distribution of differences between both measurements is shown in the bottom right figure. The red line of slope one is used for visual guidance only.}
\label{fig_WivesRV_vs_GaiaRV}
\end{figure}

In total, we measured the radial velocity for 80 HdC stars, of which 49 were also released in Gaia DR3. A comparison between the two datasets is presented in Figure~\ref{fig_WivesRV_vs_GaiaRV}, and we found excellent agreement between the two. We determined an error on RV of $\sim$20 km s$^{-1}$ for the 2.3m/WiFeS measurements, which corresponds to about 1/5th of a WiFeS pixel. 

\begin{figure}
\centering
\includegraphics[scale=0.45]{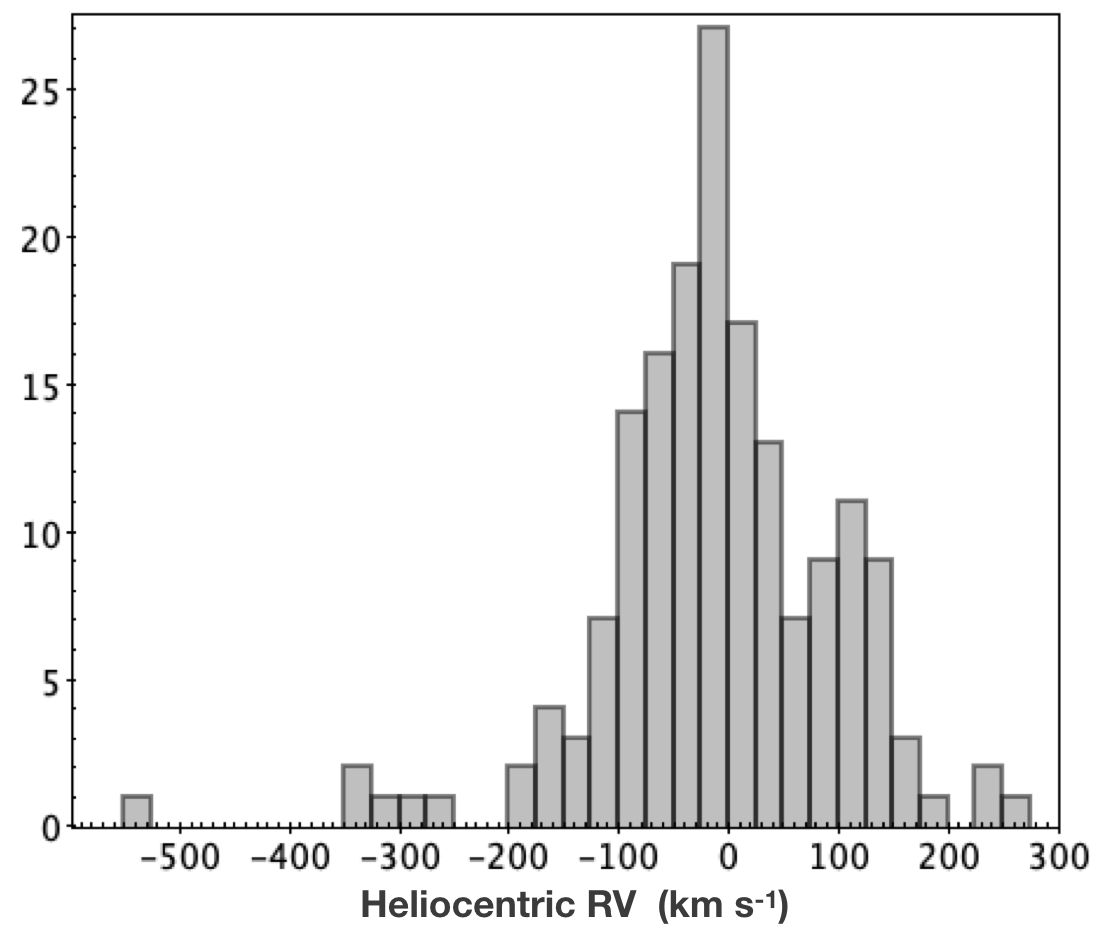}
\caption{Distribution of radial velocity measurements accumulated for Galactic HdC and EHe stars.}
\label{fig_RVDistrib}
\end{figure}

We searched for our targets in the data releases of large high-resolution spectroscopic surveys. On a few occasions, these surveys have observed some HdC stars, and the published RVs are consistent with those released by Gaia DR3 or other measurements found in the literature: B565 was observed by the GALAH survey \citep{2021MNRAS.506..150B}, A249 was observed by the RAVE DR6 survey \citep{2020AJ....160...82S}, UV Cas and R CrB were observed by the LAMOST DR7 survey \citep{2022yCat.5156....0L}, and finally, U Aqr and UV Cas were observed by Apogee DR6 \citep{2020AJ....160..120J}. The last survey has also observed two RCB stars, ES Aq and SV Sge. While their RVs were not published by Gaia DR3, Apogee's RV values are consistent with the ones we found using the 2.3m/WiFeS spectrograph.

However, we have identified some discrepancies in our study. First, the RVs published for two EHe stars, V2205 Oph by RAVE DR6 and V2244 Oph by Apogee DR6, are both very high (above 500 km s$^{-1}$), but also with high $\chi^2$ values of their respective fits, indicating issues with hot stars. Secondly, and perhaps more interestingly, the bright dLHdC star HD 182040 was observed by Apogee DR6 with an RV of -45 +/- 0.01 km s$^{-1}$, which is 10 km s$^{-1}$ lower than the median value obtained with 15 measurements by \citet{1997MNRAS.285..266L} while they were not able to measure any intrinsic RV variation and thus estimated an upper limit for RVamp of $\sim$6 km s$^{-1}$. Thus, it is possible that HD 182040 is currently going through a phase of higher RV variations. The measurement made by Apogee needs to be confirmed.

We also checked the consistency of all RV values obtained from low to mid-resolution spectroscopic surveys with the ones released in Gaia DR3. We found an excellent agreement with the measurements reported by \citet{1987MNRAS.226..317J, 1997MNRAS.285..266L, 2008A&A...481..673T}, but not with the values listed by \citet{2021ApJ...910..132K} obtained from low resolution IR spectra. All RV values listed in their table 3 should be multiplied by -1 (priv. communication). An explanation on the discrepancy will be given in Karambelkar et al. in prep.

\begin{table}[!htbp]
\caption{Heliocentric velocity of Galactic HdC stars not provided by Gaia DR3, given by 2.3m/WiFeS spectra or other references\label{tab.RV_HdC}}
\medskip
\centering
\begin{tabular}{lrrl}
\hline
Name & \multicolumn{1}{c}{Helio.} & \multicolumn{1}{c}{Error} & References \\
 & \multicolumn{2}{c}{RV (km s$^{-1}$)} &  \\
\hline
\multicolumn{4}{c}{Galactic HdC stars}\\
\hline
ASAS-RCB-1	& -85		& 20  &  \\
ASAS-RCB-3  &  -20  &  20  & \\
ASAS-RCB-7	& -332	& 20	&   \\	
ASAS-RCB-8	&  135	& 5	& P21 \\ 
ASAS-RCB-9	& 151	& 20	&  \\
ASAS-RCB-11	& -69		& 20	&  \\	
ASAS-RCB-19	& 36		& 20	&  \\	
ASAS-RCB-21	& 30		& 20	&  \\	
DY Cen & 32 & 2 & JRL20 \\
EROS2-CG-RCB-5	&  -323 &  5	&  T08  \\	
EROS2-CG-RCB-7	&  -111 & 8 	&  T08  \\	
EROS2-CG-RCB-8	&  	112	& 5 	&  T08  \\	
EROS2-CG-RCB-11	&  73		&  5	&  T08  \\	
EROS2-CG-RCB-12	&  -44		&  7	&  T08  \\	
EROS2-CG-RCB-13	&  -292	&  5	&  T08  \\	
EROS2-CG-RCB-14	&  -11		&  5	&  T08  \\	
ES Aql	& -56		& 5	& J20  \\ 
HD 137613	& 71		& 2	& LC97 \\
HD 148839	& -13			& 2 & LC97	  \\
HD 173409	& -59		& 2	& LC97  \\	
HD 175893	& 56			&  3	& LC97  \\
HD 182040	& -35			&  2	& LC97  \\   
HE 1015-2050	& 18			& 1.5 	& GA13 \\ 
MACHO 401.48170.2237	& 87		& 20 	&  \\ 
MV Sgr & -93  &  4  &  P96  \\
R CrB    	& 22	& 1	&  F19, P04, LC97\\ 
RS Tel	 & 5		& 5 	&  LC97 \\
S Aps	& -75			& 2.5 	& LC97 \\
SU Tau	& 45   & 5	&  A00, P04 \\
SV Sge	& -5		& 5 	&  J20 \\  
V1783 Sgr	& 120		& 20 	&  \\
V348 Sgr  & 145 &  5 &  B99 \\
V3795 Sgr	& -30 & 5 	& A00  \\
V4017 Sgr & -88 & 20 & \\
V854 Cen  & -25 &  5 & LC89 \\
WISE J132354.47-673720.8	&  31		& 20 	&  \\
WISE J163450.35-380218.5	& -133	& 20 	&  \\	
WISE J170343.87-385126.6	&  7			& 20 	&  \\
WISE J171815.36-341339.9	& -54		& 20 	&  \\
WISE J171908.50-435044.6	& -152	& 20 	&  \\	
WISE J172553.80-312421.1	& -113 	& 20 	&  \\	
WISE J173202.75-432906.1	& -82		& 20 	&  \\
WISE J174328.50-375029.0	& -34		& 20 	&  \\
WISE J175558.51-164744.3	& -25		& 20 	&  \\
WISE J182334.24-282957.1	& -74		& 20	&  \\	
WISE J182723.38-200830.1	& -29	    & 20 	&  \\
WISE J184246.26-125414.7	&  49		& 20 	&  \\	
\hline
\multicolumn{4}{l}{Stars without references are our measurements from 2.3m/WiFeS.}\\
\multicolumn{4}{l}{References for the other RV measurements:}\\
\multicolumn{4}{l}{A00: \citet{2000A&A...353..287A}, B99: \citet{1999ApJ...515..610B},} \\
\multicolumn{4}{l}{F19: \citet{2019MNRAS.482.4174F}, GA13: \citet{2013ApJ...763L..37G},}\\
\multicolumn{4}{l}{J20: \citet{2020AJ....160..120J}, JRL20: \citet{2020MNRAS.493.3565J},}\\
\multicolumn{4}{l}{LC89: \citet{1989MNRAS.240..689L}, LC97: \citet{1997MNRAS.285..266L},} \\
\multicolumn{4}{l}{P04: \citet{2004MNRAS.353..143P}, P21: \citet{2021ApJ...921...52P},}\\
\multicolumn{4}{l}{P96: \citet{1996MNRAS.282..889P}, T08: \citet{2008A&A...481..673T}.}\\

\end{tabular}
\end{table}

\begin{table}[!htbp]
\caption{Same as Table ~\ref{tab.RV_HdC} but for Galactic DYPer type and EHe stars\label{tab.RV_DYPer_EHe}}
\medskip
\centering
\begin{tabular}{lrrl}
\hline
Name & \multicolumn{1}{c}{Helio.} & \multicolumn{1}{c}{Error} & References \\
 & \multicolumn{2}{c}{RV (km s$^{-1}$)}  &  \\
\hline
\multicolumn{4}{c}{Galactic DYPer type stars}\\
\hline
ASAS-DYPer-1	& 5 & 20  &  \\
DY Per	& -44		& 5  & B92, Z05 DB07 \\
\hline
\multicolumn{4}{c}{Galactic EHe stars}\\
\hline
PV Tel 	&  	-171 	&  5 	&  J87\\	
V2244 Oph  	&  -8		&  5 	&  J87 \\	
V1920 Cyg 	&  -90		&   5	&  J87 \\	
V2076 Oph  	&  	77	&  5 	& J87 \\	 
V2205 Oph  	&  	-64	&  7 	&  J87 \\	
BX Cir  	&  	-89	&  5 	&  J87 \\	
V821 Cen  	&  	-68	&  5 	& J87 \\	
DN Leo 	&  155		&   5	& J87 \\	
LSS 99 	&  	109	&  5 	& J87 \\	
LS IV +06 2	&  -24		&  5 	&  J87\\	    
LSS 4357 	&  	-99	&  5 	& J87 \\	
LSE 78 	&  	-91	&   5	&  J87 \\	
\hline
\multicolumn{4}{l}{Stars without references are our measurements from 2.3m/WiFeS.}\\
\multicolumn{4}{l}{References for the other RV measurements: : B92: \citet{1992AJ....104.1585B},}\\ 
\multicolumn{4}{l}{DB07: \citet{2007A&A...473..143D}, J87: \citet{1987MNRAS.226..317J},}\\
\multicolumn{4}{l}{Z05: \citet{2005A&A...438L..13Z}}\\
\end{tabular}
\end{table}

We have compiled a list of heliocentric RVs and associated errors for 34 HdC, 12 EHe and 2 DYPer type stars that were not released by Gaia DR3. The details are provided in Tables~\ref{tab.RV_HdC} and ~\ref{tab.RV_DYPer_EHe}. About half of these measurements are from our 2.3m/WiFeS survey, while the other half were obtained from the literature. In the case of RCB stars, we set the error values to 5 km s$^{-1}$ if the RV measurements were based on only a limited number of observations to take into account the intrinsic RV variability discussed in \citet{Tiss2023a}. Furthermore, a similar increase in the error of Gaia RV measurements was applied if the stars were observed for fewer than 8 visibility periods.

Overall, we obtained RV measurements for approximately 90\% of all our targets, enabling us to calculate the three-dimensional velocity components. We are missing only 26 heliocentric RV measurements over the nearly 200 HdC, EHe and DYPer type stars we are studying. The distribution of these measurements is presented in Figure~\ref{fig_RVDistrib}. Few targets have large heliocentric RV values: being below -200 km s$^{-1}$ for AOHer, NSV11154, ASAS-RCB-7, EROS2-CG-RCB-4, -5 and -13, and above 200 km s$^{-1}$ for F152, VZ Sgr and MACHO 308.38099.66. None was found with a total velocity higher than the Galactic escape velocity estimated at $\sim$550 km s$^{-1}$ by \citet{2021A&A...649A.136K}.

\begin{figure*}
\centering
\includegraphics[scale=0.48]{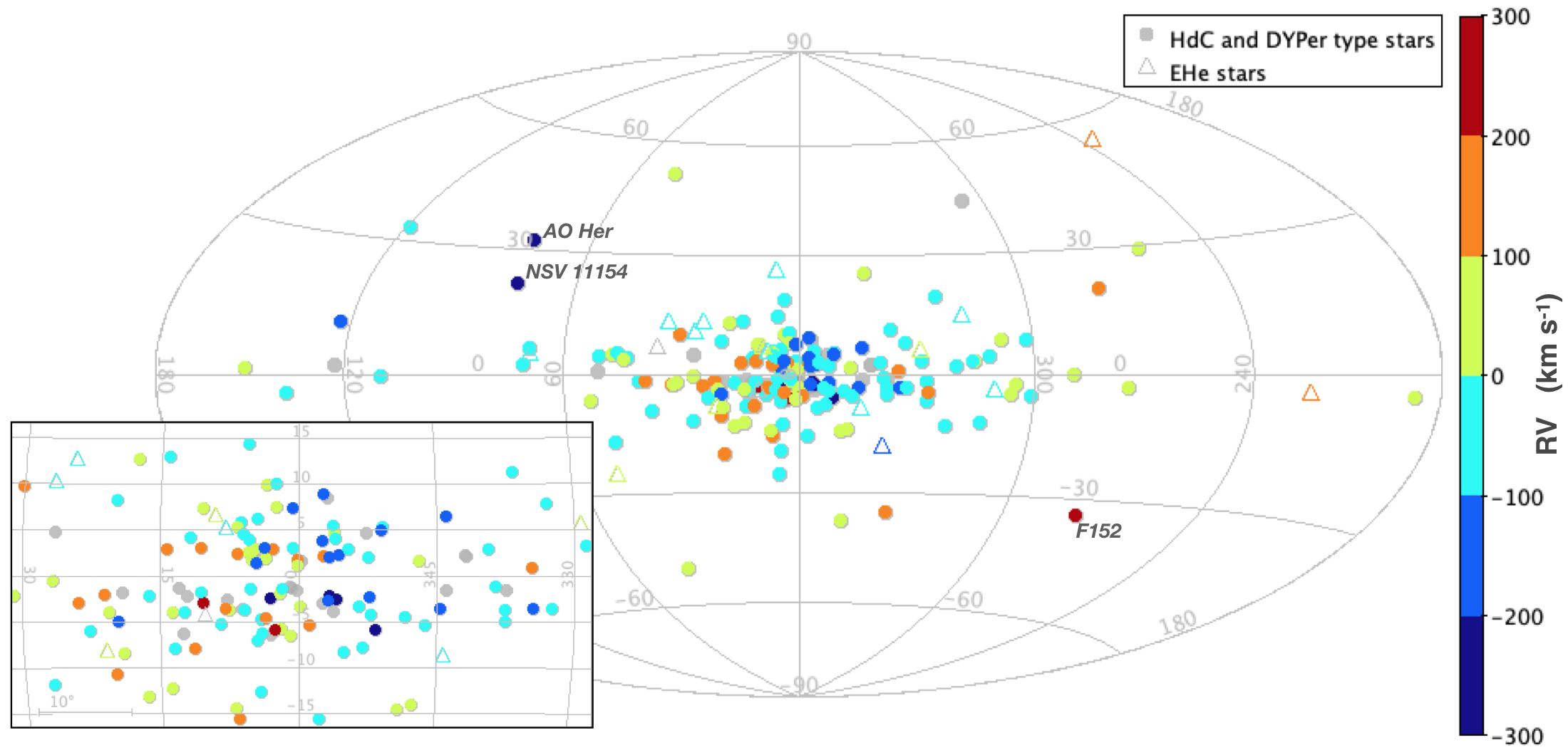}
\caption{Galactic distribution of HdC and DYPer type stars (large dots) and EHe stars (triangles) colour-coded by their mean heliocentric radial velocities. No RV values are known for stars represented in grey. Three HdC stars with large RV values are indicated by their names. Bottom left: Zoom on the Galactic centre area, $|$l$|<$30$^\circ$ and $|$b$|<$15$^\circ$.} \label{fig_RV_SkyDistrib}

\centering
\includegraphics[scale=0.48]{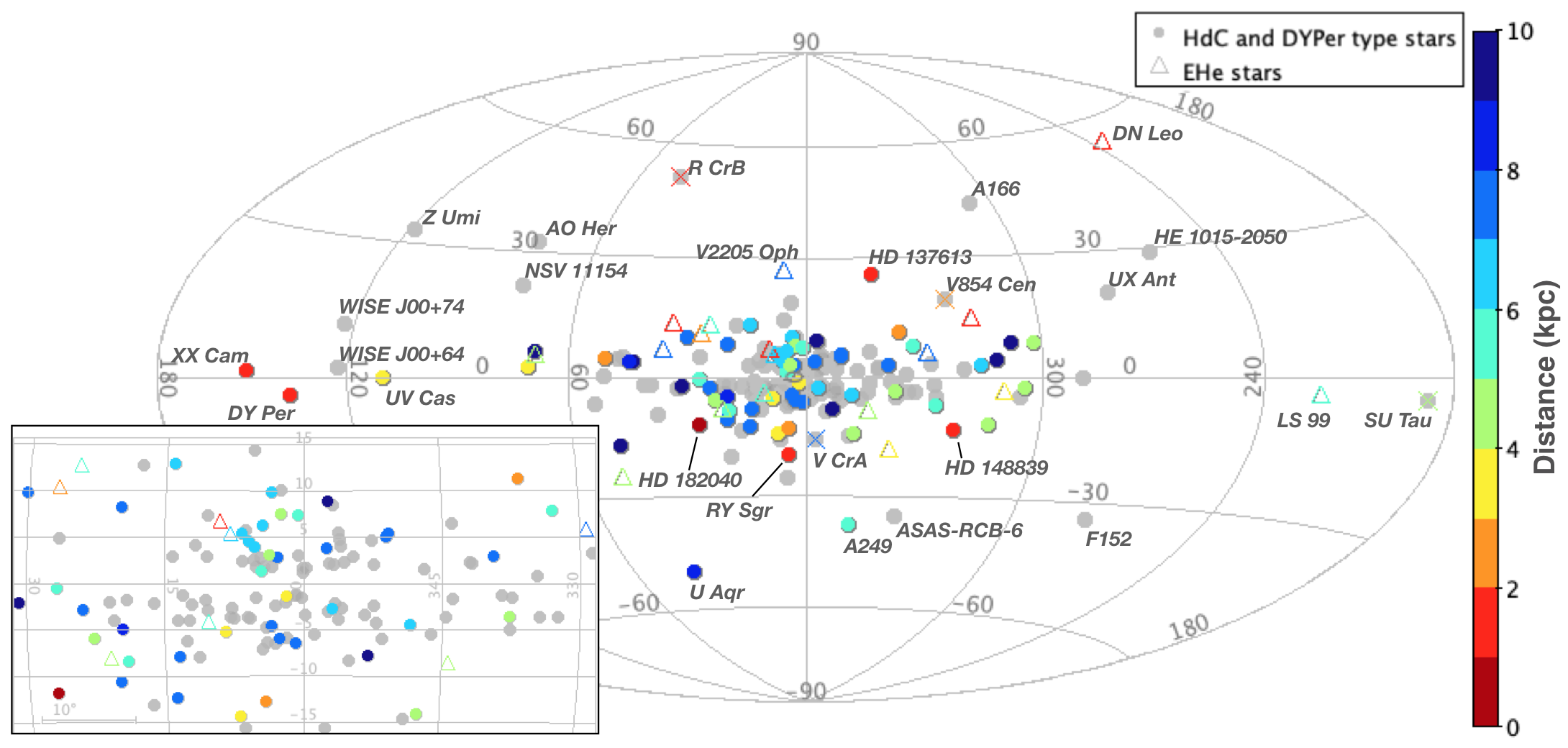}
\caption{Galactic distribution of HdC and DYPer type stars (large dots) and EHe stars (triangles) colour-coded by their mean geometric distances in kilo-parsec as determined by \citet{2021AJ....161..147B}. The coloured symbols are for targets with a valid astrometric fit and a parallax S/N higher than 3. All other stars are indicated in grey. Four bright RCB stars, R CrB, V854 Cen, V CrA and SU Tau are indicated with crosses that are coloured according to the distance estimated with an absolute magnitude M$_V\sim$-5 mag (see text). Names are shown for the stars located outside the Galactic bulge area. Bottom left: Zoom on the Galactic centre area, $|$l$|<$30$^\circ$ and $|$b$|<$15$^\circ$.} \label{fig_Galactic_Distrib}
\end{figure*}

The sky distribution of all HdC and EHe stars, colour-coded by their respective RV values, is presented in Figure~\ref{fig_RV_SkyDistrib}. A few stars, namely AO Her, NSV 11154, and F152, exhibit significantly high RV values and are positioned towards the Galactic halo.

An evident asymmetry towards the Galactic bulge is discernible. Notably, we observe a higher occurrence of positive RV for positive Galactic longitudes (orange and green dots), while the opposite trend is observed for negative longitudes, with a greater prevalence of negative RV values (dark and light blue dots). To validate this observation, we conducted a detailed analysis of the RV distribution among HdC stars situated in a region centred on the Galactic bulge ($|$l$|<$30$^\circ$ and $|$b$|<$15$^\circ$). We selected stars that met our astrometric selection criteria outlined in Section~\ref{sec_AstromSelection} and had a median geometric distance exceeding 6 kpc, and stars with a parallax S/N lower than 3 that supposedly would be located at higher distances. We determined median RV values of +40$\pm$30 km s$^{-1}$ and -80$\pm$30 km s$^{-1}$ for stars with positive and negative longitudes, respectively. This is consistent with the alternating positive-negative variations observed in the median RV sky map produced with Gaia DR3 \citep[Fig. 5]{2023A&A...674A...5K} caused by the Galactic disk rotation relative to the Sun.

\subsection{3D distribution and velocities}

\begin{figure}
\centering
\includegraphics[scale=0.35]{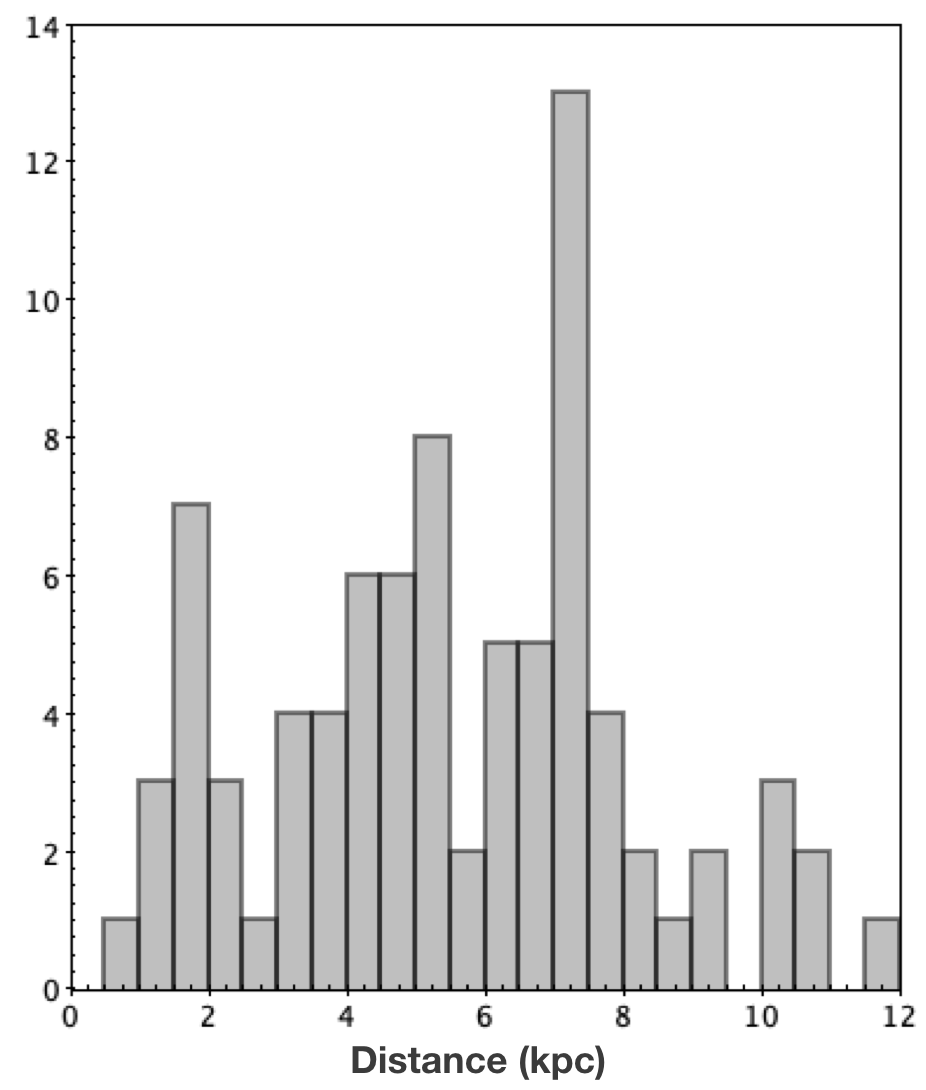}
\caption{Distribution of geometric distances inferred by \citet{2021AJ....161..147B} for Galactic HdC, EHe, and DYPer type stars.}
\label{fig_DistanceDistrib}
\end{figure}

We used the median geometric distances inferred by \citet{2021AJ....161..147B}. They were estimated from the Gaia DR3 parallax measurements using a realistic prior as a function of Galactic position. Figure~\ref{fig_Galactic_Distrib} displays the sky distribution of 83 targets, with colours indicating their respective distances. These targets presented a valid astrometric fit and a parallax S/N higher than 3. Their distance distribution is shown in Figure~\ref{fig_DistanceDistrib}. About 20\% of the targets are located within 3 kpc and $\sim$70\% have distances between 3 and 8 kpc. The remaining targets are located further away, up to the parallax resolution limit of Gaia which corresponds to about 10 kpc. Gaia allow us to probe the nearby portion of the Galactic bulge. Many more HdC stars distances remain to be estimated in that galactic region as one could see from the sky distribution on HdC stars in Figure~\ref{fig_Galactic_Distrib}.

The Galactocentric kinematics and coordinates of our targets were determined using the GALPY Python package \citep{2015ApJS..216...29B}. We adopted a distance of 8.2 kpc to the Galactic centre \citep{2017MNRAS.465...76M,2019A&A...625L..10G,2023MNRAS.519..948L}, and a distance of 17 pc above the Galactic plane for the Sun \citep{2017MNRAS.465..472K}. In a rectangular left-hand Galactocentric frame, we used the following velocity components: (V$_x$,V$_y$,V$_z$)$_{\sun} = $ (-11.1, 245.0, 7.25) km s$^{-1}$. The uncertainties were calculated by propagating the covariance parameters between the two Gaia proper motion components.

\begin{figure*}
\centering
\includegraphics[scale=0.39]{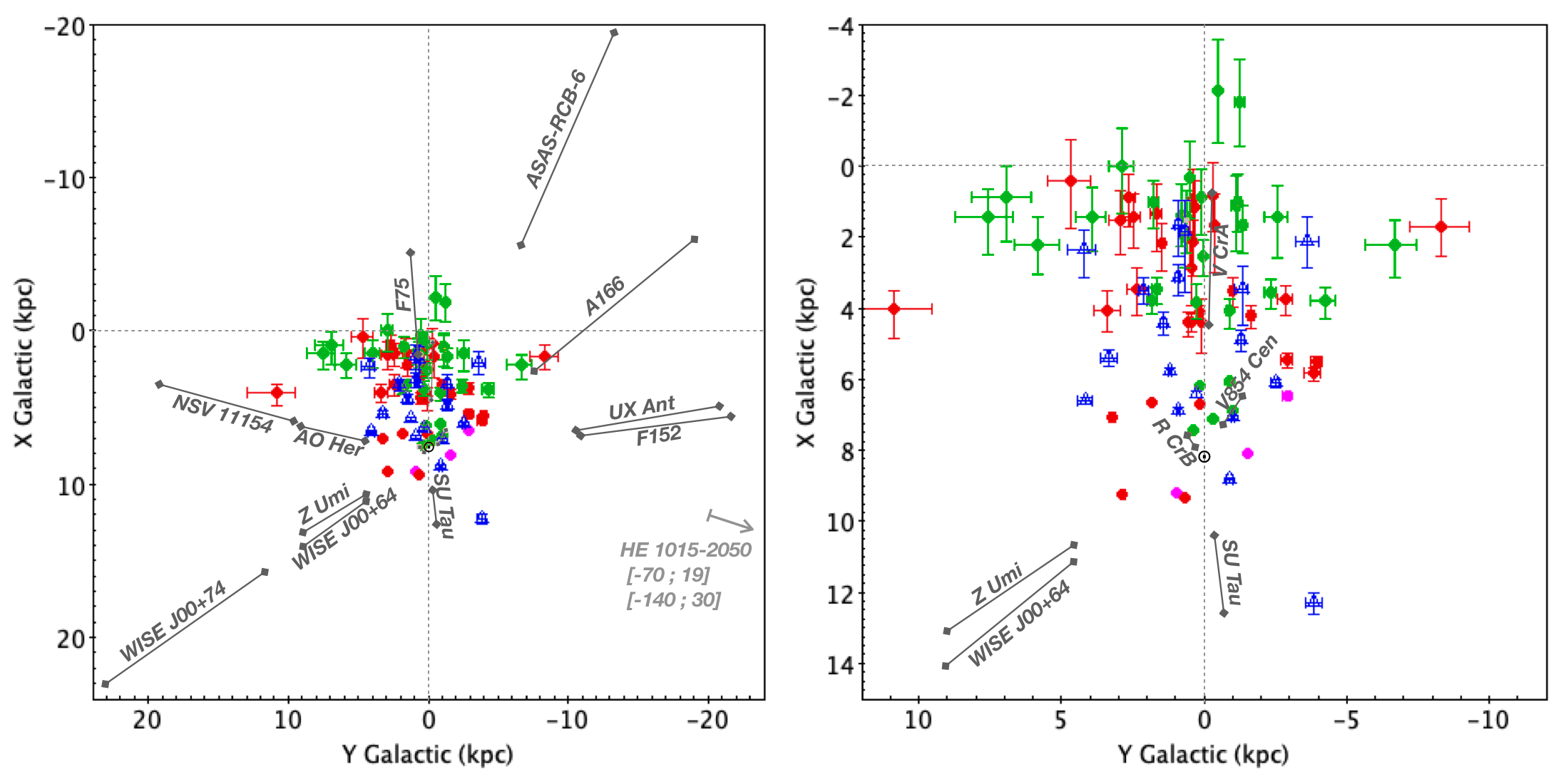}
\includegraphics[scale=0.39]{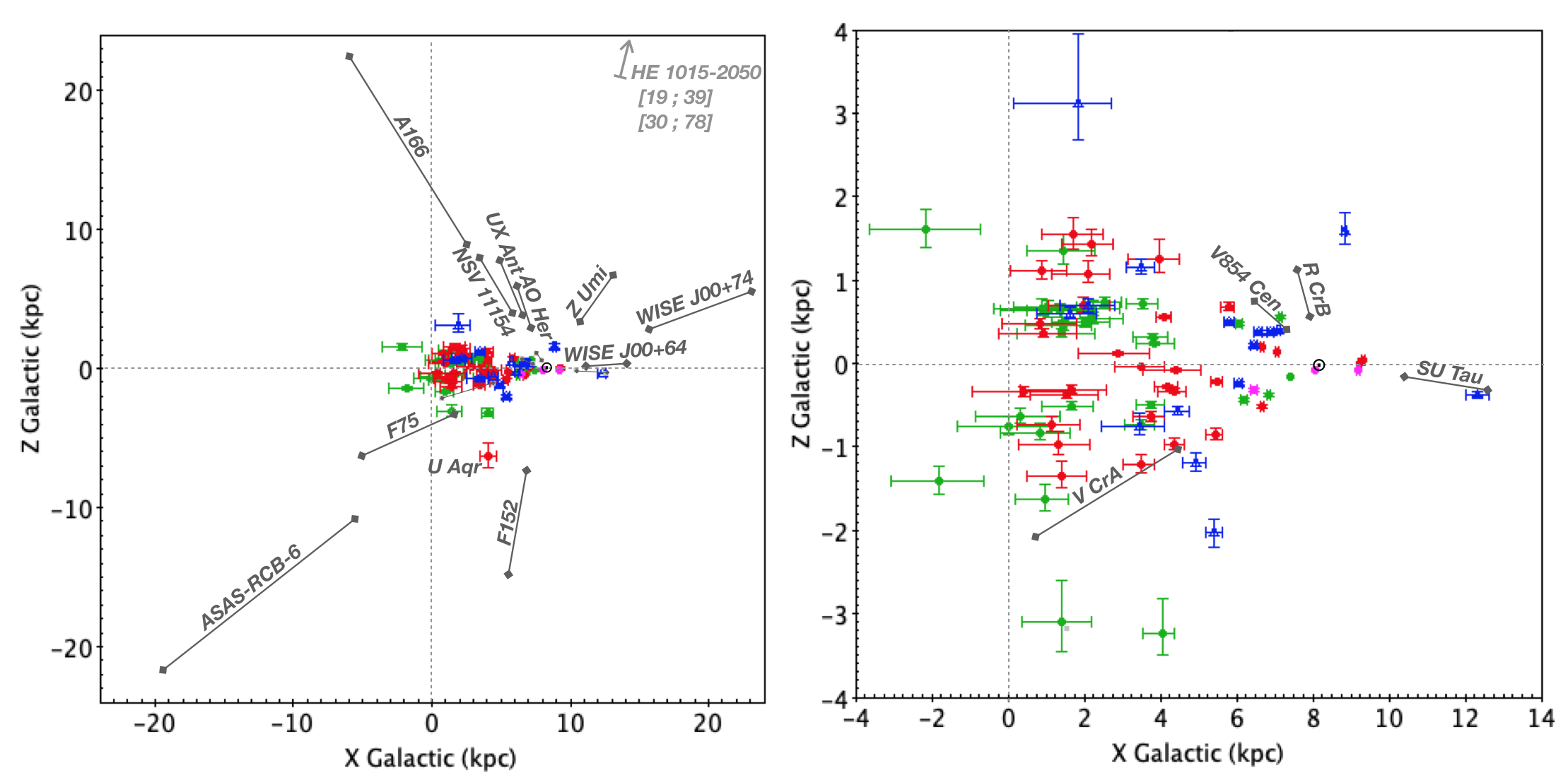}
\includegraphics[scale=0.39]{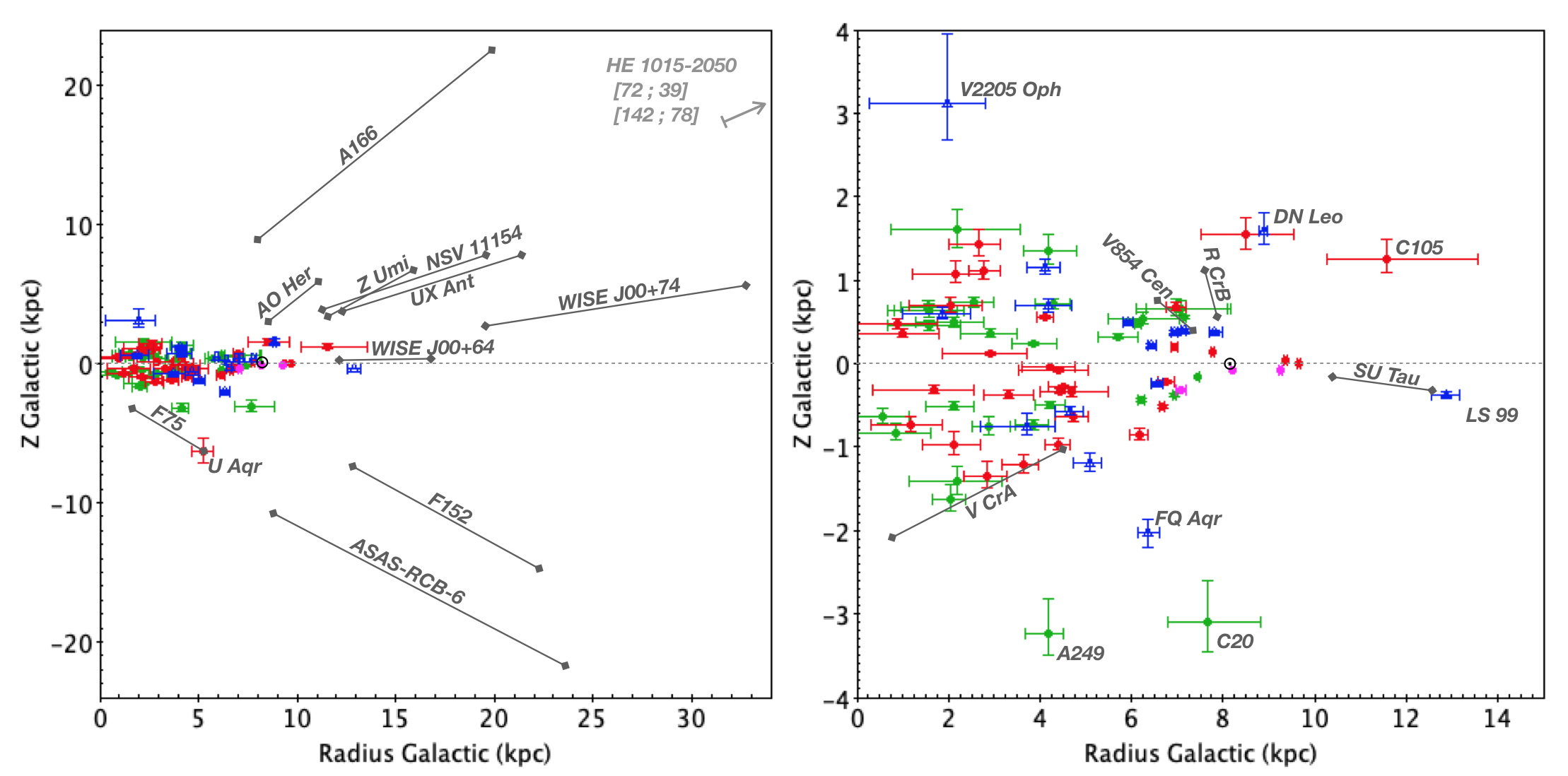}
\caption{Rectangular Galactocentric (X, Y, Z) distribution of RCB (red dots), dLHdC (green dots), DYPer type (purple dots), and EHe stars (blue triangles) with a valid astrometric fit and a parallax S/N higher than 3. From top to bottom: X versus Y, Z versus X and Z versus the Galactic radius. On the left side, large-scale views presenting stars located in the Galactic halo. On the right side, closer-in views. We have indicated with grey lines the possible distances for some HdC stars that do not have significant parallaxes measured by Gaia DR3. We used a V absolute magnitude range between -5 and -3.5 mag (resp. -4.5 and -3 mag) for the warm (resp. cold) HdC stars, i.e. the HdC stars with a temperature classification lower or equal to HdC4 (resp. higher than HdC4). The location of the far away dLHdC star, HE 1015-2050, is indicated with an arrow and the coordinates of the two distances extremities in square brackets.}
\label{fig_XYZ_Distrib}
\end{figure*}

In Figure~\ref{fig_XYZ_Distrib}, we present a top view (X vs Y) and two side views (Z vs X and Z vs radial distance, i.e., $\sqrt{X^2+Y^2}$) illustrating the spatial distribution of all our selected targets, including some HdC stars for which distance measurements are unavailable. For these HdC stars, we used an absolute V magnitude range of [-5, -3.5] mag for the warmer HdC stars (HdC index between 0 and 4) and [-4.5, -3] mag for the colder ones (HdC index greater than 4). In the diagrams on the left side, stars located in the Galactic halo can be identified, as their expected distances lie significantly above the Galactic plane. Specifically, there are seven RCB stars (U Aqr, AO Her, NSV 11154, UX Ant, Z Umi, ASAS-RCB-6, and WISE J00+74) and four dLHdC stars (F152, F75, A166, and HE 1015-2050). The last one, HE 1015-2050 is even located in the far outer rim of our Galaxy. The close-up views on the right side confirm the extensive distribution around the Galactic plane for all HdC and EHe stars.

\begin{figure*}
\centering
\includegraphics[scale=0.40]{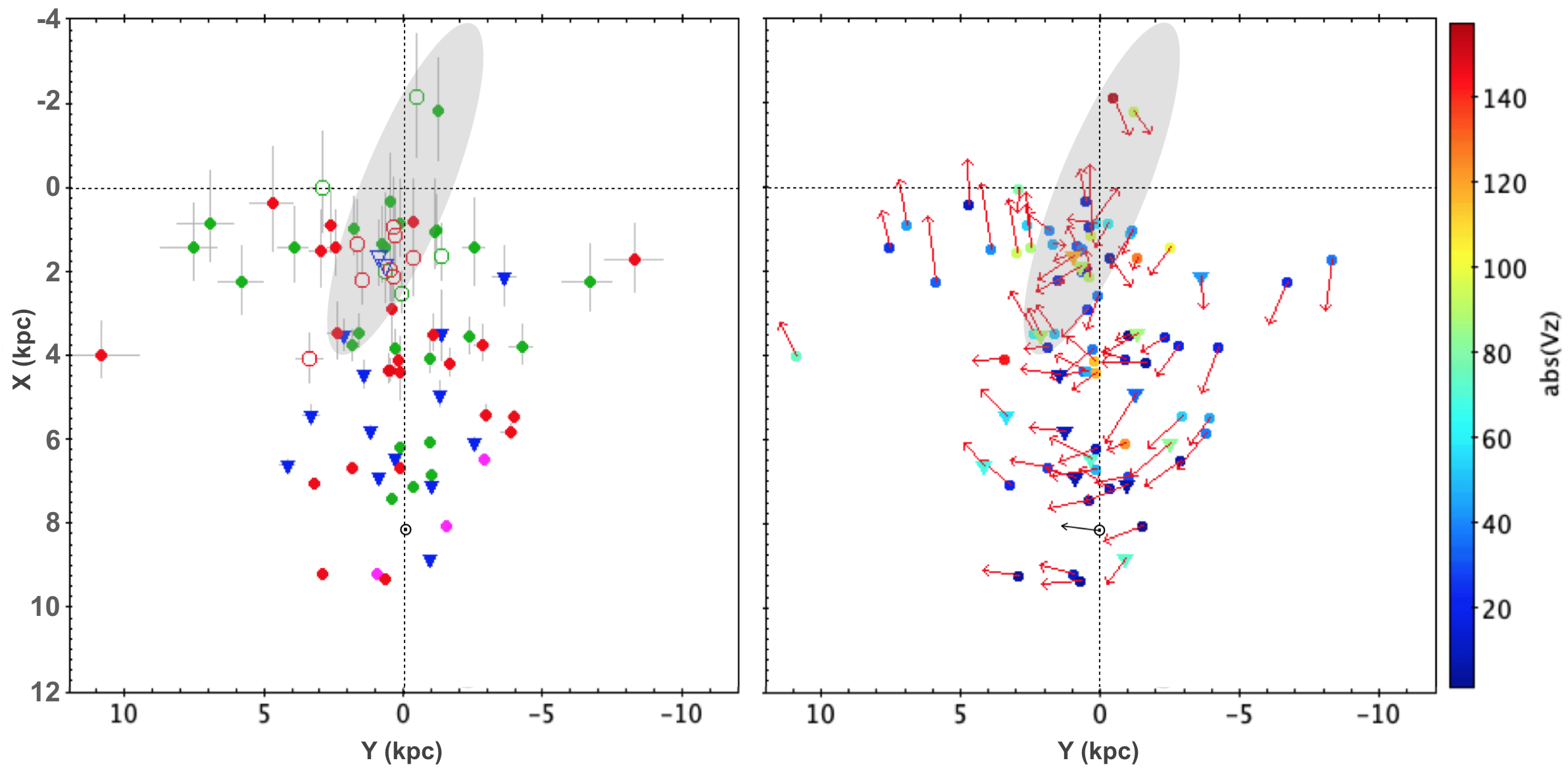}
\caption{Rectangular Galactocentric X versus Y distribution of our selected targets. The grey area represents the approximate position and shape of the Galactic bulge. The Sun is indicated at (X,Y)=(8.2, 0) kpc. Left: Selected targets with peculiar velocities higher than 200 km s$^{-1}$ relative to LSR are represented with open symbols, while the others are represented with filled symbols. Right: Red arrows represent the tangential velocities, $\sqrt{V_\phi^2+V_r^2}$, of the targets, which are colour-coded for their absolute velocities perpendicular to the plane, V$_z$. The colour code is identical to that of Figure~\ref{fig_CylindVelocities} with large dots and circles representing HdC stars and down triangles representing the EHe stars.}
\label{fig_ArrowsVelocities}
\end{figure*}

\begin{figure}
\centering
\includegraphics[scale=0.41]{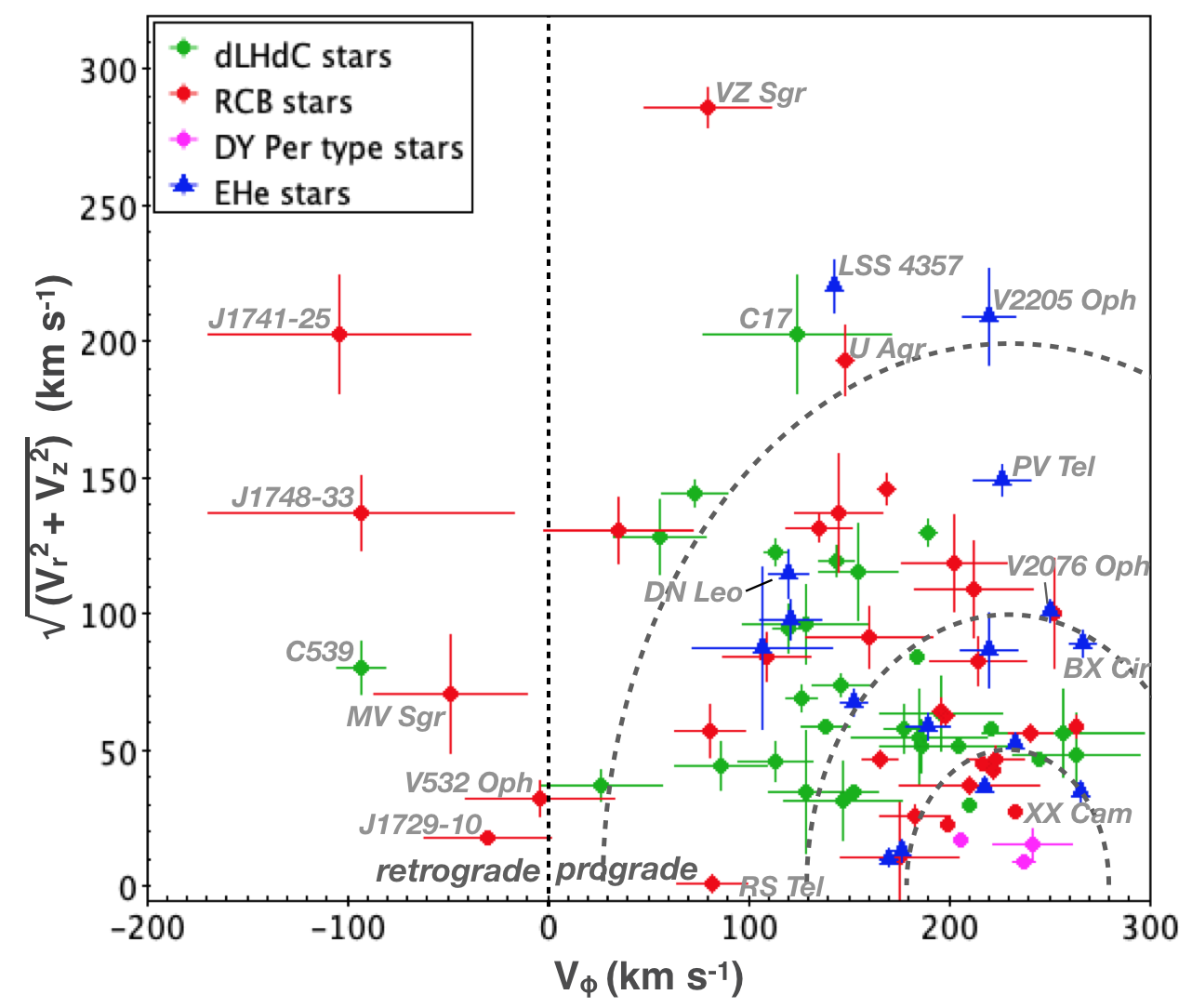}
\caption{Toomre diagram of our selected targets using cylindrical Galactocentric velocities. The LSR velocity is around (V$_r$,V$_\phi$,V$_z$) = (0, 230, 0) km s$^{-1}$. The dashed concentric ellipses represent the total peculiar velocity, 50, 100 and 200 km s$^{-1}$ compared to LSR. The dashed vertical line separates the prograde and retrograde orbits. Names of some targets are indicated, the aliases J1729-10, J1741-25 and J1748-33 are respectively for the RCB stars WISE J172951.80-101715.9, WISE J174119.57-250621.2 and WISE J174851.29-330617.0.}
\label{fig_CylindVelocities}
\end{figure}

We computed the velocities in cylindrical Galactocentric coordinates (V$_r$, V$_{\phi}$, V$_z$). In Figure~\ref{fig_ArrowsVelocities} (right side), we present a top view (X vs Y) representation of our targets, where their respective tangential velocities $\sqrt{V_{\phi}^2+V_r^2}$ are illustrated as vectors. The majority of targets exhibit circular motion in a prograde direction. This prograde rotation is also evident in the Toomre diagram (Figure~\ref{fig_CylindVelocities}), where the perpendicular velocities $\sqrt{V_r^2+V_z^2}$ are plotted against the azimuthal velocity. Stars with prograde velocities display a distribution consistent with that of a thick disk, with total velocities differing from that of the local standard of rest (LSR) by 30 to 200 km s$^{-1}$. Furthermore, by selecting only HdC stars, located outside the bulge area, whose total velocities are within 200 km s$^{-1}$ of the LSR, we determined a vertical scale height (h$_z$) consistent with 900 pc. This value is characteristic of a thick disk distribution. The scale height value obtained for EHe stars is similar to that found for HdC stars.

Among the stars we examined, 15 exhibit total velocities lagging that of the LSR by more than 200 km s$^{-1}$. Notably, 6 HdC stars, namely WISE J174119.57-250621.2, WISE J174851.29-330617.0, WISE J172951.80-101715.9, MV Sgr, V532 Oph, and the dLHdC star C539, appear to have retrograde orbits. Interestingly, we found that the majority of these 15 stars are situated within or near the Galactic bulge (as denoted by the grey area in Figure~\ref{fig_ArrowsVelocities}, left side). The Galactic bulge region is known for its high velocity dispersion \citep{2012AJ....143...57K}.
 
\subsection{Orbits \label{sec_orbits}}

The orbital characteristics of each star were determined by direct integration forwards in time for 5 Gyrs using again the GALPY package. We choose their \textit{MWPotential14} model for the Milky Way's gravitational potential, which includes a bulge modelled as a power-law density profile with an exponential cut-off, a Miyamoto-Nagai Potential disc \citep{1975PASJ...27..533M}, and a dark matter halo described by an NFW potential \citep{1996ApJ...462..563N}. From these simulations, we extracted several parameters related to the orbits of our targets, including the maximum (apocentre, R$_a$) and minimum (pericentre, R$_p$) distances from the Galactic centre, the maximum vertical amplitude (Z$_{max}$), the eccentricity (e), the energy (E), and the Z component of angular momentum (L$_z$). For each star, we conducted 1000 simulations that utilised different inputs for distance, proper motions, and radial velocity. Their values were picked randomly assuming Gaussian distributions of their respective uncertainties. Thus, we retrieved for each of these parameters the values corresponding to the maximum likelihood of their respective binned distributions. These probability distribution functions were then examined to derive asymmetric uncertainties at one sigma. These values are listed in Table~\ref{tab.OrbitGalpy} and \ref{tab.OrbitGalpy2} for each star studied.

\begin{figure}
\centering
\includegraphics[scale=0.37]{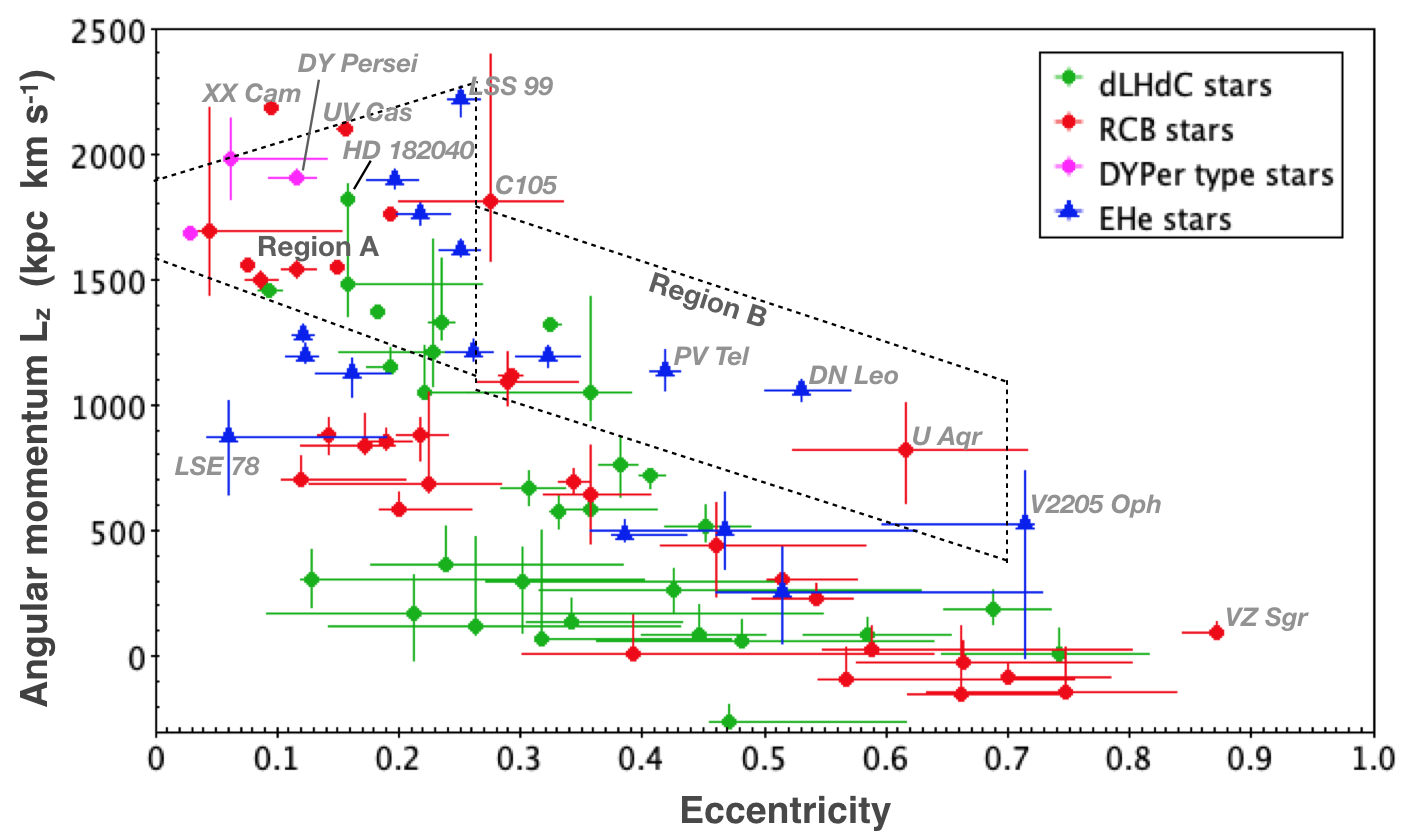}
\caption{Z component of the angular momentum, L$_z$, versus the orbit eccentricity for the 15 selected targets whose 2-sigma errors on distance are lower than 500 pc. Two regions defined by \citet{2006A&A...447..173P} are indicated by the dotted lines. They correspond to areas occupied mostly by thin disk (region A) and by thick disk (region B) WDs, all located up to 3 kpc of the Sun. The latter group of stars are of higher eccentricity (e$>$0.27) and lower angular momentum.}
\label{fig_Lzvse}
\end{figure}

In a diagram illustrating angular momentum (L$_z$) versus eccentricity, \citet{2006A&A...447..173P} defined two distinct regions using WDs located mostly within 3 kpc from the Sun: region A, characterised by low eccentricity (e $<$ 0.27) and L$_z$ around 1800 kpc km s$^{-1}$, where the majority of WDs belonging to the thin disk structure would be distributed; and region B, representing an area occupied mostly by thick-disk stars with higher eccentricities and lower angular momenta. We have reproduced this diagram in Figure~\ref{fig_Lzvse} and included our targets. Utilising this classification scheme, we observe that half of our targets reside within the two regions related to the disk structures. Meanwhile, the remaining half exhibits lower angular momentum across a broader range of eccentricities corresponding for the most part to stars belonging to the Galactic bulge. Most of the ones located within region A have peculiar velocities with respect to the local standard of rest (LSR) lower than 80 km s$^{-1}$.  XX Cam and UV Cas exhibit the most typical thin disk circular orbits characterised respectively by low eccentricities of about 0.1 and 0.16 and maximum vertical extensions, Z$_{max}$, of only 70 and 60 parsecs.

Finally, it is worth noting that a distinct group of stars emerged from our study: all three DYPer type stars, including DY Persei itself, exhibit characteristics consistent with a thin disk distribution. This finding is quite surprising as it suggests that DYPer type stars may originate from a younger stellar population group compared to HdC and EHe stars. However, the number of DYPer type stars analysed in our study is small, and further confirmation is required to solidify this observation.

\section{Discussion on population synthesis simulation and implications on galactic stellar structure membership \label{sec_discussion}}

\begin{figure*}
\centering
\includegraphics[scale=0.44]{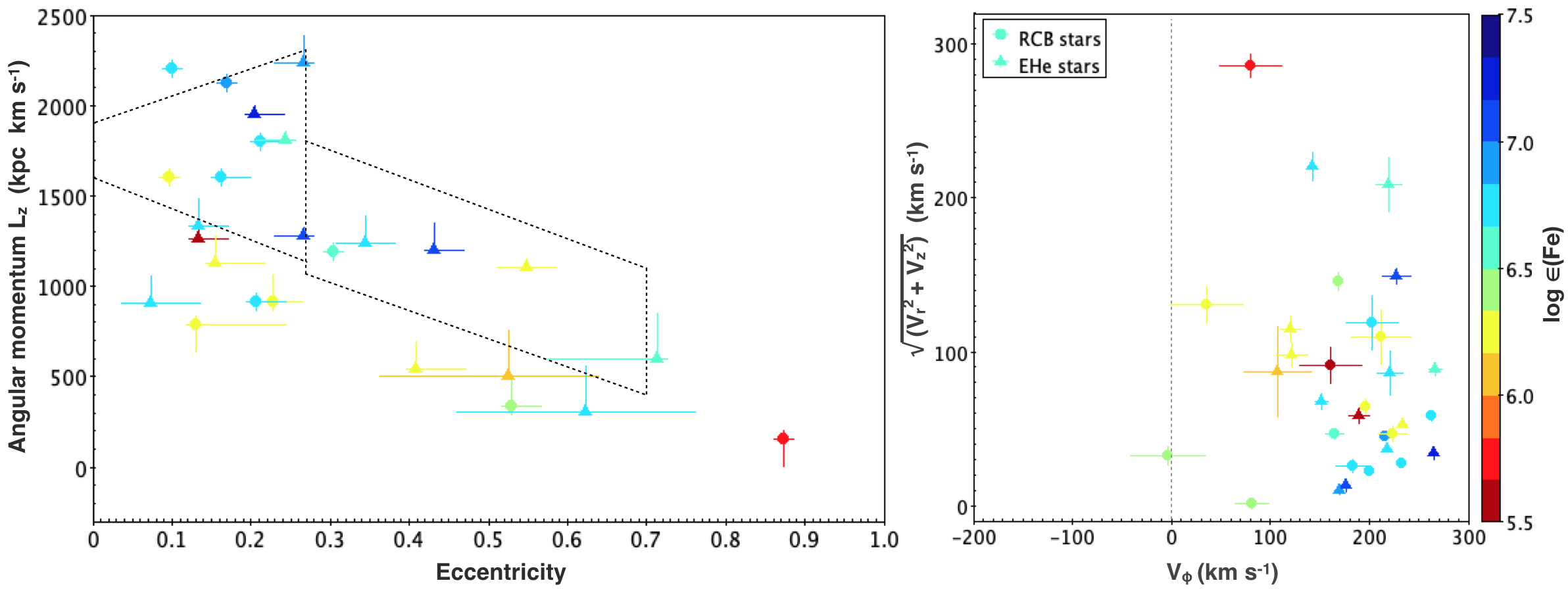}
\caption{Similar diagrams to those in Figure~\ref{fig_Lzvse} but only with RCB and EHe stars with an abundance analysis published. They are colour-coded with the iron abundances found in \citet{2011MNRAS.414.3599J} and \citet{2017PASP..129j4202H}.}
\label{fig_metallicity}
\end{figure*}

We have compelling evidence that HdC and EHe stars are distributed across all three major old stellar structures of the Galaxy, namely the thick disk, the bulge, and the halo. Interestingly, we note also that about a quarter of the stars studied exhibit orbits characteristic of the thin disk, with XX Cam and UV Cas being the most prominent examples. However, this group of stars does not display velocity dispersions clustered around the local standard of rest (LSR), but have a median velocity value with respect to LSR of about 55 km s$^{-1}$. This suggests that we may be observing the end tail of a thick disk distribution or a population with characteristics of both a thicker thin disk and a thinner thick disk \citep{2017A&A...608L...1H}. The fact that we measured a scale height of about 900 pc using all the disk stars supports the thick disk hypothesis. However, when we consider all HdC and EHe stars characterised by low eccentricity and high angular momentum L$_z$ (region A from Fig.~\ref{fig_Lzvse}), we measure a scale height that is noticeably lower with a value of about 600 pc. Furthermore, when we consider the additional information on metallicity (Figure~\ref{fig_metallicity}) obtained from atmospheric abundance analysis \citep{2011MNRAS.414.3599J,2017PASP..129j4202H}, an interesting pattern emerges. We observe that stars exhibiting low velocity dispersion, low eccentricity, and a high L$_z$ values also tend to have higher iron abundances. For instance, XX Cam and UV Cas, which are listed by \citet{2011MNRAS.414.3599J} as RCB stars with the highest iron abundance, possess characteristic thin disk-like orbits. This suggests potentially the presence of distinct groups within HdC and EHe stars, characterised by different kinematics and potentially different ages. A link between Galactic stellar structure and peculiar abundances was already noticed by \citet{2022A&A...667A..85C} with 5 Galactic HdC stars that tend to lie in the halo. Furthermore, \citet{2008A&A...481..673T} measured a scale height for some RCB stars, likely situated within the Galactic bulge, which is compatible with a thin disk distribution. Hence, the potential presence of an HdC star population distributed within a thin disk-like structure cannot be dismissed and warrants further investigation.

\subsection{Population synthesis simulation and delay time distribution}

In the context of the double degenerate scenario, HdC stars (helium shell-burning giant stage) and subsequently EHe stars (hot phases during which their atmospheres rapidly shrink to become WDs) are expected to represent short-lived phases following the merger of two WDs. Altogether, these phases typically last for approximately 100,000 years or less \citep{2002MNRAS.333..121S}. Such short duration on the scale of the Milky Way's age, gives us an opportunity to link each HdC star to its original stellar system, and thus to test the predictions made by population synthesis simulations.

\begin{figure}
\centering
\includegraphics[scale=0.6]{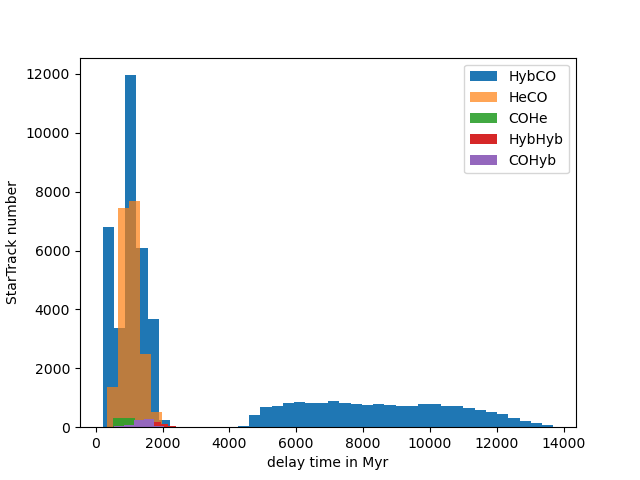}
\caption{Delay time distribution of five distinct WD merger channels in the intermediary total mass range and for a solar metallicity. The heaviest channel HybCO is the only one presenting a bimodal distribution with a long time delay.}
\label{fig_DTD}
\end{figure}

We performed such simulations of close binary systems using StarTrack \citep{2015ApJ...809..184K,2017NatAs...1E.135C,2022A&A...667A..83T}. Details on the initial stellar distributions (i.e. mass of each star in terms of initial mass function, mass ratio, orbital separations and eccentricities) are given in \citet[Sect.2]{2019MNRAS.484..698R}. Our results reveal interesting characteristics of the more massive HybCO merger channel corresponding to a merger between a `hybrid' COHe WD \citep{1996MNRAS.280.1035T,2019MNRAS.482.1135Z} and a CO WD. These types of WD mergers are also more numerous compared to the other WD merger types we consider and are formed through a stable Roche lobe overflow (RLOF) phase when the primary star fills its Roche lobe on the main sequence or in the Hertzsprung gap, continuing through to when the star is a red giant, followed later by a common-envelope phase where the secondary star loses its envelope as a red giant or an AGB star. \citet{2022A&A...667A..83T} describe a bimodal delay time distribution for this channel, where approximately 60\% of systems merge within 2 billion years after the original stellar systems formed, while the remaining 40\% take at least 5 billion years or even longer, potentially up to the age of the Universe. Mergers with shorter (resp. longer) delay times encounter a common-envelope phase when the mass-losing star is a red giant (resp. on the AGB branch). That delay time distribution is presented in Figure~\ref{fig_DTD} for the HybCO channel, along with other merger channels considered also as plausible progenitors of HdC stars, including HeCO, COHe, HybHyb, and COHyb, albeit with lower median total mass \citep[Fig. 13]{2022A&A...667A..83T}. In that notation, the WD created by the initially more massive star on the zero-age main sequence (ZAMS) is indicated first. We have excluded the least-massive HeHe channel and the heaviest COCO and CO+ONe/ONe+CO channels, as neither are favoured WD mergers for HdC star formation. It is worth noting that these results are based on simulations with solar metallicity, though the primary distribution characteristics persist at lower metallicities. A simulation conducted with a metallicity similar to that of the SMC yielded a comparable bimodal distribution for the HybCO channel, albeit with a slightly higher fraction ($\sim$70\%) of mergers occurring with a short time delay. 

Remarkably, among all the channels explored within the intermediate total mass range (between $\sim$0.6 and $\sim$1.05 solar masses), the HybCO channel stands out as the only one capable of supporting merging processes occurring long after star formation (between 5 and 13.5 billion years). This observation aligns with our findings that a significant portion of HdC stars belong to the old Galactic stellar structures.

Assuming that all HdC stars are formed exclusively through the double degenerate HybCO channel, we have convolved the HybCO delay time distribution with the Milky Way's star formation history derived by \citet[Fig. 9]{2019ApJ...887..148F}. Our analysis indicates that approximately 70\% of HdC stars likely originate from merger events with long time delays ($>$5 Gyrs), while the remaining 30\% likely arises from shorter time delays ($<$2 Gyrs). As a result, it should not be surprising to find HdC stars also located within the thin disk of the Milky Way.

Another interesting consequence of the scenario described above concerns the population of HdC stars in the LMC. Indeed, the star formation history of this galaxy reveals a significant burst of star formation within the past 2 billion years \citep[see Fig. 5]{2021MNRAS.508..245M}. This burst is primarily concentrated within the LMC's bar structure, coincidentally the preferred location for most HdC stars discovered in the LMC \citep[Fig. 2]{2009A&A...501..985T}. This alignment may suggest that, conversely to the HdC star population of the Milky Way, a majority of the LMC HdC stars originate from the shorter time delay process of the HybCO channel.

As mentioned earlier, there is a possibility that a fraction of HdC stars belongs to the thin disk. This suggests that they may originate from a relatively young stellar population, which could align with the peak at about 1 billion years in the HybCO channel's delay time distribution. On the other hand, the remaining HdC stars found in the old Galactic stellar structures would correspond to longer time delays, potentially explaining the wide range of lower metallicities observed in HdC stars \citep{2011MNRAS.414.3599J}.

\begin{figure*}
\centering
\includegraphics[scale=0.44]{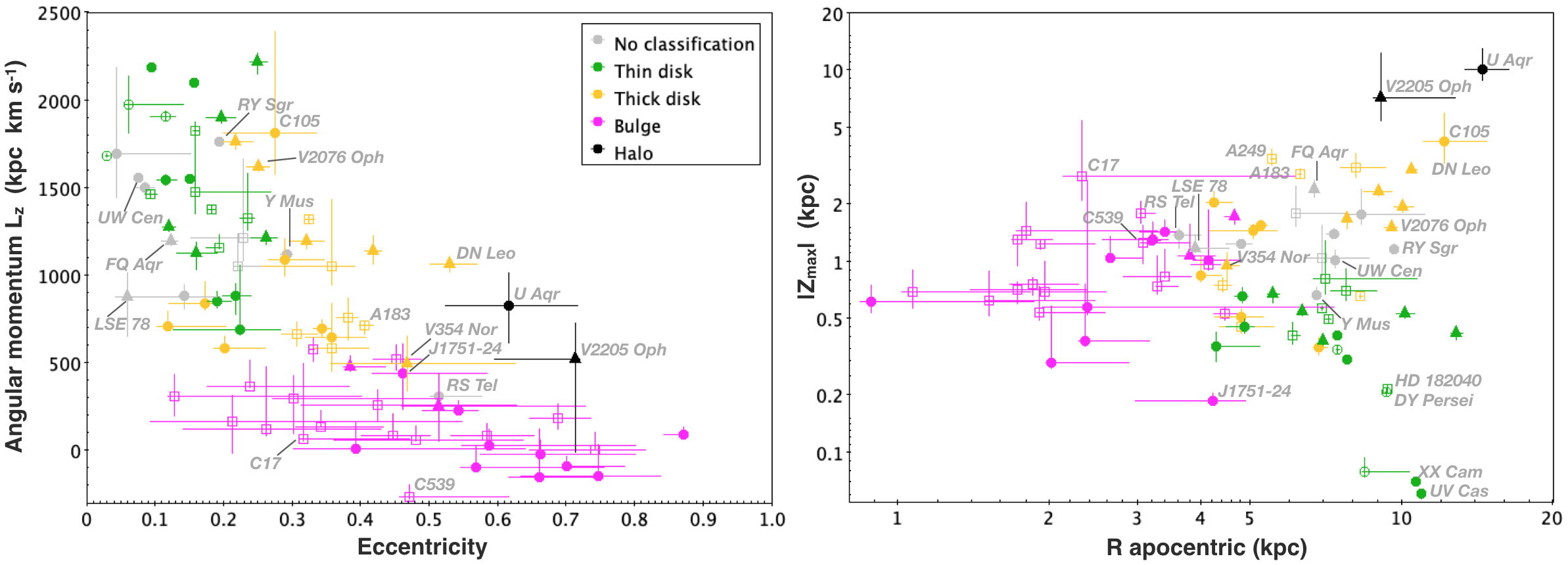}
\caption{Orbital parameters distributions for all targets, that is, EHe (up triangles), RCB (full dots), dLHdC (squares), and the DYPer type (circles) stars, classified within a Galactic structure as indicated in the legend. Left: Z component of the angular momentum, L$_z$, versus the orbit eccentricity. Right: Maximum vertical amplitude versus the apocentric distance in kilo-parsec. Names of some targets are indicated in grey, J1751-24 corresponds to the RCB star WISE J175107.12-242357.3.}
\label{fig_PerStructures}
\end{figure*}

\subsection{Classification of selected targets within a Galactic structure}

Below we present a list of the HdC, EHe, and DYPer type stars that we classified within one of the four main Galactic stellar structures and a description of the empirical criteria used for this categorisation. A visualisation of this classification can be seen in the two diagrams of figure~\ref{fig_PerStructures}, representing  the angular momentum, L$_z$, versus the orbit eccentricity and the maximum vertical amplitude, Z$_{max}$, versus the apocentric distance, R$_a$. Each Galactic substructure occupies preferential areas in each diagram.

\textbf{Thin disk:} We gave a thin disk label to all targets with an eccentricity lower than 0.27, a maximum amplitude $|$Z$_{max}|$ perpendicular to the galactic plane within 1 kpc, an angular momentum L$_z$ value higher than 700 kpc km s$^{-1}$, thus probing thin disk objects down to about 3 kpc from the Galactic centre, and a difference in total velocity with respect to LSR ($\bigtriangleup$v$_{LSR}$) lower than 80 km s$^{-1}$. That last threshold value corresponds to about 2 sigma of the total velocity dispersion measured from a large sample of local thin disk red giant stars studied by \citet{2022ApJ...932...28V}. We underline that not all of these stars belong to the Galactic thin disk structure as it is not possible to differentiate individual star between the thin disk and the end tail of the thick disk kinematic distributions. Atmospheric metallicity studies should help to disentangle that issue.
\begin{description}[font=-]
\item \textit{RCB stars:} GU Sgr, RT Nor, SV Sge, UV Cas, V482 Cyg, WISE J185525.52-025145.7 and XX Cam
\item \textit{dLHdC stars:} A814, C526, HD 148839, HD 173409, HD 182040 and M38
\item \textit{EHe stars:} LSS 99, LS IV +06 2, NO Ser, V4732 Sgr and V821 Cen
\item \textit{DYPer type stars:} DY Persei, ASAS-DYPer-1 and -2
\end{description}

\textbf{Thick disk:} We kept all targets located outside the bulge area as defined in Figure~\ref{fig_ArrowsVelocities} with a total velocity value ranging between 80$<\bigtriangleup$v$_{LSR}<$200 km s$^{-1}$ and an apocentric distance higher than 4 kpc. We note nevertheless that the dLHdC star A249 has the highest orbit inclination of all the stars listed below and a high maximum vertical amplitude of $\sim$3.4 kpc, it could be a member of the halo.

\begin{description}[font=-]
\item \textit{RCB stars:} ASAS-RCB-3, ASAS-RCB-8, C105, FH Sct, V2552 Oph, WISE J175521.75-281131.2 and WISE J184158.40-054819.2
\item \textit{dLHdC stars:} A182, A183, A249, A977, C20 and HD 137613
\item \textit{EHe stars:} BX Cir, DN Leo, PV Tel, V1920 Cyg, V2076 Oph and V354 Nor
\end{description}

\textbf{Bulge:} We kept all targets located inside the bulge area as defined in Figure~\ref{fig_ArrowsVelocities}, and whose vertical Z distances to the Galactic plane are within 2 kpc. Then, we enlarged our criteria by incorporating four targets (listed below in square brackets) located within 3 kpc from the Galactic centre. Their Z$_{max}$ and R$_a$  distances are respectively lower than 2 and 4 kpc, and their angular momentum L$_Z$ lower than 500 kpc km s$^{-1}$. Finally, we noticed that the dLHdC star, C17, that is located near the edge of the bulge area, presents a Z$_{max}$ value higher than 2 kpc by about one sigma and a large asymmetric R$_a$ distribution towards longer distances (see Fig.~\ref{fig_PerStructures}). Thus, C17 could also be a member of the thick disk structure.
\begin{description}[font=-]
\item \textit{RCB stars:} ASAS-RCB-10, MV Sgr, V532 Oph, VZ Sgr, WISE J172447.52-290418.6, WISE J172951.80-101715.9, WISE J174119.57-250621.2, WISE J174851.29-330617.0, WISE J175107.12-242357.3 and [WISE J175749.76-075314.9]
\item \textit{dLHdC stars:} A223, A226, A770, A811, B42, B563, [B564], B565, B566, B567, C17, C27, C38, C528, [C539] and [P12]
\item \textit{EHe stars:} LSS 4357 and V2244 Oph
\end{description}



\textbf{Halo:} We simply kept the two targets, U Aqr and V2205 Oph, with a maximum vertical amplitude, $|$Z$_{max}|$, higher than 5 kpc and with $\bigtriangleup$v$_{LSR}>$200 km s$^{-1}$. We also added all HdC stars whose Gaia DR3 parallaxes are not known but have been located at vertical distances away from the Galactic plane higher than 3 kpc, using plausible absolute magnitude ranges (see Figure~\ref{fig_XYZ_Distrib}).
\begin{description}[font=-]
\item \textit{RCB stars:} AO Her, ASAS-RCB-6, NSV 11154, U Aqr, UX Ant, WISE J004822.34+741757.4 and Z Umi
\item \textit{dLHdC stars:} A166, F75, F152 and HE 1015-2050
\item \textit{EHe stars:} V2205 Oph
\end{description}

\textbf{Not classified:} We did not classify the 11 targets listed below using the pragmatic criteria we have just enumerated. Among them, two have velocities $\bigtriangleup$v$_{LSR}$ between 80 and 200 km s$^{-1}$: the EHe star LSE 78 and the RCB stars, RS Tel. They are likely members of the thick disk structure, but they were not selected as such due to apocentric distance values just below the 4 kpc threshold chosen (see Fig.~\ref{fig_PerStructures}). All other 9 targets have a velocity $\bigtriangleup$v$_{LSR}$ lower than 80 km s$^{-1}$ and could thus belong either to the thin disk or the thick disk structures. However, we should notice that the 3 following targets, A980, DY Cen and FQ Aqr, have maximum vertical amplitude values higher than 1.5 kpc. Therefore, they more likely belong to the thick disk. Conversely, Y Mus is more likely within the thin disk structure as its maximum vertical amplitude value is only 0.66$^{+0.07}_{-0.04}$ kpc with an eccentricity of 0.29, that is just above the threshold we used. All targets that are more likely members of the thick disk are listed with brackets.
\begin{description}[font=-]
\item \textit{RCB stars:} (DY Cen), HD 175893, (RS Tel), RY Sgr, S Aps, UW Cen and Y Mus
\item \textit{dLHdC stars:} (A980) and C542
\item \textit{EHe stars:} (FQ Aqr) and (LSE 78)
\end{description}

Five Galactic HdC stars listed as chemically peculiar (i.e., enhanced s-processed material such as strontium and barium) were identified by \citet{2022A&A...667A..85C}. The study also observed a tendency for these stars to be located in the Galactic halo. We classified three of them in the Galactic halo (U Aqr,  HE 1015-2050 and A166),  one in the thick disk structure (A249) but with the highest orbit inclination among the stars listed as thick disk stars that it could thus well be a member of the halo, and the last one in the Galactic bulge (C539). That last dLHdC star is not located within the elliptical Bulge area defined in Fig.~\ref{fig_ArrowsVelocities}, and thus its Bulge membership is debatable with also the lowest angular momentum L$_z$ value registered among all the stars studied.

Finally, we studied two bright RCB stars, V CrA and R CrB, for which we do not have valid distance measurements. Therefore, we used a range of plausible absolute magnitudes, as discussed in Section~\ref{sec_brightHdC}. Based on the resulting distances, we found that V CrA almost certainly lies within the Galactic bulge (see Figure~\ref{fig_XYZ_Distrib}). R CrB has measured proper motions of pmra$=$-2.1$\pm$0.4 and pmdec$=$-11.5$\pm$0.5 mas/yr, according to Hipparcos \citep{1998PASA...15..179C}. Utilising these proper motion values and a heliocentric radial velocity of 22 km s$^{-1}$, we found that in all cases R CrB's total velocity $\bigtriangleup$v$_{LSR}$ is lower than 80 km s$^{-1}$ and that its orbit's eccentricity is lower than our threshold of 0.27. Two characteristics of the thin disk structure. However, in the two most favoured scenarios in which R CrB has an absolute magnitude M$_V$ ranging between -5 and -4.5 mag, its resulting maximum vertical amplitude is higher than our 1 kpc threshold, with respectively a value of $\sim$1.5 and $\sim$1.2 kpc. Therefore, R CrB is likely a member of the thick disk structure. The iron abundance values, log $\epsilon$(Fe), documented by \citet{2011MNRAS.414.3599J} for V CrA and R CrB are respectively 5.5 and 6.5.

\subsection{Comparison with \citet{2023MNRAS.tmp.3151M} classification}

During the submission process, an article related to a similar analysis focusing principally on EHe stars has been brought to our attention by our colleagues \citet{2023MNRAS.tmp.3151M}. We both used identical Galactic potential models to compute orbits using the Galpy package for EHe stars and HdC stars with the goal to classify each of them within a Galactic substructure. HdC stars were studied by \citet{2023MNRAS.tmp.3151M} as a statistical sample used only for comparison with EHe stars. We cross-matched our results and discuss below the differences we found for both groups of stars. 

\subsubsection{EHe stars sample}

\citet{2023MNRAS.tmp.3151M} have studied 27 EHe stars, including all the 16 EHe stars we focused on in our analysis. The eleven EHe stars remaining have atmospheres of hotter effective temperature than our sample. We have calculated orbits also for these stars (see Table~\ref{tab.OrbitGalpy2}) for completeness and consistency. A first direct comparison of the orbital parameter values shows that both analysis agree well on EHe stars, except in two cases: V2076 Oph and EC 20111–6902. For V2076 Oph, both teams agree that the orbital parameter values listed in \citet{2023MNRAS.tmp.3151M} are incorrect due to an error in the sign of the input radial velocity value. Consequently, V2076 Oph, with a high maximum vertical amplitude value of $\sim$1.5 kpc, has to be classified as a member of the thick disk substructure. Regarding EC 20111–6902, we did not publish any orbital parameters as we do not trust the parallax measurements provided by Gaia DR3. Indeed, its astrometric fit result was accompanied with a quality RUWE value higher than 2, invalidating the results and thus any distance estimation.

On the 25 remaining EHe stars, we agree on the Galactic substructure classification for 20 of them. We list below the 5 stars for which we have a disagreement and discuss the reasons. 

\textit{V354 Nor (= CoD-48 10153)}: it is listed as a halo star by \citet{2023MNRAS.tmp.3151M} due to its position with a low V velocity value in their Toomre diagram using a rectangular Galactic coordinate system (U, V, W). However, V354 Nor is not located in the solar neighbourhood where velocities comparison in such coordinate system should only be applied but at a position in the disk situated at about 90 degrees from the Sun. Its velocities in cylindrical Galactocentric coordinates and orbital parameters are typical of a thick disk star.

\textit{LSE 78 (= CoD-46 11775)}: \citet{2023MNRAS.tmp.3151M} classified it as a bulge member as it falls near the edge of the bulge area they have defined as $|$X$|\leq$4 kpc, $|$Y$|\leq$1.5 kpc, and $|$Z$|\leq$1.5 kpc (see their figure 1). However, this definition corresponds to a bar structure aligned with the Sun-Galactic centre direction. We have chosen a more commonly used bar major axis orientation rotated by about 20 degrees to that direction \citep{2016PASA...33...25Z} and thus, LSE 78 is not located within our bulge area. Furthermore, LSE 78 orbital parameter values present all the characteristics in eccentricity (e$<$0.2), and in pericentric and apocentric distances of a star that belong to a disk and it is more probably a member of the thick disk structure.

\textit{DN Leo (= BD+10 2179)}: we classified it as a thick disk star while \citet{2023MNRAS.tmp.3151M} listed it as a halo star despite the fact we found very similar orbital parameter values. The distinct classification is simply caused by differences in orbital parameter boundaries used in both analyses for the definition of Galactic substructures. We have a very conservative approach as we require a maximum vertical amplitude Z$_{max}$ higher than 5 kpc to classify a star as a halo star, while \citet{2023MNRAS.tmp.3151M} consider a halo membership for stars with Z$_{max}>$ 1.2 kpc and Ra$>$7 kpc or stars with high orbital inclination. We found that DN Leo's Z$_{max}$ is about 3 kpc with an orbit inclination that is on par with other thick disk stars. Nevertheless, both cases are indeed arguable. 

\textit{BD+37 1977}:  \citet{2023MNRAS.tmp.3151M} classified it as a halo star, but we think that its velocities and orbital parameter values are of a classical thick disk stars with an elliptical orbit (e$\sim$0.5) and apocentric distance of 12.5 kpc. Its low maximum vertical amplitude of Z$_{max}\sim$1.5 kpc also supports such classification. We did not find any particularly high orbital inclination for that star.

\textit{BD+37 442}: \citet{2023MNRAS.tmp.3151M} classified it as a member of the thin disk structure. We classified it as a thick disk star because of its high difference in total velocity with respect to LSR of $\sim$106 km s$^{-1}$ and an eccentricity value above our threshold of 0.27. It is located within region B in our figure~\ref{fig_Lzvse}.

Overall, we support \citet{2023MNRAS.tmp.3151M}'s halo classification for 3 EHe stars of the 6 they listed as such. They are EC 19529-4430, EC 20236-5703 and V2205 Oph.

\subsubsection{HdC stars sample}

Orbital parameter values were also published by \citet[Table A1]{2023MNRAS.tmp.3151M} for 36 Galactic RCB and 8 dLHdC stars for which they had an RV measurement. However, among the RCB stars, we do not trust the Gaia DR3 astrometric fit results for 17 of them. We did not select them in our analysis as they underwent photometric decline events and thus they have astrometric fit results that are likely affected by the PSF chromaticity effect described earlier. These RCB stars are (names used in \citet{2023MNRAS.tmp.3151M} are given in brackets): UX Ant, V854 Cen, Z Umi, R CrB, ASAS-RCB-9 (IO Nor),  ASAS-RCB-12 (IRAS 16571-5011), ASAS-RCB-4 (GV Oph), V517 Oph (HV 7863), OGLE-GC-RCB-1 (Terz V 2637), ASAS-RCB-7 (V653 Sco), WX CrA, V CrA, V1157 Sgr, ES Aql, WISE J004822.34+741757.4 (IRAS 00450+7401), WISE J180550.49-151301.7 (IRAS 18029-1513), WISE J182943.83-190246.2 (ATO J277.4326 19.0462). We do not recommend the use of their orbital parameter values. This should also explain the unusual large range of absolute magnitude M$_V$ (between -8 and 8 mag) presented by \citet{2023MNRAS.tmp.3151M} for HdC stars in their colour-magnitude diagram (see their figure 3 for G$_{BP}$-G$_{RP}>$0.8 mag). A mixed between wrong distances due to erroneous Gaia DR3 astrometric fit results and the use of the Gaia mean photometric magnitudes, biased by the multiple photometric declines encountered, is likely to have caused such variations in absolute magnitude. A colour-magnitude diagram using magnitudes at maximum brightness is presented in \citet[Fig. 10]{2022A&A...667A..83T}. 

Among the remaining 19 RCB and 8 dLHdC stars we have in common, we did not simulate orbits for the dLHdC star F75 (= CD 35 13668) due to a poor constraint on its Gaia DR3 parallax value (Plx/ePlx$\sim$2). We agree on the Galactic substructure classification for only 14 stars. We list and discuss the remaining ones below grouped by similar reasoning.

\textit{A182, UW Cen, Y Mus and RT Nor}: the classification by \citet{2023MNRAS.tmp.3151M} for these 4 stars are affected by the use of velocities  calculated in a rectangular Galactic coordinate system (U, V, W) as already discussed for the EHe star V354 Nor above. Their classification as halo stars for the first 3 HdC stars and as a thick disk star for RT Nor is thus incorrect. 

\textit{V2552 Oph, RS Tel and WISE J172951.80-101715.9 (= AC Ser)}: the first two RCB stars were classified as bulge stars by \citet{2023MNRAS.tmp.3151M} but we classified them as thick disk stars for the same reasons that the ones discussed for LSE 78 above. Conversely, WISE J172951.80-101715.9 is within our defined bulge elliptical area rotated by 20 degrees to the Sun-Galactic centre direction, while it is not in the area defined by \citet{2023MNRAS.tmp.3151M} who classified it as a halo star.  We confirm that the WISE J172951.80-101715.9 orbital parameter values have all the characteristics of a bulge star with a highly elliptical orbit and a low angular momentum. 

\textit{HD 137613}: here the disagreement can be simply explained by different definition of the velocity boundary in both analyses. The separation between the thin and the thick disk membership is set at 100 km s$^{-1}$ for \citet{2023MNRAS.tmp.3151M} while we use a more conservative threshold of 80 km s$^{-1}$. With a difference in total velocity with respect to LSR of $\sim$95 km s$^{-1}$ and an orbit eccentricity of 0.32, we gave a thick disk membership to  HD 137613.

\textit{A183 (= SOPS IV e-67)}: \citet{2023MNRAS.tmp.3151M} classified it as a halo dLHdC star, while we gave it a thick disk membership despite the fact we both calculated consistent orbital parameter values. With Z$_{max}\sim$3 kpc and a distance R$_a\sim$6.3 kpc, the elliptical orbit (e$\sim$0.4) of A183 is still within the range expected of a thick disk star. The unusually high vertical velocity that is measured, V$_Z\sim$-123 km s$^{-1}$, is due to the fact that A183 is actually located near the extremity of its orbit. Nevertheless, we note that A183's orbit inclination is substantial and thus its classification as a halo star is arguable.

\textit{WISE J174119.57-250621.2 (= UCAC4 325–115052), WISE J174851.29-330617.0 and HD 173409}: we have a clear disagreement in orbital parameter results for these three stars resulting in different classification. We suspect an error occurring in \citet{2023MNRAS.tmp.3151M} simulations, perhaps due to some issues on the input distances (private communication), as in all three cases the resulting apocentric distances found are extremely high (above 20 kpc, and even above 40 kpc for the last two). We found that the first two RCB stars are members of the Galactic bulge considering they both belong to the bulge area as defined by us and even by \citet{2023MNRAS.tmp.3151M}, at a distance lower than 2 kpc from the Galactic centre, and because their respective orbits are typical of bulge stars with high eccentricity values and retrograde velocities (see Fig.~\ref{fig_CylindVelocities}). HD 173409 is a bright dLHdC star distanced by only $\sim$2.2 kpc from the Sun whose uncertainties on Gaia DR3 astrometric measurements are low. Its velocities and orbital parameter values are typically the ones of a thin disk star.

\section{Conclusion \label{sec_Concl}}

We examined the photometric and astrometric results from Gaia DR3 for HdC, EHe, and DYPer type stars, with a specific focus on potential biases affecting the heavy dust producers, namely the RCB stars. Our analysis revealed a noticeable variation in the values of ipd\_gof\_harmonic\_amplitude for RCB stars that went into their characteristic photometric decline phases during the Gaia DR3 observations. This suggests that changes in the PSF shape (likely induced by coma) have occurred while the RCB stars were reddened by dust clouds. Additionally, through Gaia light curves, we established a connection between poor astrometric results and RCB stars' photometric declines. The computational approach employed in Gaia DR3, utilising a single PSF model derived from a single colour index for all astrometric measurements, introduces a systematic shift away from the true position when the star's colour changes. This PSF chromatic effect was anticipated \citep{2006MNRAS.367..290J}, and while the assumption of using the same colour for all observations of a given source was considered to be a sufficiently good approximation for most sources in the Gaia collaboration, it hindered accurate results for certain variable stars \citep{2021A&A...649A...2L}. In the case of an RCB star, this issue is exacerbated by the star's substantial and unpredictable changes in luminosity and colour. These stars hold potential for validating the corrections applied for the PSF chromaticity effect in future Gaia astrometric releases.

We applied both astrometric and photometric criteria to select stars suitable for kinematic and spatial distribution studies. Among these stars, we compiled a sample of 28 RCB, 31 dLHdC, 18 EHe, and 3 DYPer type stars, all with parallax signal-to-noise ratios higher than 3. Additionally, we have compiled a comprehensive list of heliocentric RV measurements, comprising 23 observations obtained through our 2.3m/WiFeS spectroscopic survey, along with data collected from the literature for an additional 38 targets. These measurements serve as a valuable addition to the RV data already published in Gaia DR3.

Our analysis unequivocally confirms the presence of HdC stars in all major old stellar structures of the Galaxy, including the thick disk, the bulge, and the halo. EHe stars were also found within these three stellar structures, but we found only two that belong to the bulge and one, V2205 Oph, to the halo dynamical structure. More work is needed to discover EHe stars in these two oldest Galactic structures. The analysis presented by \citet{2023MNRAS.tmp.3151M} has begun to unveil few of those. We concur with the identification of an extra bulge EHe star and the recognition of two more halo EHe stars.

Surprisingly, some HdC and EHe stars also showcase orbital characteristics consistent with the thin disk. The RCB stars XX Cam and UV Cas stand out as primary examples with circular orbits closely confined near the Galactic plane. We labelled these group of stars as thin disk compatible, even if they may represent the end tail of a thick disk distribution. There is a hint that these stars exhibit higher metallicities, which would then suggest an origin from a relatively younger stellar population compared to other HdC and EHe stars with lower metallicities. We are working on increasing the sample of HdC stars with measured atmospheric abundances. Furthermore, we report that all three DYPer type stars studied display orbits characteristic of the thin disk. This suggests that DYPer type stars may originate from even younger stellar populations compared to HdC and EHe stars which would support a different formation channel.

In the context of the double degenerate scenario, population synthesis simulations indicate a broad range of time delays, potentially spanning up to the age of the Universe, between the formation of the original stellar systems and the subsequent helium-burning phases following WD mergers. The favoured pathway for the formation of HdC stars is through the HybCO merger channel, uniquely positioned within the intermediate total mass range ($\sim$0.6 to $\sim$1.05 solar masses) with the essential long delay time required to align with our observations.

The delay time distribution for the HybCO channel exhibits a bimodal pattern, comprising a younger component peaking at approximately 1 billion years and an older one starting at around 5 billion years and beyond. Incorporating the Galactic star formation history data, our analysis suggests that roughly 75\% of the current population of Galactic HdC stars originate from longer delay times, consequently being associated with the ancient Galactic stellar structures. The remaining 25\% should be observable within the thin disk structure, stemming from younger stellar populations. Conversely, in the case of the LMC, the majority of HdC stars visible within the bar structure are likely products of shorter time delay merger processes as a strong burst of star formation occurred in the past 2 Gyrs. Overall, the diverse range of time delays may be manifesting observable effects, presenting an opportunity for in-depth study. Further investigations, such as high-resolution spectroscopic observations of various HdC star populations, can provide valuable insights to validate the presented scenario.
 
Future Gaia releases, coupled with advancements in astrometric accuracy and the availability of astrometric time series in Gaia DR4, will greatly aid in validating and improving our analysis. Additionally, to further enhance our understanding, it will be crucial to conduct high-resolution spectroscopic follow-up studies on a larger sample of HdC and EHe stars, especially the ones we classified within Galactic stellar substructures. Currently, the number of stars with published abundance analyses remains limited, highlighting the need for more comprehensive spectroscopic investigations.

\begin{acknowledgements}

We are most grateful to Fr\'ed\'eric Arenou for his insights on the Gaia datasets. We are also very thankful to Asish Philip Monai and Simon Jeffery for their collaborative spirit and prompt responses during the comparative analysis of our results on orbital parameters values. Our work have significantly enhanced the quality of both studies. PT personally thanks Tony Martin-Jones for his usual highly careful reading and comments. PT acknowledges also financial support from “Programme National de Physique Stellaire” (PNPS) of CNRS/INSU, France. AJR was supported by the Australian Research Council through award number FT170100243.

We acknowledge with thanks the variable star observations from the AAVSO International Database contributed by observers worldwide and used in this research.

\end{acknowledgements}

\bibliographystyle{aa}
\bibliography{RCB_Distances_GAIA}


\begin{appendix}
\setlength{\tabcolsep}{2.5pt}

\onecolumn
\section{Additional tables}

\begin{longtable}{lrrrrrrrr}
\caption{Velocities in cylindrical Galactocentric coordinates and dynamical orbital parameter values of HdC stars
\label{tab.OrbitGalpy}}\\
\hline
Name & \multicolumn{1}{c}{V$_r$} & \multicolumn{1}{c}{V$_\varphi$} & \multicolumn{1}{c}{V$_z$} & \multicolumn{1}{c}{L$_z$} & \multicolumn{1}{c}{e} & \multicolumn{1}{c}{R$_{peri}$} & \multicolumn{1}{c}{$R_{apo}$} & \multicolumn{1}{c}{Z$_{max}$} \\ 
 & \multicolumn{1}{c}{km s$^{-1}$} & \multicolumn{1}{c}{km s$^{-1}$}  & \multicolumn{1}{c}{km s$^{-1}$} & \multicolumn{1}{c}{kpc km s$^{-1}$} & & \multicolumn{1}{c}{kpc}  & \multicolumn{1}{c}{kpc}  & \multicolumn{1}{c}{kpc} \\
\hline
\endfirsthead 

\caption{continued}\\
\hline
Name & \multicolumn{1}{c}{V$_r$} & \multicolumn{1}{c}{V$_\varphi$} & \multicolumn{1}{c}{V$_z$} & \multicolumn{1}{c}{L$_z$} & \multicolumn{1}{c}{e} & \multicolumn{1}{c}{R$_{peri}$} & \multicolumn{1}{c}{$R_{apo}$} & \multicolumn{1}{c}{Z$_{max}$} \\ 
 & \multicolumn{1}{c}{km s$^{-1}$} & \multicolumn{1}{c}{km s$^{-1}$}  & \multicolumn{1}{c}{km s$^{-1}$} & \multicolumn{1}{c}{kpc km s$^{-1}$} & & \multicolumn{1}{c}{kpc}  & \multicolumn{1}{c}{kpc}  & \multicolumn{1}{c}{kpc} \\
\hline
\endhead 
\hline
\multicolumn{9}{c}{\emph{Known Galactic RCB stars}}\\


\hline
ASAS-RCB-3  &  -100.1$\pm$20.0  &  253.2$\pm$1.9  &  -3.7$\pm$1.8  &  1090$^{+119}_{-96}$ & 0.29$^{+0.06}_{-0.02}$ & 3.47$^{+0.47}_{-0.39}$ & 6.86$^{+0.26}_{-0.15}$ & 0.35$^{+0.05}_{-0.03}$  \\
ASAS-RCB-8  &  -79.8$\pm$24.6  &  202.2$\pm$26.2  &  -87.6$\pm$9.2  &  582$^{+70}_{-18}$ & 0.20$^{+0.06}_{-0.02}$ & 2.51$^{+0.36}_{-0.08}$ & 4.25$^{+0.38}_{-0.14}$ & 2.02$^{+0.29}_{-0.1}$  \\
ASAS-RCB-10$^{\star}$  &  -94.7$\pm$1.4  &  35.8$\pm$36.4  &  -90.3$\pm$16.9  &  25$^{+93}_{-2}$ & 0.59$^{+0.21}_{-0.04}$ & 0.14$^{+0.51}_{-0.02}$ & 3.21$^{+0.24}_{-0.66}$ & 1.29$^{+0.31}_{-0.05}$  \\
C105  &  53.6$\pm$1.3  &  160.5$\pm$31.9  &  -74.3$\pm$14.0  &  1808$^{+580}_{-238}$ & 0.28$^{+0.06}_{-0.07}$ & 6.32$^{+3.14}_{-1.1}$ & 12.18$^{+2.62}_{-1.11}$ & 4.19$^{+1.67}_{-0.98}$  \\
DY Cen  &  3.1$\pm$16.8  &  210.5$\pm$34.8  &  37.1$\pm$3.3  &  1695$^{+492}_{-252}$ & 0.04$^{+0.11}_{-0.01}$ & 9.01$^{+0.69}_{-2.53}$ & 8.35$^{+2.8}_{-0.61}$ & 1.75$^{+0.63}_{-0.2}$  \\
FH Sct  &  -57.1$\pm$24.6  &  212.4$\pm$29.4  &  -93.1$\pm$13.7  &  704$^{+90}_{-17}$ & 0.12$^{+0.09}_{-0.02}$ & 2.7$^{+0.47}_{-0.07}$ & 4.0$^{+0.4}_{-0.08}$ & 0.84$^{+0.22}_{-0.06}$  \\
GU Sgr  &  54.6$\pm$5.0  &  195.8$\pm$4.2  &  -34.1$\pm$5.1  &  881$^{+68}_{-105}$ & 0.22$^{+0.02}_{-0.02}$ & 3.21$^{+0.2}_{-0.38}$ & 4.89$^{+0.23}_{-0.39}$ & 0.45$^{+0.04}_{-0.03}$  \\
HD 175893$^{\star}$  &  -40.8$\pm$3.0  &  198.5$\pm$4.3  &  -47.7$\pm$3.1  &  884$^{+64}_{-73}$ & 0.14$^{+0.01}_{-0.01}$ & 3.56$^{+0.18}_{-0.17}$ & 4.81$^{+0.25}_{-0.32}$ & 1.23$^{+0.06}_{-0.07}$  \\
MV Sgr  &  -52.9$\pm$28.2  &  -48.6$\pm$37.7  &  46.9$\pm$4.2  &  -95$^{+123}_{-13}$ & 0.57$^{+0.19}_{-0.02}$ & 0.73$^{+0.02}_{-0.31}$ & 2.64$^{+0.43}_{-0.01}$ & 1.03$^{+0.31}_{-0.05}$  \\
RS Tel  &  -0.8$\pm$5.0  &  81.8$\pm$17.4  &  -0.9$\pm$1.5  &  308$^{+70}_{-83}$ & 0.51$^{+0.06}_{-0.01}$ & 1.24$^{+0.08}_{-0.24}$ & 3.62$^{+0.38}_{-0.09}$ & 1.36$^{+0.12}_{-0.2}$  \\
RT Nor  &  14.5$\pm$5.0  &  183.1$\pm$17.3  &  -21.1$\pm$3.7  &  852$^{+54}_{-38}$ & 0.19$^{+0.02}_{-0.01}$ & 3.15$^{+0.2}_{-0.14}$ & 4.84$^{+0.12}_{-0.15}$ & 0.66$^{+0.07}_{-0.05}$  \\
RY Sgr  &  33.8$\pm$4.8  &  263.1$\pm$0.9  &  -47.8$\pm$3.1  &  1762$^{+9}_{-13}$ & 0.19$^{+0.01}_{-0.01}$ & 6.53$^{+0.04}_{-0.12}$ & 9.68$^{+0.06}_{-0.04}$ & 1.15$^{+0.08}_{-0.06}$  \\
S Aps  &  19.2$\pm$3.9  &  241.0$\pm$11.1  &  53.2$\pm$2.7  &  1496$^{+31}_{-37}$ & 0.09$^{+0.01}_{-0.01}$ & 6.17$^{+0.04}_{-0.05}$ & 7.34$^{+0.2}_{-0.24}$ & 1.38$^{+0.1}_{-0.06}$  \\
SV Sge  &  33.4$\pm$4.1  &  222.5$\pm$3.0  &  26.9$\pm$0.8  &  1541$^{+34}_{-33}$ & 0.12$^{+0.02}_{-0.01}$ & 5.89$^{+0.21}_{-0.2}$ & 7.47$^{+0.06}_{-0.04}$ & 0.41$^{+0.01}_{-0.01}$  \\
U Aqr  &  129.7$\pm$14.3  &  148.2$\pm$3.9  &  -142.6$\pm$12.0  &  823$^{+188}_{-210}$ & 0.62$^{+0.10}_{-0.09}$ & 3.53$^{+0.62}_{-0.79}$ & 14.52$^{+1.88}_{-1.1}$ & 10.09$^{+2.81}_{-1.24}$  \\
UV Cas  &  45.1$\pm$0.2  &  216.4$\pm$1.2  &  3.5$\pm$0.2  &  2096$^{+12}_{-15}$ & 0.16$^{+0.01}_{-0.01}$ & 7.95$^{+0.03}_{-0.04}$ & 10.95$^{+0.09}_{-0.18}$ & 0.06$^{+0.003}_{-0.004}$  \\
UW Cen  &  26.3$\pm$6.1  &  223.3$\pm$14.4  &  -38.7$\pm$3.9  &  1554$^{+21}_{-6}$ & 0.076$^{+0.005}_{-0.003}$ & 6.35$^{+0.09}_{-0.04}$ & 7.39$^{+0.17}_{-0.04}$ & 1.01$^{+0.12}_{-0.09}$  \\
V2552 Oph  &  -89.2$\pm$5.0  &  168.8$\pm$4.1  &  -115.0$\pm$6.4  &  694$^{+47}_{-46}$ & 0.34$^{+0.01}_{-0.01}$ & 2.44$^{+0.15}_{-0.06}$ & 5.26$^{+0.15}_{-0.24}$ & 1.54$^{+0.1}_{-0.05}$  \\
V482 Cyg  &  -11.0$\pm$2.3  &  199.7$\pm$2.2  &  -19.4$\pm$0.8  &  1547$^{+10}_{-7}$ & 0.15$^{+0.004}_{-0.004}$ & 5.76$^{+0.05}_{-0.05}$  & 7.79$^{+0.01}_{-0.01}$ & 0.31$^{+0.01}_{-0.01}$ \\
V532 Oph  &  -31.8$\pm$6.2  &  -3.3$\pm$37.2  &  5.5$\pm$0.7  &  -24$^{+84}_{-7}$ & 0.66$^{+0.14}_{-0.09}$ & 0.42$^{+0.04}_{-0.16}$ & 2.38$^{+4.96}_{-0.43}$ & 0.57$^{+2.06}_{-0.05}$  \\
VZ Sgr  &  -268.9$\pm$5.6  &  80.2$\pm$31.6  &  96.4$\pm$14.8  &  91$^{+46}_{-4}$ & 0.87$^{+0.003}_{-0.03}$ & 0.3$^{+0.08}_{-0.02}$ & 4.13$^{+0.7}_{-0.27}$ & 1.0$^{+0.83}_{-0.02}$ \\
WISE J172447.52-290418.6  &  59.3$\pm$9.5  &  109.1$\pm$22.2  &  -59.9$\pm$8.4  &  9$^{+148}_{-5}$ & 0.39$^{+0.25}_{-0.09}$ & 0.6$^{+0.28}_{-0.3}$ & 0.89$^{+0.98}_{-0.04}$ & 0.61$^{+0.13}_{-0.07}$  \\
WISE J172951.80-101715.9$^{\star}$  &  -3.4$\pm$11.8  &  -29.6$\pm$31.7  &  -17.4$\pm$0.8  &  -88$^{+54}_{-5}$ & 0.70$^{+0.08}_{-0.01}$ & 0.62$^{+0.01}_{-0.22}$ & 3.39$^{+0.18}_{-0.37}$ & 1.42$^{+0.23}_{-0.03}$  \\
WISE J174119.57-250621.2$^{\star}$  &  -202.0$\pm$21.3  &  -103.5$\pm$65.5  &  14.1$\pm$0.8  &  -152$^{+272}_{-12}$ & 0.66$^{+0.08}_{-0.04}$ & 0.29$^{+0.32}_{-0.02}$ & 2.36$^{+0.81}_{-0.07}$ & 0.38$^{+0.37}_{-0.03}$  \\
WISE J174851.29-330617.0$^{\star}$  &  136.8$\pm$13.2  &  -92.7$\pm$76.0  &  4.2$\pm$1.8  &  -145$^{+181}_{-14}$ & 0.75$^{+0.09}_{-0.11}$ & 0.57$^{+0.11}_{-0.19}$ & 2.02$^{+0.86}_{-0.05}$ & 0.29$^{+0.29}_{-0.02}$  \\
WISE J175107.12-242357.3$^{\star}$  &  129.2$\pm$5.1  &  135.5$\pm$16.4  &  24.3$\pm$4.0  &  438$^{+171}_{-206}$ & 0.46$^{+0.12}_{-0.04}$ & 1.3$^{+0.43}_{-0.56}$ & 4.24$^{+0.69}_{-1.26}$ & 0.19$^{+0.02}_{-0.01}$  \\
WISE J175521.75-281131.2  &  69.9$\pm$5.1  &  145.5$\pm$22.0  &  117.9$\pm$25.0  &  646$^{+194}_{-192}$ & 0.36$^{+0.05}_{-0.04}$ & 2.35$^{+0.52}_{-0.59}$ & 5.09$^{+0.6}_{-0.94}$ & 1.43$^{+0.12}_{-0.13}$  \\
WISE J175749.76-075314.9  &  -21.2$\pm$22.6  &  80.8$\pm$17.1  &  52.7$\pm$5.3  &  226$^{+55}_{-5}$ & 0.54$^{+0.03}_{-0.05}$ & 0.89$^{+0.17}_{-0.02}$ & 3.19$^{+0.12}_{-0.06}$ & 1.27$^{+0.15}_{-0.07}$  \\
WISE J184158.40-054819.2  &  -53.1$\pm$5.9  &  214.5$\pm$23.5  &  -62.8$\pm$10.5  &  841$^{+123}_{-37}$ & 0.17$^{+0.02}_{-0.05}$ & 3.38$^{+0.19}_{-0.09}$ & 4.82$^{+0.52}_{-0.55}$ & 0.51$^{+0.05}_{-0.03}$  \\
WISE J185525.52-025145.7  &  -5.2$\pm$40.6  &  175.4$\pm$29.4  &  9.3$\pm$1.0  &  689$^{+366}_{-34}$ & 0.22$^{+0.06}_{-0.1}$ & 2.37$^{+1.47}_{-0.14}$ & 4.3$^{+0.94}_{-0.05}$ & 0.36$^{+0.07}_{-0.06}$  \\
XX Cam  &  27.0$\pm$0.4  &  233.1$\pm$0.4  &  3.0$\pm$0.1  &  2184$^{+3}_{-3}$ & 0.095$^{+0.002}_{-0.002}$ & 8.82$^{+0.01}_{-0.01}$ & 10.68$^{+0.03}_{-0.03}$ & 0.07$^{+0.002}_{-0.001}$ \\
Y Mus  &  11.6$\pm$3.8  &  165.3$\pm$9.0  &  -45.6$\pm$3.0  &  1120$^{+16}_{-13}$ & 0.29$^{+0.01}_{-0.01}$ & 3.73$^{+0.07}_{-0.07}$ & 6.79$^{+0.03}_{-0.01}$ & 0.66$^{+0.07}_{-0.04}$  \\
\hline
\multicolumn{9}{c}{\emph{Known Galactic dLHdC stars}}\\
\hline
A182$^{\star}$  &  -30.8$\pm$1.2  &  153.1$\pm$11.5  &  16.5$\pm$1.2  &  666$^{+70}_{-71}$ & 0.31$^{+0.03}_{-0.02}$ & 2.35$^{+0.21}_{-0.24}$ & 4.42$^{+0.18}_{-0.18}$ & 0.75$^{+0.07}_{-0.06}$  \\
A183$^{\star}$  &  8.6$\pm$0.5  &  113.7$\pm$5.6  &  -122.5$\pm$5.2  &  716$^{+26}_{-48}$ & 0.41$^{+0.01}_{-0.01}$ & 2.65$^{+0.06}_{-0.11}$ & 6.3$^{+0.03}_{-0.07}$ & 2.84$^{+0.18}_{-0.12}$  \\
A223  &  -53.2$\pm$2.3  &  146.7$\pm$14.5  &  -51.5$\pm$4.6  &  573$^{+63}_{-66}$ & 0.33$^{+0.002}_{-0.01}$ & 2.09$^{+0.15}_{-0.17}$ & 4.14$^{+0.24}_{-0.31}$ & 0.96$^{+0.06}_{-0.06}$  \\
A226  &  85.7$\pm$9.9  &  120.0$\pm$8.1  &  -40.0$\pm$4.3  &  259$^{+87}_{-92}$ & 0.42$^{+0.2}_{-0.11}$ & 1.25$^{+0.34}_{-0.47}$ & 3.04$^{+0.2}_{-0.01}$ & 1.78$^{+0.28}_{-0.2}$  \\
A249  &  -57.3$\pm$2.2  &  138.3$\pm$11.7  &  14.0$\pm$2.7  &  584$^{+77}_{-120}$ & 0.36$^{+0.06}_{-0.03}$ & 2.69$^{+0.18}_{-0.35}$ & 5.54$^{+0.12}_{-0.03}$ & 3.41$^{+0.43}_{-0.26}$  \\
A770  &  5.0$\pm$14.7  &  186.5$\pm$20.7  &  51.2$\pm$9.8  &  306$^{+122}_{-112}$ & 0.13$^{+0.27}_{-0.01}$ & 1.67$^{+0.18}_{-0.68}$ & 1.96$^{+0.63}_{-0.03}$ & 0.69$^{+0.31}_{-0.06}$  \\
A811  &  67.9$\pm$5.0  &  126.7$\pm$7.3  &  14.1$\pm$0.7  &  518$^{+82}_{-63}$ & 0.45$^{+0.04}_{-0.03}$ & 1.68$^{+0.23}_{-0.21}$ & 4.46$^{+0.25}_{-0.21}$ & 0.53$^{+0.04}_{-0.04}$  \\
A814  &  -46.6$\pm$19.0  &  256.9$\pm$39.9  &  32.1$\pm$3.2  &  1477$^{+399}_{-123}$ & 0.16$^{+0.11}_{-0.01}$ & 6.42$^{+0.43}_{-0.91}$ & 7.08$^{+3.68}_{-0.38}$ & 0.8$^{+0.47}_{-0.13}$  \\
A977  &  124.0$\pm$5.0  &  190.0$\pm$4.2  &  -37.9$\pm$4.2  &  760$^{+114}_{-124}$ & 0.38$^{+0.01}_{-0.02}$ & 2.23$^{+0.37}_{-0.26}$ & 4.81$^{+0.78}_{-0.45}$ & 0.46$^{+0.02}_{-0.03}$  \\
A980  &  -26.4$\pm$24.2  &  263.2$\pm$31.8  &  -40.7$\pm$9.7  &  1047$^{+187}_{-16}$ & 0.22$^{+0.06}_{-0.01}$ & 3.96$^{+0.53}_{-0.13}$ & 6.17$^{+1.55}_{-0.08}$ & 1.77$^{+0.69}_{-0.26}$  \\
B42  &  9.2$\pm$2.1  &  177.8$\pm$9.6  &  -57.0$\pm$9.2  &  167$^{+150}_{-183}$ & 0.21$^{+0.34}_{-0.12}$ & 0.32$^{+0.96}_{-0.14}$ & 1.73$^{+0.3}_{-0.04}$ & 1.29$^{+0.06}_{-0.35}$  \\
B563  &  139.5$\pm$5.0  &  73.3$\pm$15.8  &  -36.3$\pm$2.4  &  184$^{+78}_{-63}$ & 0.69$^{+0.05}_{-0.04}$ & 0.61$^{+0.19}_{-0.18}$ & 3.4$^{+0.43}_{-0.58}$ & 0.83$^{+0.33}_{-0.02}$  \\
B564  &  19.5$\pm$15.6  &  55.7$\pm$23.0  &  127.0$\pm$13.8  &  86$^{+67}_{-16}$ & 0.58$^{+0.07}_{-0.05}$ & 0.54$^{+0.16}_{-0.11}$ & 1.92$^{+0.55}_{-0.02}$ & 1.24$^{+0.06}_{-0.08}$  \\
B565  &  -33.2$\pm$10.4  &  86.2$\pm$23.1  &  -29.3$\pm$5.2  &  85$^{+122}_{-10}$ & 0.45$^{+0.05}_{-0.05}$ & 0.85$^{+0.17}_{-0.26}$ & 1.73$^{+0.65}_{-0.34}$ & 0.71$^{+0.07}_{-0.11}$  \\
B566  &  -14.2$\pm$6.0  &  26.5$\pm$30.5  &  -34.4$\pm$5.6  &  5$^{+99}_{-2}$ & 0.74$^{+0.07}_{-0.1}$ & 0.42$^{+0.12}_{-0.1}$ & 1.52$^{+0.95}_{-0.03}$ & 0.62$^{+0.27}_{-0.11}$  \\
B567  &  112.8$\pm$5.3  &  144.2$\pm$9.0  &  -38.2$\pm$6.1  &  296$^{+133}_{-201}$ & 0.3$^{+0.22}_{-0.03}$ & 0.66$^{+0.66}_{-0.27}$ & 1.91$^{+0.66}_{-0.02}$ & 0.54$^{+0.2}_{-0.05}$  \\
C17  &  -127.8$\pm$17.1  &  124.1$\pm$46.8  &  157.6$\pm$24.4  &  68$^{+429}_{-14}$ & 0.32$^{+0.15}_{-0.004}$ & 0.39$^{+2.21}_{-0.07}$ & 2.32$^{+3.81}_{-0.19}$ & 2.77$^{+2.62}_{-0.7}$  \\
C20$^{\star}$  &  19.8$\pm$20.5  &  147.7$\pm$29.6  &  -24.7$\pm$7.3  &  1052$^{+378}_{-106}$ & 0.36$^{+0.03}_{-0.1}$ & 3.88$^{+1.71}_{-0.53}$ & 8.11$^{+1.18}_{-0.56}$ & 3.09$^{+0.57}_{-0.34}$  \\
C27  &  -17.8$\pm$30.5  &  128.9$\pm$31.9  &  -94.3$\pm$13.5  &  118$^{+358}_{-36}$ & 0.26$^{+0.17}_{-0.12}$ & 0.82$^{+1.63}_{-0.28}$ & 1.8$^{+1.47}_{-0.08}$ & 1.44$^{+0.59}_{-0.14}$  \\
C38  &  -32.4$\pm$24.0  &  128.9$\pm$19.0  &  11.7$\pm$2.1  &  59$^{+84}_{-4}$ & 0.48$^{+0.16}_{-0.12}$ & 0.57$^{+0.17}_{-0.18}$ & 1.07$^{+0.76}_{-0.05}$ & 0.69$^{+0.2}_{-0.12}$  \\
C526  &  -59.1$\pm$14.2  &  196.3$\pm$30.6  &  -22.9$\pm$4.4  &  1327$^{+256}_{-71}$ & 0.24$^{+0.01}_{-0.01}$ & 4.63$^{+0.89}_{-0.27}$ & 7.76$^{+1.19}_{-0.54}$ & 0.69$^{+0.21}_{-0.08}$  \\
C528  &  17.7$\pm$15.3  &  114.0$\pm$17.9  &  -42.0$\pm$5.0  &  135$^{+92}_{-2}$ & 0.34$^{+0.09}_{-0.04}$ & 1.07$^{+0.13}_{-0.26}$ & 1.85$^{+0.43}_{-0.19}$ & 0.76$^{+0.07}_{-0.09}$  \\
C539  &  8.2$\pm$29.3  &  -92.6$\pm$11.7  &  80.0$\pm$9.4  &  -263$^{+71}_{-9}$ & 0.47$^{+0.14}_{-0.01}$ & 1.15$^{+0.03}_{-0.3}$ & 3.07$^{+0.72}_{-0.03}$ & 1.24$^{+0.29}_{-0.28}$  \\
C542  &  32.6$\pm$27.2  &  185.5$\pm$34.0  &  -43.7$\pm$8.7  &  1211$^{+447}_{-132}$ & 0.23$^{+0.03}_{-0.08}$ & 4.43$^{+2.02}_{-0.63}$ & 6.96$^{+1.66}_{-0.54}$ & 1.03$^{+0.51}_{-0.2}$  \\
HD 137613  &  -83.6$\pm$1.8  &  184.3$\pm$1.4  &  8.6$\pm$1.4  &  1319$^{+19}_{-19}$ & 0.33$^{+0.01}_{-0.01}$ & 4.21$^{+0.08}_{-0.07}$ & 8.29$^{+0.07}_{-0.06}$ & 0.66$^{+0.02}_{-0.01}$  \\
HD 148839$^{\star}$  &  -4.0$\pm$1.8  &  210.5$\pm$1.7  &  29.3$\pm$0.7  &  1459$^{+19}_{-18}$ & 0.09$^{+0.01}_{-0.01}$ & 5.74$^{+0.11}_{-0.11}$ & 6.97$^{+0.03}_{-0.03}$ & 0.57$^{+0.01}_{-0.01}$  \\
HD 173409  &  56.9$\pm$2.0  &  221.5$\pm$0.6  &  -9.5$\pm$1.1  &  1375$^{+19}_{-20}$ & 0.18$^{+0.01}_{-0.01}$ & 4.99$^{+0.06}_{-0.09}$ & 7.17$^{+0.09}_{-0.09}$ & 0.49$^{+0.02}_{-0.02}$  \\
HD 182040  &  46.2$\pm$1.8  &  244.7$\pm$0.9  &  -6.8$\pm$0.6  &  1820$^{+8}_{-7}$ & 0.16$^{+0.004}_{-0.003}$ & 6.78$^{+0.05}_{-0.05}$ & 9.38$^{+0.02}_{-0.02}$ & 0.21$^{+0.01}_{-0.01}$  \\
M38  &  47.3$\pm$2.3  &  205.1$\pm$27.8  &  -20.4$\pm$3.0  &  1155$^{+78}_{-8}$ & 0.19$^{+0.003}_{-0.02}$ & 4.1$^{+0.31}_{-0.03}$ & 6.09$^{+0.22}_{-0.02}$ & 0.41$^{+0.07}_{-0.04}$  \\
P12  &  -58.1$\pm$17.7  &  154.7$\pm$19.5  &  -99.6$\pm$17.8  &  364$^{+152}_{-8}$ & 0.24$^{+0.14}_{-0.06}$ & 1.46$^{+0.6}_{-0.07}$ & 3.27$^{+0.33}_{-0.02}$ & 0.74$^{+0.32}_{-0.07}$  \\
\hline
\multicolumn{9}{l}{$^{\star}$ Other names used in \citet{2023MNRAS.tmp.3151M}:}\\
\multicolumn{9}{l}{A182 = C* 2277, A183 = SOPS IV e-67, ASAS-RCB-10 = ASAS J171710-2043.3, C20 = C* 2891,}\\
\multicolumn{9}{l}{HD 175893 = V4152 Sgr, HD 148839 = LV TrA, WISE J172951.80-101715.9 = AC Ser,}\\ 
\multicolumn{9}{l}{WISE J174119.57-250621.2 = UCAC4 325-115052, WISE J174851.29-330617.0 = 2MASS J17485129-3306172,}\\
\multicolumn{9}{l}{WISE J175107.12-242357.3 = 2MASS J17510712-2423573.}\\
\hline
\end{longtable}


\begin{table*}[!htbp]
\caption{Velocities in cylindrical Galactocentric coordinates and dynamical orbital parameter values of DYPer type and EHe stars
\label{tab.OrbitGalpy2}}
\medskip
\centering
\begin{tabular}{lrrrrrrrr}
\hline
Name & \multicolumn{1}{c}{V$_r$} & \multicolumn{1}{c}{V$_\varphi$} & \multicolumn{1}{c}{V$_z$} & \multicolumn{1}{c}{L$_z$} & \multicolumn{1}{c}{e} & \multicolumn{1}{c}{R$_{peri}$} & \multicolumn{1}{c}{$R_{apo}$} & \multicolumn{1}{c}{Z$_{max}$} \\ 
 & \multicolumn{1}{c}{km s$^{-1}$} & \multicolumn{1}{c}{km s$^{-1}$}  & \multicolumn{1}{c}{km s$^{-1}$} & \multicolumn{1}{c}{kpc km s$^{-1}$} & & \multicolumn{1}{c}{kpc}  & \multicolumn{1}{c}{kpc}  &\multicolumn{1}{c}{ kpc} \\
\hline
\multicolumn{9}{c}{}\\
\multicolumn{9}{c}{\emph{Known Galactic DYPer type stars}}\\
\hline
ASAS-DYPer-1  &  15.6$\pm$5.8  &  241.6$\pm$19.2  &  0.8$\pm$1.2  &  1976$^{+163}_{-159}$ & 0.06$^{+0.08}_{-0.01}$ & 8.13$^{+0.04}_{-0.76}$ & 8.46$^{+1.9}_{-0.08}$ & 0.08$^{+0.01}_{-0.01}$  \\
ASAS-DYPer-2  &  5.9$\pm$3.8  &  237.0$\pm$5.4  &  7.0$\pm$0.3  &  1681$^{+10}_{-7}$ & 0.029$^{+0.003}_{-0.002}$ & 7.06$^{+0.02}_{-0.02}$ & 7.48$^{+0.06}_{-0.04}$ & 0.34$^{+0.02}_{-0.01}$ \\
DY Per  &  -11.7$\pm$3.3  &  206.1$\pm$3.8  &  11.9$\pm$0.4  &  1902$^{+33}_{-24}$ & 0.12$^{+0.01}_{-0.02}$ & 7.37$^{+0.3}_{-0.19}$ & 9.34$^{+0.04}_{-0.04}$ & 0.21$^{+0.01}_{-0.01}$ \\
\hline
\multicolumn{9}{c}{}\\
\multicolumn{9}{c}{\emph{Known Galactic EHe stars}}\\
\hline
BX Cir  &  31.8$\pm$5.4  &  266.9$\pm$7.0  &  82.8$\pm$4.3  &  1761$^{+36}_{-41}$  &  0.22$^{+0.03}_{-0.02}$  &  6.48$^{+0.03}_{-0.05}$  &  10.08$^{+0.51}_{-0.41}$  &  1.91$^{+0.12}_{-0.12}$ \\
DN Leo$^{\star}$  &  88.7$\pm$7.8  &  119.8$\pm$9.8  &  73.1$\pm$9.4  &  1060$^{+39}_{-45}$  &  0.53$^{+0.04}_{-0.03}$  &  3.19$^{+0.18}_{-0.26}$  &  10.48$^{+0.27}_{-0.2}$  &  3.01$^{+0.26}_{-0.13}$ \\
FQ Aqr  &  -6.1$\pm$2.0  &  189.9$\pm$10.5  &  -58.0$\pm$4.6  &  1193$^{+47}_{-23}$  &  0.12$^{+0.01}_{-0.01}$  &  5.26$^{+0.13}_{-0.1}$  &  6.71$^{+0.04}_{-0.01}$  &  2.36$^{+0.25}_{-0.2}$ \\
LSE 78$^{\star}$  &  4.0$\pm$6.0  &  220.5$\pm$13.7  &  86.2$\pm$13.9  &  876$^{+140}_{-228}$  &  0.06$^{+0.13}_{-0.02}$  &  4.09$^{+0.25}_{-1.11}$  &  3.90$^{+0.65}_{-0.1}$  &  1.17$^{+0.21}_{-0.06}$ \\
LS IV +06 2  &  -10.4$\pm$4.8  &  176.8$\pm$4.5  &  -7.4$\pm$1.3  &  1212$^{+47}_{-37}$  &  0.26$^{+0.02}_{-0.02}$  &  4.06$^{+0.22}_{-0.17}$  &  6.98$^{+0.05}_{-0.08}$  &  0.38$^{+0.02}_{-0.02}$ \\
LSS 99  &  7.8$\pm$2.7  &  170.0$\pm$4.4  &  -5.5$\pm$0.6  &  2216$^{+51}_{-68}$  &  0.25$^{+0.01}_{-0.01}$  &  7.79$^{+0.22}_{-0.32}$  &  12.85$^{+0.36}_{-0.29}$  &  0.42$^{+0.03}_{-0.03}$ \\
LSS 4357  &  186.1$\pm$5.1  &  143.2$\pm$2.3  &  117.9$\pm$15.8  &  251$^{+181}_{-199}$  &  0.52$^{+0.21}_{-0.06}$  &  0.91$^{+0.5}_{-0.46}$  &  3.81$^{+0.47}_{-0.21}$  &  1.07$^{+0.48}_{-0.03}$ \\
NO Ser  &  34.7$\pm$1.7  &  218.4$\pm$1.0  &  -10.7$\pm$0.8  &  1277$^{+35}_{-24}$  &  0.12$^{+0.01}_{-0.01}$  &  4.90$^{+0.15}_{-0.12}$  &  6.35$^{+0.05}_{-0.1}$  &  0.54$^{+0.02}_{-0.03}$ \\
PV Tel  &  145.1$\pm$5.4  &  227.1$\pm$14.2  &  34.5$\pm$3.4  &  1138$^{+87}_{-78}$  &  0.42$^{+0.01}_{-0.01}$  &  3.63$^{+0.19}_{-0.17}$  &  9.03$^{+0.56}_{-0.63}$  &  2.32$^{+0.13}_{-0.12}$ \\
V354 Nor$^{\star}$  &  75.8$\pm$33.5  &  107.3$\pm$34.3  &  -42.9$\pm$6.8  &  497$^{+155}_{-156}$  &  0.47$^{+0.16}_{-0.11}$  &  1.66$^{+0.56}_{-0.58}$  &  4.51$^{+0.26}_{-0.03}$  &  0.94$^{+0.16}_{-0.16}$ \\
V821 Cen$^{\star}$  &  33.9$\pm$4.3  &  265.7$\pm$3.4  &  5.7$\pm$1.5  &  1899$^{+38}_{-34}$  &  0.20$^{+0.02}_{-0.02}$  &  6.95$^{+0.05}_{-0.05}$  &  10.15$^{+0.48}_{-0.39}$  &  0.53$^{+0.03}_{-0.03}$ \\
V1920 Cyg  &  -6.3$\pm$6.3  &  152.4$\pm$6.7  &  -67.4$\pm$4.6  &  1196$^{+44}_{-46}$  &  0.32$^{+0.03}_{-0.03}$  &  4.08$^{+0.21}_{-0.27}$  &  7.81$^{+0.08}_{-0.04}$  &  1.66$^{+0.13}_{-0.19}$ \\
V2076 Oph  &  -71.1$\pm$4.9  &  250.4$\pm$0.8  &  72.1$\pm$2.9  &  1619$^{+26}_{-28}$  &  0.25$^{+0.02}_{-0.02}$  &  5.71$^{+0.08}_{-0.08}$  &  9.57$^{+0.24}_{-0.31}$  &  1.49$^{+0.08}_{-0.06}$ \\
V2205 Oph  &  188.8$\pm$13.0  &  219.7$\pm$13.2  &  89.5$\pm$29.6  &  521$^{+208}_{-532}$  &  0.71$^{+0.01}_{-0.12}$  &  1.92$^{+0.75}_{-0.04}$  &  9.12$^{+3.67}_{-0.32}$  &  7.18$^{+5.18}_{-1.73}$ \\
V2244 Oph  &  -42.4$\pm$4.9  &  121.4$\pm$15.2  &  -88.3$\pm$7.8  &  479$^{+61}_{-28}$  &  0.39$^{+0.05}_{-0.01}$  &  2.06$^{+0.05}_{-0.28}$  &  4.68$^{+0.07}_{-0.2}$  &  1.71$^{+0.12}_{-0.15}$ \\
V4732 Sgr$^{\star}$  &  51.5$\pm$0.6  &  233.6$\pm$1.0  &  8.6$\pm$0.5  &  1123$^{+62}_{-96}$  &  0.16$^{+0.03}_{-0.03}$  &  4.00$^{+0.37}_{-0.33}$  &  5.57$^{+0.16}_{-0.15}$  &  0.67$^{+0.07}_{-0.07}$ \\
\hline
\multicolumn{9}{c}{}\\
\multicolumn{9}{c}{\emph{Other Galactic EHe stars studied by \citet{2023MNRAS.tmp.3151M}}}\\
\hline
BD$+$37 442               &  -103.9 $\pm$ 16.1  &  206.7 $\pm$ 17.0  &  0.3 $\pm$ 10.1  &  1851$^{+131}_{-136}$  &  0.34$^{+0.07}_{-0.06}$  &  5.81$^{+0.79}_{-0.7}$  &  12.02$^{+0.41}_{-0.27}$  &  0.47$^{+0.11}_{-0.04}$ \\
BD$+$37 1977              &  -133.3 $\pm$ 10.0  &  167.8 $\pm$ 9.3  &  1.3 $\pm$ 9.2  &  1574$^{+39}_{-76}$  &  0.48$^{+0.04}_{-0.03}$  &  4.54$^{+0.17}_{-0.31}$  &  12.53$^{+0.63}_{-0.3}$  &  1.52$^{+0.34}_{-0.14}$ \\
PG 1415$+$492             &  -47.0 $\pm$ 5.4  &  236.7 $\pm$ 4.5  &  78.3 $\pm$ 2.1  &  1973$^{+13}_{-14}$  &  0.18$^{+0.004}_{-0.003}$  &  8.46$^{+0.07}_{-0.04}$  &  12.20$^{+0.16}_{-0.15}$  &  4.84$^{+0.35}_{-0.33}$ \\
V652 Her                &  35.5 $\pm$ 1.9  &  243.1 $\pm$ 0.3  &  41.4 $\pm$ 2.5  &  1728$^{+25}_{-20}$  &  0.15$^{+0.01}_{-0.01}$  &  6.67$^{+0.15}_{-0.12}$  &  9.08$^{+0.04}_{-0.04}$  &  1.4$^{+0.07}_{-0.13}$ \\
LSS 5121                &  116.3 $\pm$ 5.9  &  191.4 $\pm$ 1.2  &  61.7 $\pm$ 4.4  &  714$^{+76}_{-98}$  &  0.35$^{+0.04}_{-0.04}$  &  2.48$^{+0.23}_{-0.34}$  &  4.98$^{+0.21}_{-0.24}$  &  1.01$^{+0.09}_{-0.06}$ \\
GALEX J184559.8-413827  &  -5.4 $\pm$ 5.8  &  259.5 $\pm$ 1.7  &  19.1 $\pm$ 2.4  &  840$^{+174}_{-392}$  &  0.19$^{+0.23}_{-0.02}$  &  3.40$^{+ 0.53}_{-1.29}$  &  5.08$^{+0.44}_{-0.06}$  &  2.0$^{+1.26}_{-0.32}$ \\
GALEX J191049.5-441713  &  -18.5 $\pm$ 2.8  &  219.7 $\pm$ 3.1  &  17.7 $\pm$ 1.8  &  1521$^{+27}_{-63}$  &  0.08$^{+0.01}_{-0.01}$  &  6.08$^{+0.14}_{-0.31}$  &  7.13$^{+0.09}_{-0.19}$  &  0.58$^{+0.07}_{-0.03}$ \\
EC 19529-4430           &  -30.4 $\pm$ 8.9  &  -90.7 $\pm$ 53.5  &  92.9 $\pm$ 14.4  &  -514$^{+288}_{-13}$  &  0.31$^{+0.32}_{-0.01}$  &  0.45$^{+1.45}_{-0.1}$  &  5.04$^{+0.49}_{-0.04}$  &  3.06$^{+0.57}_{-0.42}$ \\
EC 20236-5703           &  129.0 $\pm$ 8.2  &  60.9 $\pm$ 42.1  &  34.8 $\pm$ 3.2  &  397$^{+125}_{-183}$  &  0.68$^{+0.1}_{-0.07}$  &  0.91$^{+0.45}_{-0.33}$  &  6.74$^{+0.13}_{-0.06}$  &  3.5$^{+1.19}_{-0.47}$ \\
BPS CS 22940-0009       &  1.4 $\pm$ 3.4  &  135.0 $\pm$ 7.8  &  -22.3 $\pm$ 1.4  &  896$^{+60}_{-61}$  &  0.40$^{+0.03}_{-0.03}$  &  2.75$^{+0.24}_{-0.18}$  &  6.75$^{+0.08}_{-0.09}$  &  1.31$^{+0.09}_{-0.08}$ \\
\hline
\multicolumn{9}{l}{$^{\star}$ Other names used in \citet{2023MNRAS.tmp.3151M}: V4732 Sgr = LS IV-14 109, V821 Cen = HD 124448,}\\ 
\multicolumn{9}{l}{V354 Nor =  CoD-48 10153, DN Leo = BD+10 2179, LSE 78 = CoD-46 11775.}\\ 
\hline
\end{tabular}
\end{table*}

\end{appendix}

\end{document}